\documentclass[pdflatex, sn-standardnature, referee]{sn-jnl}
\theoremstyle{thmstyleone}

\theoremstyle{thmstyletwo}

\theoremstyle{thmstylethree}

\usepackage{amsmath}
\usepackage[nomarkers,figuresonly, noheads, nofiglist]{}
\DeclareUnicodeCharacter{2009}{\,}
\usepackage{pdfpages}

\begin{document}

\title[ ]{Non-Drude THz conductivity of graphene due to structural distortions}

\author[1,2]{\fnm{Tan-Phat} \sur{Nguyen}}
\equalcont{These authors contributed equally to this work.}

\author[3,4]{\fnm{Mykhailo} \sur{Klymenko}}
\equalcont{These authors contributed equally to this work.}

\author[1,2]{\fnm{Gary} \sur{Beane}}
\author[1,2]{\fnm{Mitko} \sur{Oldfield}}
\author[1]{\fnm{Kaijian} \sur{Xing}}
\author[1,2]{\fnm{Matthew} \sur{Gebert}}
\author[1,2,5]{\fnm{Semonti} \sur{Bhattacharyya}}
\author[1,2]{\fnm{Michael S.} \sur{Fuhrer}}
\author*[3,4,6]{\fnm{Jared H.} \sur{Cole}}\email{jared.cole@rmit.edu.au}
\author*[1,2]{\fnm{Agustin} \sur{Schiffrin}}\email{agustin.schiffrin@monash.edu}

\affil[1]{\orgdiv{School of Physics and Astronomy}, \orgname{Monash University}, \orgaddress{\city{Clayton},\state{VIC 3800}, \country{Australia}}}

\affil[2]{\orgdiv{ARC Centre of Excellence in Future Low-Energy Electronics Technologies}, \orgname{Monash University}, \orgaddress{\city{Clayton}, \state{VIC 3800}, \country{Australia}}}

\affil[3]{\orgdiv{School of Science}, \orgname{RMIT University}, \orgaddress{\city{Melbourne},  \state{VIC 3001}, \country{Australia}}}

\affil[4]{\orgdiv{ARC Centre of Excellence in Exciton Science}, \orgname{RMIT University}, \orgaddress{\city{Melbourne}, \postcode{3001}, \state{VIC 3001}, \country{Australia}}}

\affil[5]{\orgdiv{Huygens-Kamerlingh Onnes Laboratory}, \orgname{Leiden University}, \orgaddress{\city{PO Box 9504, 2300RA Leiden}, \country{The Netherlands}}}

\affil[6]{\orgdiv{ARC Centre of Excellence in Future Low-Energy Electronics Technologies}, \orgname{RMIT University}, \orgaddress{\city{Melbourne}, \state{VIC 3001}, \country{Australia}}}

\abstract{
The remarkable electrical, optical and mechanical properties of graphene make it a desirable material for electronics, optoelectronics and quantum applications. A fundamental understanding of the electrical conductivity of graphene across a wide frequency range is required for the development of such technologies. In this study, we use terahertz (THz) time-domain spectroscopy to measure the complex dynamic conductivity of electrostatically gated graphene, in a broad $\sim$0.1 - 7 THz frequency range. The conductivity of doped graphene follows the conventional Drude model, and is predominantly governed by intraband processes. In contrast, undoped charge-neutral graphene exhibits a THz conductivity that significantly deviates from Drude-type models. Via quantum kinetic equations and density matrix theory, we show that this discrepancy can be explained by additional interband processes, that can be exacerbated by electron backscattering. We propose a mechanism where such backscattering -- which involves flipping of the electron pseudo-spin -- is mediated by the substantial vector scattering potentials that are associated with structural deformations of graphene. Our findings highlight the significant impact that structural distortions and resulting electrostatic vector scattering potentials can have on the THz conductivity of charge-neutral graphene. Our results emphasise the importance of the planar morphology of graphene for its broadband THz electronic response.
}


\maketitle


Atomically thin, single-layer graphene can host a profusion of exotic electronic phenomena, owing to its two-dimensional (2D) morphology and unique gapless electronic band structure composed of two overlapping, linearly dispersive Dirac bands \cite{Novoselov_2004, CastroNeto_2009}. Large-scale manufacturing processes \cite{Kim_2009, Bae_2010} (e.g., chemical vapour deposition (CVD), epitaxial growth), and the ability to tune charge carrier density and electrical conductivity by an applied gate voltage \cite{Novoselov_2004}, have enabled graphene as an active material in many solid-state technologies, such as high-mobility flexible electronic devices, spintronic systems and super-capacitors for energy storage \cite{Chen_2007, Kamalakar_2015, El-Kady_2016}. In particular, its zero-energy bandgap and tunable carrier density make graphene suitable for applications in optics and optoelectronics, within an extensive spectral range from the visible \cite{Falkovsky_2008} (wavelengths ${\sim}380-750$ nm) to the far infrared \cite{Hendry_2010} ($\sim$1 mm). Notably, its tunable electronic response to an external electromagnetic field oscillating at frequencies in the terahertz (THz) spectral window (${\sim}0.1-10$ THz, with wavelengths ${\sim}0.01 - 1$ mm), makes graphene attractive for THz technologies, i.e., THz modulators, detectors, sensors and biomedical imaging systems \cite{Zheng_2020, Jin_2022}.  

The electrical response of graphene to an incident electric field, $\mathbf{F}_{\textrm{THz}}^{(\textrm{inc})}(t) = \frac {1}{\sqrt{2\pi}} \int \Tilde{\mathbf{F}}_{\textrm{THz}}^{(\textrm{inc})}(\omega) \mathrm{e}^{i\omega t} \mathrm{d}\omega$, oscillating at THz frequencies is encoded in its frequency-dependent complex dynamic conductivity $\Tilde{\sigma}(\omega)$, which is governed by the frequency-dependent polarisation and absorption of the material, with the resulting complex current density (assuming the medium is isotropic) $\Tilde{\mathbf{J}}(\omega)= \Tilde{\sigma}(\omega) \cdot \Tilde{\mathbf{F}}_{\textrm{THz}}^{(\textrm{inc})}(\omega)$. Conventionally, $\Tilde{\sigma}(\omega)$ can be approximated within the few-THz spectral range (i.e., ${\sim}0.1 - 2$ THz) \cite{Buron_2012, Buron_2015, Scarfe_2021, Frenzel_2013, Jnawali_2013, Pistore_2022} -- and even up to the infrared when only the real part of $\Tilde{\sigma}(\omega)$ is considered \cite{Horng_2011, Ren_2012} -- by the Drude model, as $\Tilde{\sigma}_\textrm{Drude} (\omega)= \sigma_0/(1 - i\omega\tau)$, where $\sigma_0$ is the zero-frequency (DC) conductivity and $\tau$ is the average time between electron scattering events in the material (due to, e.g., defects, impurities, phonons, other charge carriers). In this Drude model, it is assumed that $\Tilde{\sigma}(\omega)$ is determined solely by intraband transitions of free electrons, that is, within a quadratically dispersive conduction band \cite{Buron_2012, Frenzel_2013, Jnawali_2013, Buron_2015, Scarfe_2021}. The Drude model omits the actual band structure of the material -- in the case of graphene, its linearly dispersive Dirac bands -- and transitions between different bands (i.e., interband). Extensions to the Drude model, such as the Drude-Smith \cite{Smith_2001, Buron_2014, Cocker_2017}, Drude-Lorentz \cite{Schubert_2004, Chen_2019} or localisation-modified Drude models \cite{Lee_1993, Lee_1995}, include phenomenological corrections. These corrections can account for carrier backscattering due to disorder in the case of Drude-Smith and localisation-modified Drude models (where disorder causes weak localisation, and the suppression of both long-range carrier transport and low-frequency conductivity), or for resonances at specific frequencies (due to, e.g., phonons, interband transitions) in the case of the Drude-Lorentz model. Further modifications to the Drude model can also take into account the specific graphene band structure \cite{Ando_2006}. These Drude-type models can describe the dynamic conductivity of graphene in specific cases, e.g., metallic graphene (i.e., significantly doped, with the Fermi level $E_\textrm{f}$ within the valence or conduction bands, far from the Dirac point), when only the real part of $\Tilde{\sigma}(\omega)$ is considered \cite{Horng_2011, Ren_2012, Buron_2012, Buron_2015, Scarfe_2021}, or when full complex conductivity is measured but within a narrow THz bandwidth (typically not larger than ${\sim}2$ THz) \cite{Frenzel_2013, Jnawali_2013, Pistore_2022}. However, they do not provide accurate predictions for the full complex conductivity (i.e., real and imaginary components) of charge-neutral graphene (i.e., with $E_\textrm{f}$ at the Dirac point) within a broad THz spectral window (see Supplementary Note 4 in Supplementary Information [SI]). It is specifically with $E_\textrm{f}$ at the Dirac point that the actual band structure of graphene (i.e., linear Dirac dispersion instead of free-electron quadratic dispersion) and interband transitions play a significant role \cite{Hafez_2020, Gallagher_2019, Sensale-Rodríguez_2013}. To our knowledge, both the real and imaginary parts of graphene's $\Tilde{\sigma}(\omega)$, within the full ${\sim}0.1 - 10$ THz range, for different charge carrier densities (i.e., from doped to charge neutral), have not yet been measured or quantitatively modelled. 

In this work, we retrieved $\Tilde{\sigma}\left( \omega \right)$ -- both real and imaginary parts -- of graphene at room temperature, for different Fermi levels $E_\textrm{f}$, i.e., from $n-$ to $p-$doped metallic behaviour via the charge neutrality Dirac point (CNP), within the broad ${\sim}0.1 - 7$ THz spectral range (i.e., characteristic energies $\hbar\omega \approx 1 - 30$ meV). To do so, we measured via THz time-domain spectroscopy \cite{Tonouchi_2007, Jepsen_2011} (THz-TDS; see Methods and SI Supplementary Note 2) the time-dependent THz waveform $\mathbf{F}_{\textrm{THz}}^{(\textrm{trans})}(t)$ transmitted through a sample consisting of CVD-grown single-layer graphene on a SiO$_2$/$p-$doped Si substrate. We controlled the graphene Fermi level $E_\textrm{f}$ (and hence the carrier density) by a gate voltage $V_\textrm{g}$ applied between graphene and doped Si (Fig. \ref{fig1}a). We found that when graphene is substantially doped, with $E_\textrm{f}$ far from the CNP (i.e., metallic behaviour), $\Tilde{\sigma}\left( \omega \right)$ follows the conventional Drude model. Conversely, when graphene is charge-neutral with $E_\textrm{f}$ at the CNP, $\Tilde{\sigma}\left( \omega \right)$ exhibits a very significant deviation from Drude-type models, with a notable suppression of its imaginary part, Im$\left[\Tilde{\sigma}\left(\omega \right)\right]$. To explain this deviation, we developed a model for $\Tilde{\sigma}\left( \omega \right)$ based on density matrix theory \cite{Haug_2004} and quantum kinetic equations \cite{Kitamura_2015, Culcer_2017, Culcer_2020}, including both intraband and interband electronic processes, with the latter being accentuated by scattering of carriers. We propose a mechanism where such scattering -- involving flipping of the electron pseudo-spin -- is mediated by the substantial electrostatic vector scattering potentials that originate from structural deformations of graphene. The very good agreement between theory and experiment, for both Re$\left[\Tilde{\sigma}\left(\omega\right)\right]$ and Im$\left[\Tilde{\sigma}\left(\omega\right)\right]$, within a broad ${\sim}0.1 - 7$ THz frequency range, highlights the impact that structural distortions can have on the broadband THz conductivity of charge-neutral graphene. Our results provide compelling evidence of a previously unreported phenomenon, whose observation and understanding are enabled by our ability to combine both tuning of graphene's Fermi level and full retrieval of $\Tilde{\sigma}\left(\omega\right)$ within the broad ${\sim}0.1 - 7$ THz window.
\section*{Results}

\textbf{Retrieval of complex THz conductivity $\Tilde{\sigma}\left(\omega \right)$ of graphene} 

\begin{figure}
    \centering
    \includegraphics[width=0.75\linewidth]{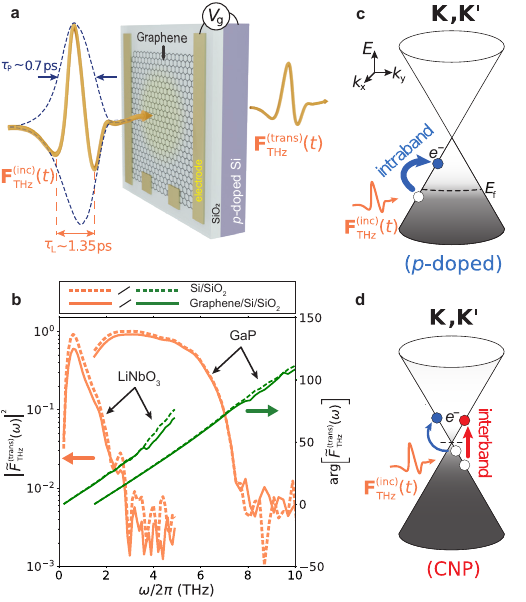}
    \caption{
        \textbf{THz time-domain spectroscopy of gated graphene.} 
        \textbf{a,} Schematic of device consisting of single-layer CVD-grown graphene on 300 nm thick SiO$_2$ on $p-$doped Si, irradiated with a THz waveform with instantaneous electric field $\mathbf{F}_{\textrm{THz}}^{\textrm{(inc)}}(t)$ (generated via optical rectification, here in LiNbO$_{3}$; pulse duration $\tau_\textrm{P} \approx 0.7$ ps; period $\tau_\textrm{L} \approx 1.35$ ps). Gold electrodes allow for tuning of the graphene Fermi level $E_\textrm{f}$ by a gate voltage $V_\textrm{g}$, and for DC electrical conductivity characterisation via four-point-probe measurements. 
        \textbf{b,} Spectra $\vert \Tilde{F}_{\textrm{THz}}^{\textrm{(trans)}}(\omega)\vert ^2$ and spectral phases arg$\left[  \Tilde{F}_{\textrm{THz}}^{\textrm{(trans)}}(\omega) \right]$ [$\Tilde{F}_{\textrm{THz}}^{\textrm{(trans)}}(\omega)$: Fourier transform of $F_{\textrm{THz}}^{\textrm{(trans)}}(t)$] of THz waveforms transmitted through SiO$_{2}$/Si (bare substrate; reference) or graphene/SiO$_{2}$/Si heterostructure areas, for both LiNbO$_3$- and GaP-generated THz. These spectra enable the retrieval of the complex THz conductivity, $\tilde{\sigma}(\omega)$, as a function of $\omega$ and $V_\textrm{g}$. \textbf{c, d,} Schematic of graphene band structure (Dirac cones) at \textbf{K} and \textbf{K'} points in reciprocal space, for different values of $E_\textrm{f}$ tuned via $V_\textrm{g}$: $E_\textrm{f} (V_\textrm{g}) < E_\textrm{CNP}$ [i.e., $p-$doped; (c)] and $E_\textrm{f} (V_\textrm{g}) = E_\textrm{CNP}$ [i.e., charge neutrality point (CNP) with charge carrier density $n (V_\textrm{g}) \approx 0$; (d)]. Different THz-induced electronic excitations (e.g., intraband and/or interband) affecting $\tilde{\sigma}(\omega)$ can occur depending on $E_\textrm{f}$.
    }
    \label{fig1}
\end{figure}

The spectra, $\vert \Tilde{F}_{\textrm{THz}}^{\textrm{(trans)}}(\omega)\vert ^2$, and spectral phases, arg$\left[  \Tilde{F}_{\textrm{THz}}^{\textrm{(trans)}}(\omega) \right]$, of the time-domain THz waveform $\mathbf{F}_{\textrm{THz}}^{\textrm{(trans)}}(t)$ transmitted through SiO$_{2}$/Si (bare substrate reference) or graphene/SiO$_{2}$/Si areas are shown in Fig. \ref{fig1}b  (see Methods and SI Supplementary Note 1 for sample fabrication and characterisation). These measurements rely on THz waveforms generated via optical rectification \cite{Hirori_2011} in lithium niobate (LiNbO$_3$) and gallium phosphide (GaP) nonlinear crystals, resulting in electric field Fourier components within the frequency range ${\sim} 0.1 - 1.5$ THz and up to ${\sim} 7$ THz, respectively. By using both LiNbO$_3$ and GaP THz generation configurations in turn, we can measure the complex transmission $\Tilde{T}(\omega)$ of the incident THz waveform $\mathbf{F}_{\textrm{THz}}^{\textrm{(inc)}}(t)$ through graphene within a broad continuous ${\sim} 0.1 - 7$ THz band. The measurement of $\Tilde{T}(\omega)$ allows for the calculation of $\Tilde{\sigma}\left(\omega \right)$ for different values of $E_\textrm{f}$ controlled via $V_\textrm{g}$ (see Methods).

\bigskip\noindent
\textbf{Gate-voltage-dependence of $\Tilde{\sigma}\left(\omega \right)$: from Drude to non-Drude behaviour}  

\begin{figure}
    \centering
    \includegraphics[width=1.0\textwidth]{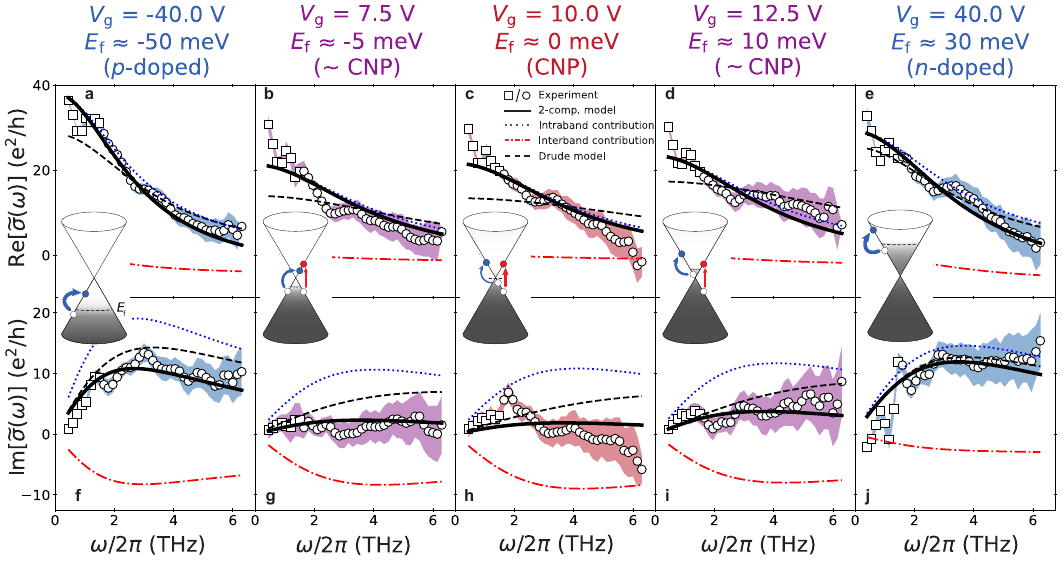}
    \caption{\textbf{Gate-controlled complex THz conductivity of graphene: from Drude to non-Drude behaviour.} \textbf{a - e,} Real part of complex THz conductivity, $\tilde{\sigma}(\omega)$, for different gate voltages $V_\textrm{g}$ tuning the Fermi level $E_{\textrm{f}}$. \textbf{f - j,} Imaginary part of $\tilde{\sigma}(\omega)$, for different gate voltages $V_\textrm{g}$. Square and circle markers: experimental data from LiNbO$_3$ and GaP THz generation configurations, respectively. Solid curves: two-component model fit including contributions from intra- (dotted blue) and interband (dash-dotted red) transitions. Black dashed curves: Drude model fit. Filled areas: $\pm$ experimental standard deviation. Insets: schematic of graphene Dirac cones with different $E_\textrm{f}$ values, with intra- and interband transitions triggered by incident THz waveform. As $E_\textrm{f}$ approaches the charge neutrality point (i.e., $\vert V_\textrm{g} - V_\textrm{CNP} \vert \rightarrow 0$), contributions of interband transitions to $\tilde{\sigma}(\omega)$ increase: graphene's behaviour changes from metal- (Drude) to semiconductor (non-Drude)-like.
}
    \label{fig2}
\end{figure}

The real, $\textrm{Re}\left[\Tilde{\sigma}\left( \omega \right)\right]$, and imaginary, $\textrm{Im}\left[\Tilde{\sigma}\left( \omega \right)\right]$, components of  $\Tilde{\sigma}\left( \omega \right)$ of graphene are shown in Fig. \ref{fig2}, within the spectral range ${\sim} 0.1 - 7$ THz. We applied different gate voltages, varying from heavily $p-$doped ($V_\textrm{g}=-40$ V, i.e., $E_\textrm{f}\approx-50$ meV) to charge-neutral ($V_\textrm{g}=10$ V; $E_\textrm{f}\approx0$ meV) to heavily $n-$doped ($V_\textrm{g}=40$ V; $E_\textrm{f}\approx30$ meV) graphene ($E_\textrm{f}$ given relative to CNP; see SI Supplementary Note 1 for the relationship between $V_\textrm{g}$ and $E_\textrm{f}$). Note that, the values of $\textrm{Re}\left[\Tilde{\sigma}\left( \omega \right)\right]$ and $\textrm{Im}\left[\Tilde{\sigma}\left( \omega \right)\right]$ in this frequency range are consistent with previous THz-TDs studies of graphene \cite{Buron_2014, Whelan_2020}. 

\begin{figure}
    \centering
    \includegraphics[width=1.0\linewidth]{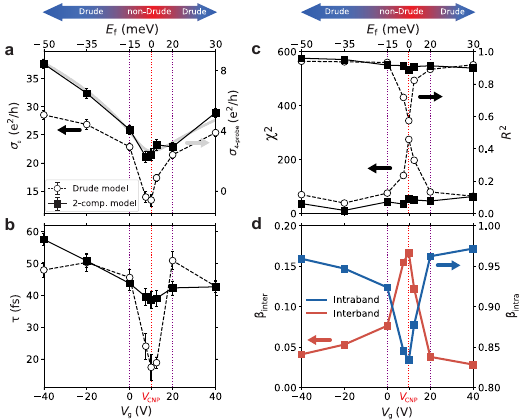}
    \caption{\textbf{Two-component model: substantial contribution of interband transitions at the CNP.} \textbf{a, b,} DC conductivity, $\sigma_0$ (left axis), and carrier intraband relaxation time constant, $\tau$, as a function of $V_{\textrm{g}}$, determined from fitting experimental $\tilde{\sigma}(\omega)$ with $\tilde{\sigma}_{\textrm{Drude}}(\omega)$ (circles) and $\tilde{\sigma}_{\textrm{2-comp}}(\omega)$ (squares). Error bars: $\pm$ fit standard deviation. Solid grey curve: $\sigma_0$ measured via four-point-probe configuration (right axis). \textbf{c,} $\chi^2$ (left axis) and $R^2$ (right axis) as a function of $V_{\textrm{g}}$, for Drude (circles) and two-component (squares) model fits. \textbf{d,} $\beta_{\textrm{intra}}$ (blue) and $\beta_{\textrm{inter}}$ (red) as a function of $V_{\textrm{g}}$, quantifying the relative contributions of intra- and interband transitions to $\Tilde{\sigma} (\omega)$. 
    }
    \label{fig3}
\end{figure}

We first fit $\textrm{Re}\left[\Tilde{\sigma}\left( \omega \right)\right]$ and $\textrm{Im}\left[\Tilde{\sigma}\left( \omega \right)\right]$ simultaneously with $\Tilde{\sigma}_\textrm{Drude} (\omega)$ (Drude model; black dashed curves in Fig. \ref{fig2}), with $\sigma_0$ and $\tau$ as fitting parameters (see Fig. \ref{fig3}a, b). When graphene is significantly $p-$ or $n-$doped (i.e., $\vert E_{\textrm{F}} \vert \gtrsim 10$ meV with $\vert V_{\textrm{g}}-V_{\textrm{CNP}}\vert \gtrsim 10$ V), $\Tilde{\sigma}\left( \omega \right)$ is well described by the conventional Drude model: $\textrm{Re}\left[\Tilde{\sigma}\left( \omega \right)\right]$ decreases monotonically with $\omega$; $\textrm{Im}\left[\Tilde{\sigma}\left( \omega \right)\right]$ increases for ${\sim}0.1 < \omega/2\pi \lesssim 2$ THz, then plateaus and decreases for $\omega/2\pi \gtrsim 2$ THz. This is consistent with previous studies \cite{Jnawali_2013, Ivanov_2015, Gallagher_2019, Cocker_2017}. For $E_{\textrm{f}}$ close to the Dirac point (i.e., $V_{\textrm{g}} \approx V_{\textrm{CNP}}$), while the retrieved $\textrm{Re}\left[\Tilde{\sigma}\left( \omega \right)\right]$ remains qualitatively similar, $\textrm{Im}\left[\Tilde{\sigma}\left( \omega \right)\right]$ is significantly suppressed (see SI Supplementary Note 7 for similar trend on additional devices). The Drude fit fails to adequately capture both $\textrm{Re}\left[\Tilde{\sigma}\left( \omega \right)\right]$ and $\textrm{Im}\left[\Tilde{\sigma}\left( \omega \right)\right]$ simultaneously, underestimating the former and overestimating the latter (see Fig. \ref{fig2}). That is, for charge-neutral graphene, the Drude model fails to accurately describe the full complex conductivity $\Tilde{\sigma}(\omega)$ within ${\sim} 0.1 - 7$ THz spectral range.

The inadequacy of the Drude model to explain $\Tilde{\sigma}(\omega)$ for charge-neutral graphene is further corroborated by the Pearson's $\chi^2$ test and the coefficient of determination $R^2$ of these Drude fits (Fig. \ref{fig3}c, d). Indeed, $\chi^2$ increases and $R^2$ decreases significantly for $V_{\textrm{g}} \approx V_{\textrm{CNP}}$ ($E_{\textrm{F}} \approx 0$) in comparison with doped graphene, with $\chi^2 [V_{\textrm{g}} \approx V_{\textrm{CNP}}] / \chi^2[\vert V_{\textrm{g}} - V_{\textrm{CNP}} \vert \gtrsim 20\textrm{ V}] > 400$\%, and $R^2$ dropping from $\geq$ 0.9 to $<0.6$. Note that modified versions of the Drude model -- accounting for, e.g., disorder-induced localisation \cite{Lee_1993, Lee_1995, Chen_2019}, molecular vibrations \cite{Patterson_2018}, electrons scattering off defects \cite{Smith_2001, Buron_2014} or charged impurities \cite{Ando_2006}) -- also fail at quantitatively explaining our measurements of $\Tilde{\sigma}(\omega)$ for charge-neutral graphene within the full broad ${\sim} 0.1 - 7$ THz spectral range (see SI Supplementary Note 4). 

\bigskip
\noindent

\textbf{Two-component quantum model of graphene THz conductivity}

The dynamic complex conductivity of pristine graphene is governed by intra- and interband electronic transitions \cite{CastroNeto_2009, Hafez_2020}. The former involves the optically induced acceleration of charge carriers (Fig. \ref{fig1}c). The latter is associated with the optically induced creation or annihilation of electron-hole pairs, and with displacement currents of bound charges (Fig. \ref{fig1}d). Note that, in the linear regime, intra- and interband transitions are uncorrelated, resulting in independent contributions to the linear optical response of pristine graphene \cite{DasSarma, Liu_2018}.



We, therefore, developed a two-component model (see Methods and SI Supplementary Note 5 for more details) of the linear optical response of graphene using density matrix theory \cite{Haug_2004} and quantum kinetic equations \cite{Kitamura_2015, Culcer_2017, Culcer_2020}, accounting quantitatively for both intra- and interband transitions, and allowing us to calculate $\Tilde{\sigma}\left( \omega \right)$:

\begin{equation}
    \Tilde{\sigma}_\textrm{2-comp} \left( \omega \right) = \Tilde{\sigma}_\textrm{intra}\left( \omega \right) + \Tilde{\sigma}_\textrm{inter}\left( \omega \right) 
\label{eq_2comp}    
\end{equation}

Here, $\Tilde{\sigma}_\textrm{intra}\left( \omega \right)$ and $\Tilde{\sigma}_\textrm{inter}\left( \omega \right)$ are the contributions to $\Tilde{\sigma}\left( \omega \right)$ given by intra- and interband electronic transitions, with: 

\begin{equation}
    \Tilde{\sigma}_\textrm{intra}\left( \omega \right) = \frac{e^2v_{\textrm{F}}}{4\pi\hbar} \int\limits_0^{\infty} d k  \frac{k \tau}{1 - i\omega \tau } \left( \partial_{k} \Bar{f}_{c,k}^{(0)} - \partial_{k} \Bar{f}_{v,k}^{(0)} \right)
    \label{cond1}
\end{equation}
where $e$ is the electron charge, $v_\textrm{F}$ is the graphene Fermi velocity, $\hbar$ is the reduced Planck's constant, $\tau$ is the carrier population intraband relaxation time constant (resulting from scattering of electrons off, e.g., defects, impurities, as defined in the Drude model), $\Bar{f}_{n,k}^{(0)}$ are the zeroth-order diagonal density matrix elements given by the Fermi-Dirac distribution for the valence ($n=v$) and conduction ($n=c$) bands, and $k=\vert \mathbf{k} \vert$ is the electron wavevector modulus. 

\begin{figure}
    \centering
    \includegraphics[width=1.0\linewidth]{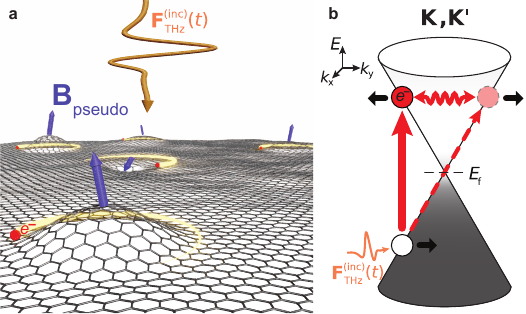}
    \caption{\textbf{Interband transitions in charge-neutral graphene enhanced by scattering of electrons due to structural deformations.} \textbf{a,} Real-space schematic of graphene, with structural deformations inducing substantial electrostatic vector scattering potentials associated with pseudo-magnetic fields (purple arrows). \textbf{b,} Interband transitions (solid red arrow) can be enhanced via acceleration of electrons (dashed red arrow) by the incident THz electric field and subsequent intravalley backscattering (oscillating red double arrow) flipping the electron pseudo-spin (black arrows).
    }
    \label{fig4}
\end{figure}

The interband contribution is given by:

\begin{equation}
    \Tilde\sigma_\textrm{inter} \left( \omega \right) = \Tilde\sigma_\textrm{inter}^{(\textrm{o})}(\omega) + \Tilde\sigma_\textrm{inter}^{(\textrm{s})}(\omega)
    \label{cond3}
\end{equation}
where $\Tilde\sigma_\textrm{inter}^{(\textrm{o})}(\omega)$ and $\Tilde\sigma_\textrm{inter}^{(\textrm{s})}(\omega)$ are attributed to direct (momentum-conserving; Fig. \ref{fig1}d) and scattering-assisted indirect (involving momentum transfer) interband transitions (Fig. \ref{fig4}b), respectively. The contribution of direct interband transitions to the dynamic conductivity of pristine graphene has an upper limit of $e^2/4\hbar$ \cite{Liu_2018}, significantly smaller than the THz intraband conductivity, and only accounting for direct interband transitions -- that is, where $\Tilde\sigma_\textrm{inter}^{(\textrm{s})}(\omega)=0$  -- is not able to reproduce our experimental $\Tilde{\sigma}(\omega)$ (see SI Supplementary Note 6). We, therefore, hypothesise that additional indirect interband processes, mediated by scattering of electrons, contribute to $\Tilde{\sigma}\left( \omega \right)$, with:

\begin{equation}
    \Tilde\sigma_\textrm{inter}^{(\textrm{o})}(\omega) = \frac{ie^2 v_\textrm{F}^2}{4 \pi \hbar} 
    \int\limits_0^{\infty} dk\, k \,  \frac{\left(\Bar{f}_{c,k}^{(0)} - \Bar{f}_{v,k}^{(0)} \right)}{ \Delta \omega_k^2 - \left(\omega  + i\gamma  \right)^2}
    \label{cond3a}
\end{equation}

\begin{equation}
    \Tilde\sigma_\textrm{inter}^{(\textrm{s})}(\omega) = \frac{ie^2 v_\textrm{F}^2}{4 \pi \hbar} 
    \int\limits_0^{\infty} dk\, k^2 \,  \frac{\Bar{v}^2 \cdot \Gamma(\omega, k, \tau) \cdot \left( \partial_{k} \Bar{f}_{c,k}^{(0)} - \partial_{k} \Bar{f}_{v,k}^{(0)} \right)}{ \Delta \omega_k^2 - \left(\omega  + i\gamma  \right)^2}
    \label{cond3b}
\end{equation}

Here, $\Delta \omega_{k}=2v_\textrm{F} k$ is the transition angular frequency between valence and conduction bands, $\gamma$ is the interband polarisation dephasing rate, $\Bar{v}$ is an energy associated with the spatially averaged square of the scattering potential of disordered graphene, and $\Gamma \left(\omega, k, \tau\right)$ is defined as:

\begin{equation}
    \Gamma  \left(\omega, k, \tau\right) = \frac{k L^2}{\hbar^2 v_\textrm{F}}\text{Im} \left(\frac{\tau}{1-i \omega\tau} \right) 
    \label{cond3c}
\end{equation}
where $L^2$ is the graphene area irradiated by the THz waveform. Note that the aforementioned scattering-assisted interband processes do not affect the DC conductivity, i.e., $\Tilde\sigma_\textrm{inter}^{(\textrm{s})}(\omega = 0) = 0$.



We fit the measured $\textrm{Re}\left[\Tilde{\sigma}\left( \omega \right)\right]$ and $\textrm{Im}\left[\Tilde{\sigma}\left( \omega \right)\right]$ with $\Tilde{\sigma}_\textrm{2-comp} \left( \omega \right)$ given by Eqs. \eqref{eq_2comp}-\eqref{cond3c}, for different gate voltages $V_\textrm{g}$ (solid black curves in Fig. \ref{fig2}). We assumed $\tau(V_\textrm{g})=\alpha \sqrt{n(V_\textrm{g})}$ \cite{Das_2011}, where $n(V_\textrm{g})$ is the $V_\textrm{g}$-dependent carrier concentration measured by four-point-probe. We used $\Bar{v}$ and $\alpha$ as global fit parameters (i.e., same for all $V_\textrm{g}$), and $\gamma$ as a local fit parameter (i.e., varying as a function of $V_\textrm{g}$); see Methods for details. The proposed two-component model is in excellent agreement with our measurements, for all considered gate voltages, and within the full ${\sim}0.1 - 7$ THz spectral range. In particular, it provides a significantly better fit than all Drude-type models at the CNP (see SI Supplementary Notes 4, 6), as shown by the fit goodness coefficients $\chi ^2$ and $R^2$ in Fig. \ref{fig3}c, yielding $\chi ^{2}_{\textrm{2-comp}} / \chi ^{2}_{\textrm{Drude}} \lesssim 20\%$ and $R^{2}_{\textrm{2-comp}} \gtrsim 90\%$ for $V_\textrm{g} \approx V_\textrm{CNP}$.

The intra- and interband components $\Tilde{\sigma}_\textrm{intra}\left( \omega \right)$ and $\Tilde{\sigma}_\textrm{inter}\left( \omega \right)$ of $\Tilde{\sigma}_\textrm{2-comp}\left( \omega \right)$ vary for different gate voltages $V_\textrm{g}$ (Fig. \ref{fig2}). This is emphasised by parameters $\beta_\textrm{intra} (V_{\textrm{g}})$ and $\beta_\textrm{inter} (V_{\textrm{g}})$ in Fig. \ref{fig3}d, defined as $\beta_\textrm{inter; intra} (V_{\textrm{g}}) = \int_{{\omega_{1}}}^{{\omega_{2}}}\vert \Tilde{\sigma}_\textrm{inter; intra}\left( \omega \right)\vert \, d\omega / \int_{{\omega_{1}}}^{{\omega_{2}}}\vert \Tilde{\sigma}_\textrm{2-comp}\left( \omega \right)\vert \, d\omega $ (with $\omega_{1}/2\pi = 0.1$ THz and $\omega_{2}/2\pi = 7$ THz). For doped graphene, $\beta_\textrm{intra} >> \beta_\textrm{inter}$; the intraband component dominates $\Tilde{\sigma}_{\textrm{2-comp}}(\omega)$ and the dynamic conductivity is Drude-like (Fig. \ref{fig2}a, e, f, j). As $\vert V_{\textrm{g}} - V_{\textrm{CNP}}  \vert$ decreases and graphene approaches the CNP, $\beta_\textrm{inter}$ increases and reaches a maximum; the dynamic conductivity differs substantially from a Drude-like behaviour.

The DC conductivity, $\sigma_0 = \textrm{Re}\left[\Tilde{\sigma}_{\textrm{2-comp}}\left( \omega = 0 \right)\right]$, retrieved from the two-component fit (Fig. \ref{fig3}a) is minimum at the CNP, similar to the Drude fit, and consistent with four-point-probe measurements (see SI Supplementary Note 1). Moreover, the two-component model agrees with experiments while assuming a Drude-like $\tau(V_{\textrm{g}})$ (Fig. \ref{fig3}b). That is, the non-Drude behaviour of $\Tilde{\sigma}\left( \omega \right)$ at the CNP manifests itself mostly at non-zero frequencies, via the two-component model parameters $\gamma$ and $\Bar{v}$.


\section*{Discussion}

The Drude-type time constant $\tau$ is associated with the intraband relaxation of electron momentum via scattering processes. In the proposed two-component model, $\tau$ affects $\Tilde\sigma_\textrm{inter}^{(\textrm{s})}(\omega)$ via the $\Gamma  \left(\omega, k, \tau\right)$ function [see Eqs. \eqref{cond3b}-\eqref{cond3c}]; that is, energy dissipated via scattering can lead to supplementary interband transitions, linked to $\Tilde\sigma_\textrm{inter}^{(\textrm{s})}(\omega)$. A scalar scattering potential, however, cannot give rise to -- though they affect (via the interband polarisation dephasing rate $\gamma$) -- such scattering-assisted interband transitions (see SI Supplementary Note 5). That is, the two-component model is in agreement with our experimental observations when $\Tilde\sigma_\textrm{inter}^{(\textrm{s})}(\omega)$ is associated with scattering given by a vector potential, characterised by parameter $\Bar{v}$.



Structural deformations of strained graphene on a substrate can lead to effective electrostatic vector potentials arising from changes in amplitude of electron-hopping between carbon atoms \cite{Levy_2010, Guinea_2010, GuineaF_2010}. These vector potentials are associated with pseudo-magnetic fields that can be very substantial (${\sim}100$ T). In particular, the transfer of CVD-grown graphene onto SiO$_2$ can result in significant structural deformations \cite{Kun_2019, Bhatt_2022, Buron_2014, Cullen_2010}, such as bumps or even crumpled areas, resulting in significant vector scattering potentials and intravalley backscattering of carriers \cite{Kun_2019}. Raman spectroscopy measurements of our graphene/SiO$_2$/Si samples revealed significant structural deformations of graphene \cite{Kun_2019} (see SI Supplementary Note 1), consistent with pseudo-magnetic fields that can be on the order of ${\sim}200$ T (see SI Supplementary Note 5). We therefore propose a plausible explanation of our measurements of $\Tilde{\sigma}(\omega)$ at the CNP, differing from a Drude-type behaviour, as the result of strain-induced vector scattering potentials associated with pseudo-magnetic fields. These pseudo-magnetic fields can couple to the graphene pseudo-spin (sub-lattice degree of freedom), enabling backscattering of Dirac electrons (forbidden for scalar scattering potentials) \cite{Kun_2019}, and potentially leading to enhanced interband transitions encoded in $\Tilde{\sigma}_\textrm{inter}^{(\textrm{s})}(\omega)$ (Fig. \ref{fig4}). Notably, the amount of structural deformations potentially associated with such substantial pseudo-magnetic fields (e.g., crumpled areas) -- amount that we estimated via Raman spectroscopy mapping and optical microscopy of our graphene sample -- is consistent with the value of $\Bar{v}$ determined via the two-component model fit (see SI Supplementary Note 5); this agreement further validates our proposal. Moreover, vector scattering potentials and pseudo-magnetic fields can arise from a wide range of structural deformations, such as those resulting from the SiO$_2$ substrate roughness\cite{Kun_2019, Bhatt_2022, Buron_2014, Cullen_2010}; we therefore expect that the effect is ubiquitous in graphene on rough substrates.



We assert that the deviation of our measurements from the Drude model cannot be interpreted within the framework of a Dirac fluid \cite{Crossno_2016, Gallagher_2019, Ku_2020, Sun_2018}; this phenomenon has been observed in ultra-clean exfoliated graphene encapsulated in hexagonal boron nitride and vanishes in CVD-grown graphene on SiO$_2$ due to impurities and disorder \cite{Ku_2020}. We also emphasise that we attempted -- without success -- to explain such deviation through an effective medium approximation modelling accounting for conductor-dielectric inhomogeneities (that is, charge puddles) of our sample \cite{Meera_2008}. 


Our work underscores the significant impact that structural deformations can have on the THz conductivity of charge-neutral graphene, in particular via the emergence of electrostatic vector scattering potentials (associated with pseudo-magnetic fields). Both the real and imaginary parts of $\Tilde{\sigma}(\omega)$ in the ${\sim}0.1-7$ THz range (Fig. \ref{fig2}) cannot be simultaneously explained by solely invoking Drude-like intraband transitions; it requires the consideration of interband excitations which can be further enhanced by scattering of electrons due to such vector scattering potentials. It is remarkable that the effects of such interband transitions, here with characteristic energies $\hbar\omega \approx 1 - 30$ meV, are observable at room temperature ($k_\textrm{B}T \approx 25$ meV). Our findings have important implications for the development of graphene-based THz technologies. Further studies can be envisioned to establish a quantitative relationship between specific structural deformations, vector scattering potentials, associated local pseudo-magnetic fields and broadband THz conductivity of charge-neutral graphene.

\section*{Methods}

\noindent
\textbf{Sample fabrication and electrical characterisation} 

The samples were prepared following the procedure reported previously \cite{Gebert_2023}. Briefly, we spin-coated a thin layer of polymethyl methacrylate (PMMA, 6$\%$wt in anisole from Microchem) onto a commercial single-layer graphene sample (Graphene Supermarket), grown on copper foil via chemical vapour deposition (CVD). We subsequently placed the graphene sample in a 0.1 M ammonium persulfate (APS) solution to dissolve the copper. We then collected the hydrophobic PMMA-coated graphene, floating at the surface of the solution, with a SiO$_2$/$p$-doped-Si(100) substrate (SiO$_2$ layer thickness: 300 nm; resistivity $> 10$ $\Omega\cdot$cm; MTI Corporation). After the transfer, the sample was soaked in acetone and was then annealed in argon/hydrogen (900:100 sccm in 3 hours at 340$^0$C) to remove PMMA and any polymer residue. We fabricated electrical contacts, necessary for four-point-probe measurements of DC electrical conductivity of graphene, via photo-lithography: we spin-coated (3000 rpm in 1 minute, per layer) two layers of positive photoresist (LOR-1A and AZ1512HS, which have different sensitivities to the photolithography process, to create an undercut effect \cite{Park_2008}) onto the graphene/SiO$_2$/Si sample, exposed the electrode areas to UV light with a mask and then removed the exposed photoresist with a developer solvent (AZ400K). Photoresist residue was removed via a UV/ozone treatment. We then deposited a 5-nm-thick titanium (Ti) adhesion layer and 50 nm of gold (Au) via e-beam deposition. The unexposed photoresist area was removed from the substrate with dimethyl sulfoxide (DMSO, at 60$^0$C), leaving only the desired Ti/Au contacts. The sample was finally attached to a chip carrier with silver epoxy and wire-bonded with aluminium-silicon (AlSi) thin wires. The prepared samples were further characterised via Raman spectroscopy (details in SI Supplementary Note 1). In the experiments, we tuned graphene's Fermi level $E_\textrm{f}$ by applying a gate voltage $V_\textrm{g}$ between graphene and the Si(100) substrate (see SI Supplementary Note 1 for further details, including the relationship between $V_\textrm{g}$ and $E_\textrm{f}$). Throughout the text we report the graphene $E_\textrm{f}$ relative to the Dirac point (charge neutrality point).

\bigskip\noindent
\textbf{Terahertz time-domain spectroscopy (THz-TDS)} 

We generated THz waveforms (with instantaneous electric field $\mathbf{F}_{\textrm{THz}}^{\textrm{(inc)}}(t)$ in Fig. \ref{fig1}a) in a LiNbO$_3$ (0.4 $\%$ MgO-doped, purchased from Egorov Scientific) nonlinear crystal via optical rectification \cite{Hirori_2011}, using laser pulses produced by a ytterbium-doped potassium gadolinium tungstate (Yb:KGW) laser system (Carbide, Light Conversion; central wavelength: 1030 nm central wavelength; duration: ${\sim} 290$ fs; maximum pulse energy: 400 $\mu$J), at an effective repetition rate of 200/3 kHz (i.e., laser repetition rate 200 kHz, used with a pulse picker 3). We also generated THz waveforms via optical rectification in a GaP nonlinear crystal (400 $\mu$m thick), using laser pulses (central wavelength: 870 nm central wavelength; duration: ${\sim} 44$ fs; pulse energy: 0.1 $\mu$J; repetition rate: 200 kHz, pulse picker 1) generated by an optical parametric amplifier (OPA; Orpheus-F, Light Conversion) pumped by the Yb:KGW laser. Both types of THz waveforms were detected via electro-optical sampling \cite{Saleh_1991} using another similar GaP crystal (see SI Supplementary Note 2). Waveforms generated with LiNbO$_3$ (GaP) had a duration of ${\sim} 0.7$ ps (${\sim} 0.2$ ps), a spectral bandwidth of ${\sim} 0.1 - 2$ THz (${\sim} 1.5 - 7$ THz), and a maximum peak electric field of ${\sim} 10$ kV/cm ($\sim 2$ kV/cm, respectively). Note that for these THz peak electric fields we can omit nonlinear THz processes in graphene.

The THz waveform $\mathbf{F}_{\textrm{THz}}^{\textrm{(trans)}}(t)$ transmitted through the sample generally consists of a directly transmitted transient, $\mathbf{F}_{\textrm{THz}}^{\textrm{(dir)}}(t)$, followed by subsequent transients $\mathbf{F}_{\textrm{THz}}^{(n\textrm{th})}(t)$ resulting from reflections within the substrate (see SI Fig. S9b). Based on the transmission $ \Tilde{T}\left(\omega \right) =  \Tilde{F}_{\textrm{graphene/SiO}_2/\textrm{Si}}^\textrm{(trans)}\left(\omega \right) / \Tilde{F}_{\textrm{SiO}_2/\textrm{Si}}^\textrm{(trans)}\left(\omega \right)$ of the THz waveform through graphene -- where $\Tilde{F}_{\textrm{graphene/SiO}_2/\textrm{Si}}^\textrm{(trans)}\left(\omega \right)$ and $\Tilde{F}_{\textrm{SiO}_2/\textrm{Si}}^\textrm{(trans)}\left(\omega \right)$ are the Fourier transforms of the THz waveforms transmitted, respectively, through the graphene-covered and bare Si/SiO$_2$ areas -- we can calculate the complex dynamic conductivity of graphene, $\Tilde{\sigma}\left( \omega \right)$ \cite{Whelan_2021, Whelan_2020}. Note that, when referring to experimental conductivity, we mean sheet conductivity, which has units of conductance. 

In the case of THz generation with LiNbO$_3$ (spectral range of ${\sim} 0.1 - 1.5$ THz), where strong THz peak electric fields can be obtained and the transmitted transient $\mathbf{F}_{\textrm{THz}}^{(1\textrm{st})}(t)$ resulting from the 1$^\textrm{st}$ reflection within the substrate can be measured with good signal-to-noise ratio (see SI Supplementary Note 8 for more detail), we obtained $\Tilde{\sigma}\left( \omega \right)$ via \cite{Whelan_2020, Whelan_2021}:  

\begin{equation} 
    \label{SI_eq_cond2}
    \Tilde{\sigma}^{\left(\textrm{1st}\right)} \left(\omega \right) = \frac{ \Tilde{n}_\textrm{A} \sqrt{\Tilde{n}_\textrm{A}^2 + 4\Tilde{n}_\textrm{B} \left(\Tilde{n}_\textrm{A} + \Tilde{n}_\textrm{B} \right) \Tilde{T}^{\textrm{(1st)}}\left(\omega \right)} \\
    -\Tilde{n}_\textrm{A}^2 - 2 \Tilde{n}_\textrm{A} \Tilde{n}_\textrm{B} \Tilde{T}^{\textrm{(1st)}}\left(\omega \right) }{2\Tilde{n}_\textrm{B} Z_{0} \Tilde{T}^{\textrm{(1st)}}\left(\omega \right)} 
\end{equation}
where $\Tilde{T}^{\textrm{(1st)}}\left(\omega \right)$ is the transmission based on the measurement of $\mathbf{F}_{\textrm{THz}}^{(1\textrm{st})}(t)$ through graphene/SiO$_2$/Si and SiO$_2$/Si, $Z_0=377$ $\Omega$ is the vacuum impedance, $\Tilde{n}_{\textrm{A}} = \Tilde{n}_{\textrm{SiO}_2\textrm{/Si}}\left(\omega \right) + 1$ and $\Tilde{n}_{\textrm{B}} = \Tilde{n}_{\textrm{SiO}_2\textrm{/Si}}\left(\omega \right) - 1$, with $\Tilde{n}_{\textrm{SiO}_2\textrm{/Si}}\left(\omega \right)$ being the bare SiO$_2$/Si substrate complex index of refraction (see SI Supplementary Note 3). Note that, in general, the calculation of $\Tilde{\sigma}\left( \omega \right)$ via transients $\mathbf{F}_{\textrm{THz}}^{(n\textrm{th})}(t)$ resulting from reflections within the substrate is more accurate \cite{Whelan_2017, Whelan_2021} than via directly transmitted transients $\mathbf{F}_{\textrm{THz}}^{\textrm{(dir)}}(t)$.

In the case of THz generation with GaP (spectral range of ${\sim} 1.5 - 7$ THz), where THz peak electric fields are weaker than in the LiNbO$_3$ configuration and $\mathbf{F}_{\textrm{THz}}^{(1\textrm{st})}(t)$ cannot be resolved reliably, we obtained $\Tilde{\sigma}\left( \omega \right)$ via \cite{Whelan_2020, Whelan_2021}:

\begin{equation} \label{SI_eq_cond1}
    \Tilde{\sigma}^{\textrm{(dir)}}\left( \omega \right) = \frac{\Tilde{n}_\textrm{A}}{Z_0} \left[\frac{1}{\Tilde{T}^{\textrm{(dir)}}\left(\omega \right)} - 1 \right]
\end{equation}
where $\Tilde{T}^{\textrm{(dir)}}\left(\omega \right)$ is the transmission based on the measurement of $\mathbf{F}_{\textrm{THz}}^{(\textrm{dir})}(t)$.     

All THz-TDS measurements were performed in a nitrogen environment with the sample at room temperature.


\bigskip\noindent
\textbf{Two-component dynamic conductivity model}

We consider a two-band model for a single Dirac cone, with the Hamiltonian is given by:
\begin{equation}
    H = H_0 + H_\textrm{I} + H_\textrm{scatt}
    \label{eq_hamiltonian}
\end{equation}
where $H_0$ is the equilibrium Hamiltonian for pristine graphene, which in the low-energy regime in the vicinity of the Dirac cone is given by $H_0 = v_\textrm{F}\left( \boldsymbol{\sigma} \cdot \mathbf{p}\right)$, where $v_\textrm{F}$ is the Fermi velocity, $\boldsymbol{\sigma}$ is the Pauli matrix vector and $\mathbf{p}$ is the kinetic momentum operator. The eigenstates of $H_0$ are \cite{Liu_2018}:

\begin{equation}
       \varphi_{\mathbf{k},n} (\mathbf{r}) = \langle r \vert \mathbf{k}, n \rangle  =  \frac{1}{\sqrt{2}} \frac{1}{\sqrt{L^2}} \begin{pmatrix}
    e^{-i \theta} \\
    \lambda 
    \end{pmatrix} e^{i \mathbf{k} \cdot \mathbf{r}}
    \label{wf}
\end{equation}
where $L^2$ is the real-space area of graphene considered (i.e., irradiated by the THz waveform), $n \in \{v, c\}$ is the band index ($v$: valence band; $c$: conduction band), $\mathbf{k} =  k \: \cos \theta \: \boldsymbol{\kappa}_x + k \: \sin \theta \: \boldsymbol{\kappa}_y $ is the wavevector, $\{ \boldsymbol{\kappa}_x, \boldsymbol{\kappa}_y \}$ are unit vectors defining a 2D Cartesian coordinate system of the reciprocal space with the origin at the Dirac point $\boldsymbol{K}$, $\theta$ represents the polar angle, and $\lambda=1$ if $n=c$ and $\lambda=-1$ if $n=v$. 

We employ the basis set given by Eq. \eqref{wf} to express $H$ in second-quantisation representation, where $H_0=\sum_{n,\mathbf{k}} E_{n, \mathbf{k}} a_{n, \mathbf{k}}^{\dagger}a_{n, \mathbf{k}}$, with $E_{n, \mathbf{k}}=\lambda \hbar v_\textrm{F} k$, and where $a_{n, \mathbf{k}}^{\dagger}$ and $a_{n, \mathbf{k}}$ are creation and annihilation operators. 

In Eq. \eqref{eq_hamiltonian}, $H_\textrm{I}$ and $H_\textrm{scatt}$, account for interactions between graphene electrons and an incident electromagnetic field (in our specific case, a THz waveform), and for the scattering of electrons given by a (here, both scalar and vector) scattering potential with matrix elements $V_{\mathbf{k}, \mathbf{k}'}^{(n,m)}$:

\begin{equation}    
    H_\textrm{I} = \sum_{n, m, \mathbf{k}} 
  D_{\mathbf{k}}^{(n,m)}(t)  
   a_{m, \mathbf{k}}^{\dagger}a_{n, \mathbf{k}}
\end{equation}

\begin{equation}    
    H_\textrm{scatt} = \sum_{n, m, \mathbf{k}, \mathbf{k}'} 
 V_{\mathbf{k}, \mathbf{k}'}^{(n,m)}  a_{m, \mathbf{k}'}^{\dagger}a_{n, \mathbf{k}}
\end{equation}

Here, the scattering potential can result from the combination of, e.g., defects, impurities. In the length gauge, the optical transition matrix element $D_{\mathbf{k}}^{(n,m)}(t)$ can be written as:

\begin{equation}
    D_{\mathbf{k}}^{(n,m)}(t) = F(t) \left[ie \delta_{nm} \left(\mathbf{e} \cdot \nabla_{\mathbf{k}} \right) + \left(1 - \delta_{nm} \right) \left( \mathbf{e} \cdot \mathbf{d}_{\mathbf{k}} \right) \right] 
\end{equation}
where $F(t)=F^{\textrm{(inc)}}_{\textrm{THz}}(t)$ is the incident THz waveform electric field, $\mathbf{e} = e_x\boldsymbol{\kappa}_x + e_y\boldsymbol{\kappa}_y$ is the polarisation vector, $e$ is the electron charge and  $\delta_{nm}$ is the Kronecker delta. The scalar product between $\mathbf{e}$ and the interband dipole moment $\mathbf{d}_{\mathbf{k}}$ can be expressed as \cite{Binder_2017}:
\begin{equation}
    \mathbf{e} \cdot \mathbf{d}_{\mathbf{k}}=(e/k)\left(e_x \sin \theta + e_y \cos \theta\right)
    \label{dipole}
\end{equation}

The surface current density in the time domain can be expressed in terms of the density matrix $\rho$ and velocity operator $\mathbf{v}$ (with matrix elements defined in  SI Supplementary Note 5):
\begin{equation}
    \mathbf{J}(t)=  \frac{\mathbf{e}}{\sqrt{2\pi}} \int d \omega \: \mathrm{e}^{i\omega t} \: \Tilde{F}(\omega) \: \Tilde{\sigma}(\omega) =-(e/L^2)\text{Tr} \left[\rho \: \mathbf{v} \right]
    \label{eq_current_density}
\end{equation}


The two-component model of complex dynamic conductivity, Eqs. \eqref{cond1}-\eqref{cond3b}, is based on the time evolution of $\rho$ given by the Liouville-von-Neumann equation, $i\hbar \partial_t \rho = \left[H, \rho \right]$. The density matrix can be split into diagonal terms $\Bar{f}$ corresponding to carrier populations of valence and conduction bands, non-diagonal terms $f$ accounting for intraband polarisations, and non-diagonal terms $\pi$ accounting for interband polarisations: $\rho = \Bar{f} + f + \pi$. Under the assumption of a weak applied electromagnetic field (i.e., linear regime), while considering a relatively strong scattering potential, the Liouville-von-Neumann equation results in the following set of quantum kinetic equations (see SI Supplementary Note 5 for more details):

\begin{subequations}
\begin{align}
     &\partial_t f_{n,\mathbf{k}} = \frac{eF(t)}{\hbar} \left( \mathbf{e} \cdot \nabla_{\mathbf{k}}f_{n,\mathbf{k}} \right)  +  \Pi_{\mathbf{k},\mathbf{k}}^{(n,n)} \\
    &\partial_t \pi_{\mathbf{k}} = i \Delta \omega_{\mathbf{k}} \pi_{\mathbf{k}} +i \left(\Bar{f}_{c,\mathbf{k}} - \Bar{f}_{v,\mathbf{k}} \right)  \frac{ e F(t)\left(\mathbf{e} \cdot \mathbf{d}_{\mathbf{k}} \right)}{\hbar}  + \Pi_{\mathbf{k},\mathbf{k}}^{(c,v)}\\
    &\Bar{f}_{n,\mathbf{k}} = \text{Im} \left[ f_{n,\mathbf{k}} \right] 
\end{align}
\label{sbe}
\end{subequations}
where $\Delta \omega_{\mathbf{k}} = \left( E_{c, \mathbf{k}} - E_{v,\mathbf{k'}} \right) / \hbar$ 
is the transition angular frequency, and $\Pi_{\mathbf{k}',\mathbf{k}}^{(n,m)}$ is the scattering term associated with the scattering potential $V$:

\begin{equation}
    \Pi_{\mathbf{k}',\mathbf{k}}^{(n,m)}=-\frac{i}{\hbar}\sum_{j,\mathbf{k}''} V_{\mathbf{k}',\mathbf{k}''}^{(n,j)}\pi_{\mathbf{k}'',\mathbf{k}}^{(j,m)} - V_{\mathbf{k}'',\mathbf{k}}^{(j,n)}\pi_{\mathbf{k}',\mathbf{k}''}^{(m,j)}. 
\end{equation}

The scattering term $\Pi_{\mathbf{k},\mathbf{k}}^{(n,n)}$ in Eq. {\eqref{sbe}a} can be approximated \cite{Kitamura_2015, Kim_2008} as $\Pi_{\mathbf{k},\mathbf{k}}^{(n,n)} \approx -f_{n,\mathbf{k}} / \tau$, where $\tau$ is the intraband momentum relaxation time constant. Note that, with this approximation, solutions to Eq. \eqref{sbe}a give rise to the conventional Drude model. 

The non-diagonal scattering term $\Pi_{\mathbf{k},\mathbf{k}}^{(c,v)}$ in Eq. {\eqref{sbe}}b can be approximated as (see SI Supplementary Note 5 for more details):

\begin{equation}
    \Pi_{\mathbf{k},\mathbf{k}}^{(c,v)} \approx -\frac{i}{\hbar} \sum_{\mathbf{k}'} \left[ V_{\mathbf{k},\mathbf{k}'}^{(c,c)} \pi_{\mathbf{k}', \mathbf{k}} - V_{\mathbf{k}',\mathbf{k}}^{(v,v)} \pi_{\mathbf{k}, \mathbf{k}'} \right] - \gamma \pi_{\mathbf{k}, \mathbf{k}},
    \label{eq_scat_nondiagonal}
\end{equation}
where $\gamma$ is the dephasing rate of the interband polarisation \cite{Haug_2004}.

The sum in Eq. \eqref{eq_scat_nondiagonal} describes coupling between non-momentum-conserving interband processes and direct momentum-conserving interband transitions. We show in the SI Supplementary Note 5 that such coupling does not take place  in graphene in which there are no vector scattering potentials.

Solutions of Eqs. \eqref{sbe}a-c with the aforementioned approximations of $\Pi_{\mathbf{k},\mathbf{k}}^{(n,n)}$ and $\Pi_{\mathbf{k},\mathbf{k}}^{(c,v)}$,  together with Eq. \eqref{eq_current_density}, result in the complex dynamic conductivity:

\begin{equation}
    \Tilde{\sigma}(\omega)= \Tilde{\sigma}_\textrm{intra}\left( \omega \right)  + \Tilde\sigma_\textrm{inter}^{(\textrm{o})} \left( \omega \right) + \Tilde{\sigma}_\textrm{inter}^{(\textrm{s})}\left( \omega \right)
    \label{eq_sigma}
\end{equation}
where

\begin{equation}
    \Tilde{\sigma}_\textrm{intra}(\omega) = 
    \frac{e^2 v_\textrm{F} }{2\hbar L^2} \sum_{n,\mathbf{k}} \frac{\tau  \partial_{k}\Bar{f}_{n,k}^{(0)}}{1 - i\omega \tau}  \cos^2 \theta
\label{sbe1_sol2}
\end{equation}

\begin{equation}
    \Tilde{\sigma}_\textrm{inter}^{(\textrm{o})}(\omega)  = 
    \frac{ie^2 v_\textrm{F} }{2 \hbar  L^2} \sum_{\mathbf{k}, \eta=\pm 1}  \frac{\eta\left(\Bar{f}_{c,k}^{(0)} - \Bar{f}_{v,k}^{(0)} \right)}{k \left(\omega - \eta \Delta \omega_{k} + i\gamma  \right)} \sin^2 \theta
    \label{curr_ter_o}
\end{equation}

\begin{equation}
    \Tilde{\sigma}_\textrm{inter}^{(\textrm{s})}(\omega)  = 
    \frac{ie^2 v_\textrm{F} }{2 \hbar  L^2} \sum_{\mathbf{k}, \eta=\pm 1}  \frac{\eta\bar{v}^2 \Gamma(\omega,k, \tau) \left( \partial_{k} \Bar{f}_{c,k}^{(0)} - \partial_{k} \Bar{f}_{v,k}^{(0)} \right)}{\left(\omega - \eta\Delta \omega_{k} + i\gamma  \right)} \sin^2 \theta
\label{curr_ter_s}
\end{equation}
with

\begin{equation}
    \Gamma(\omega, k, \tau) = \frac{k L^2}{\hbar^2 v_\textrm{F}}\text{Im} \left[\frac{\tau}{1-i \omega\tau} \right]
    \label{eq_gamma}
\end{equation}

Here, $\Bar{f}_{n,k}^{(0)}$ is the thermal equilibrium Fermi-Dirac distribution, the sum for $\eta = +1$ and $\eta = -1$ accounts for the complex-conjugate of the density matrix non-diagonal elements, and $ \Bar{v}$ corresponds to an energy associated with the spatially averaged square of the vector scattering potential in disordered graphene. We hypothesise that such vector scattering potential can be the result of severe structural deformations of graphene giving rise to substantial pseudo-magnetic fields; see SI Supplementary Note 5 for details. 

The final expressions of Eqs. \eqref{cond1}-\eqref{cond3b} are obtained from Eqs. \eqref{sbe1_sol2}-\eqref{eq_gamma} by transforming sums over wavevectors into integrals using the transformation:

\begin{equation}
    \sum_{\mathbf{k}} \rightarrow \frac{L^2}{(2 \pi)^2} \int\limits_0^{2 \pi} d \theta \int\limits_0^{\infty} dk \, k,
    \label{k_sum}
\end{equation}

We fit our experimental measurements of $\Tilde{\sigma}(\omega)$ with Eqs. \eqref{eq_2comp}-\eqref{cond3c}, with the Fermi-Dirac distribution $\Bar{f}_{n,k}^{(0)}$ at room temperature determined using the Fermi level $E_\textrm{f}$ obtained from four-point-probe measurements for different gate voltages $V_\textrm{g}$ (see SI Supplementary Note 1; we allowed for a 10\% variation of $E_\textrm{f}$ across different values of $V_\textrm{g}$ to obtain a best fit). We assumed $\tau(V_\textrm{g})=\alpha \sqrt{n(V_\textrm{g})}$ \cite{Das_2011}, where $n(V_\textrm{g})$ is the $V_\textrm{g}$-dependent carrier concentration determined via four-point-probe measurements. We used $\Bar{v}$ and $\alpha$ as global fit parameters (i.e., same for all $V_\textrm{g}$, allowing for a 10\% variation of $\alpha$ across different values of $V_\textrm{g}$), and $\gamma$ as a local fit parameter (i.e., varying as a function of $V_\textrm{g}$). A best fit was obtained for $\Bar{v} \approx 4.63 \times 10^{-6}$ eV, consistent with the estimate of the overall effective perimeter of severely distorted graphene areas (see SI Supplementary Note 5).

\bigskip\section*{Acknowledgements}
This work was supported by the ARC Centre of Excellence in Future Low-Energy Electronics Technologies (FLEET, CE170100039), the ARC Centre of Excellence in Exciton Science (CE170100026) and the Australian Government Research Training
Program (RTP) Scholarship. This work was performed in part at the Melbourne Centre for Nanofabrication (MCN) in the Victorian Node of the Australian National Fabrication Facility (ANFF). T-P.N., M.O. and M.G. acknowledge partial financial support from FLEET via PhD top-up scholarship. K.X. and M.S.F. acknowledge support from ARC grant DP200101345. M.K. and J.H.C acknowledge the support of the National Computational Infrastructure (NCI), which is supported by the Australian Government. We thank Dimi Culcer, Meera M. Parish and Shaffique Adam for fruitful discussions.

\bigskip\section*{Additional Information}
\textbf{Competing financial interests:} The authors declare no competing financial interests. 

\noindent \textbf{Data availability:} The codes corresponding to numerical computations are available at  \href{https://gitlab.com/freude1/linear-thz-response-in-graphene}{https://gitlab.com/freude1/linear-thz-response-in-graphene}.

\bibliography{bib.bib}

\begin{thebibliography}{10}
\expandafter\ifx\csname url\endcsname\relax
  \def\url#1{\burl{#1}}\fi
\expandafter\ifx\csname urlprefix\endcsname\relax\def\urlprefix{URL }\fi
\providecommand{\bibinfo}[2]{#2}
\providecommand{\eprint}[2][]{\url{#2}}
\providecommand{\doi}[1]{\url{https://doi.org/#1}}
\bibcommenthead

\bibitem{Novoselov_2004}
\bibinfo{author}{Novoselov, K.~S.} \emph{et~al.}
\newblock \bibinfo{title}{Electric field effect in atomically thin carbon films}.
\newblock \emph{\bibinfo{journal}{Science}} \textbf{\bibinfo{volume}{306}}~(5696), \bibinfo{pages}{666--669} (\bibinfo{year}{2004}) .

\bibitem{CastroNeto_2009}
\bibinfo{author}{Castro~Neto, A.~H.}, \bibinfo{author}{Guinea, F.}, \bibinfo{author}{Peres, N. M.~R.}, \bibinfo{author}{Novoselov, K.~S.} \& \bibinfo{author}{Geim, A.~K.}
\newblock \bibinfo{title}{The electronic properties of graphene}.
\newblock \emph{\bibinfo{journal}{Rev. Mod. Phys.}} \textbf{\bibinfo{volume}{81}}, \bibinfo{pages}{109--162} (\bibinfo{year}{2009}) .

\bibitem{Kim_2009}
\bibinfo{author}{Kim, K.~S.} \emph{et~al.}
\newblock \bibinfo{title}{Large-scale pattern growth of graphene films for stretchable transparent electrodes}.
\newblock \emph{\bibinfo{journal}{Nature}} \textbf{\bibinfo{volume}{457}}, \bibinfo{pages}{706--710} (\bibinfo{year}{2009}) .

\bibitem{Bae_2010}
\bibinfo{author}{Bae, S.} \emph{et~al.}
\newblock \bibinfo{title}{Roll-to-roll production of 30-inch graphene films for transparent electrodes}.
\newblock \emph{\bibinfo{journal}{Nature Nanotechnology}} \textbf{\bibinfo{volume}{5}}, \bibinfo{pages}{574--578} (\bibinfo{year}{2010}) .

\bibitem{Chen_2007}
\bibinfo{author}{Chen, J.-H.} \emph{et~al.}
\newblock \bibinfo{title}{Printed graphene circuits}.
\newblock \emph{\bibinfo{journal}{Advanced Materials}} \textbf{\bibinfo{volume}{19}}~(21) (\bibinfo{year}{2007}) .

\bibitem{Kamalakar_2015}
\bibinfo{title}{Long distance spin communication in chemical vapour deposited graphene}.
\newblock \emph{\bibinfo{journal}{Nature Communications}} \textbf{\bibinfo{volume}{6}}, \bibinfo{pages}{6766} (\bibinfo{year}{2015}) .

\bibitem{El-Kady_2016}
\bibinfo{author}{El-Kady, M.~F.}, \bibinfo{author}{Shao, Y.} \& \bibinfo{author}{Kaner, R.~B.}
\newblock \bibinfo{title}{Graphene for batteries, supercapacitors and beyond}.
\newblock \emph{\bibinfo{journal}{Nature Reviews Materials}} \textbf{\bibinfo{volume}{1}}, \bibinfo{pages}{16033} (\bibinfo{year}{2016}) .

\bibitem{Falkovsky_2008}
\bibinfo{author}{Falkovsky, L.~A.}
\newblock \bibinfo{title}{Optical properties of graphene}.
\newblock \emph{\bibinfo{journal}{Journal of Physics: Conference Series}} \textbf{\bibinfo{volume}{129}}, \bibinfo{pages}{012004} (\bibinfo{year}{2008}) .

\bibitem{Hendry_2010}
\bibinfo{author}{Hendry, E.}, \bibinfo{author}{Hale, P.~J.}, \bibinfo{author}{Moger, J.}, \bibinfo{author}{Savchenko, A.~K.} \& \bibinfo{author}{Mikhailov, S.~A.}
\newblock \bibinfo{title}{Coherent nonlinear optical response of graphene}.
\newblock \emph{\bibinfo{journal}{Phys. Rev. Lett.}} \textbf{\bibinfo{volume}{105}}, \bibinfo{pages}{097401} (\bibinfo{year}{2010}) .

\bibitem{Zheng_2020}
\bibinfo{author}{Zheng, Q.}, \bibinfo{author}{Xia, L.}, \bibinfo{author}{Tang, L.}, \bibinfo{author}{Du, C.} \& \bibinfo{author}{Cui, H.}
\newblock \bibinfo{title}{Low voltage graphene-based amplitude modulator for high efficiency terahertz modulation}.
\newblock \emph{\bibinfo{journal}{Nanomaterials}} \textbf{\bibinfo{volume}{10}}~(3) (\bibinfo{year}{2020}) .

\bibitem{Jin_2022}
\bibinfo{author}{Jin, M.} \emph{et~al.}
\newblock \bibinfo{title}{Terahertz detectors based on carbon nanomaterials}.
\newblock \emph{\bibinfo{journal}{Advanced Functional Materials}} \textbf{\bibinfo{volume}{32}}~(11), \bibinfo{pages}{2107499} .

\bibitem{Buron_2012}
\bibinfo{author}{Buron, J.~D.} \emph{et~al.}
\newblock \bibinfo{title}{Graphene conductance uniformity mapping}.
\newblock \emph{\bibinfo{journal}{Nano Letters}} \textbf{\bibinfo{volume}{12}}, \bibinfo{pages}{5074--5081} (\bibinfo{year}{2012}) .

\bibitem{Buron_2015}
\bibinfo{author}{Buron, J.~D.} \emph{et~al.}
\newblock \bibinfo{title}{Terahertz wafer-scale mobility mapping of graphene on insulating substrates without a gate}.
\newblock \emph{\bibinfo{journal}{Opt. Express}} \textbf{\bibinfo{volume}{23}}~(24), \bibinfo{pages}{30721--30729} (\bibinfo{year}{2015}) .

\bibitem{Scarfe_2021}
\bibinfo{title}{Systematic thz study of the substrate effect in limiting the mobility of graphene}.
\newblock \emph{\bibinfo{journal}{Scientific Reports}} \textbf{\bibinfo{volume}{11}}, \bibinfo{pages}{8729} (\bibinfo{year}{2021}) .

\bibitem{Frenzel_2013}
\bibinfo{author}{Frenzel, A.~J.} \emph{et~al.}
\newblock \bibinfo{title}{Observation of suppressed terahertz absorption in photoexcited graphene}.
\newblock \emph{\bibinfo{journal}{Applied Physics Letters}} \textbf{\bibinfo{volume}{102}}~(11), \bibinfo{pages}{113111} (\bibinfo{year}{2013}) .

\bibitem{Jnawali_2013}
\bibinfo{author}{Jnawali, G.}, \bibinfo{author}{Rao, Y.}, \bibinfo{author}{Yan, H.} \& \bibinfo{author}{Heinz, T.~F.}
\newblock \bibinfo{title}{Observation of a transient decrease in terahertz conductivity of single-layer graphene induced by ultrafast optical excitation}.
\newblock \emph{\bibinfo{journal}{Nano Letters}} \textbf{\bibinfo{volume}{13}}~(2), \bibinfo{pages}{524--530} (\bibinfo{year}{2013}) .

\bibitem{Pistore_2022}
\bibinfo{author}{Pistore, V.} \emph{et~al.}
\newblock \bibinfo{title}{Mapping the complex refractive index of single layer graphene on semiconductor or polymeric substrates at terahertz frequencies}.
\newblock \emph{\bibinfo{journal}{2D Materials}} \textbf{\bibinfo{volume}{9}}~(2), \bibinfo{pages}{025018} (\bibinfo{year}{2022}) .

\bibitem{Horng_2011}
\bibinfo{author}{Horng, J.} \emph{et~al.}
\newblock \bibinfo{title}{Drude conductivity of dirac fermions in graphene}.
\newblock \emph{\bibinfo{journal}{Phys. Rev. B}} \textbf{\bibinfo{volume}{83}}, \bibinfo{pages}{165113} (\bibinfo{year}{2011}) .

\bibitem{Ren_2012}
\bibinfo{author}{Ren, L.} \emph{et~al.}
\newblock \bibinfo{title}{Terahertz and infrared spectroscopy of gated large-area graphene}.
\newblock \emph{\bibinfo{journal}{Nano Letters}} \textbf{\bibinfo{volume}{12}}~(7), \bibinfo{pages}{3711--3715} (\bibinfo{year}{2012}) .

\bibitem{Smith_2001}
\bibinfo{author}{Smith, N.~V.}
\newblock \bibinfo{title}{Classical generalization of the drude formula for the optical conductivity}.
\newblock \emph{\bibinfo{journal}{Phys. Rev. B}} \textbf{\bibinfo{volume}{64}}, \bibinfo{pages}{155106} (\bibinfo{year}{2001}) .

\bibitem{Buron_2014}
\bibinfo{author}{Buron, J.~D.} \emph{et~al.}
\newblock \bibinfo{title}{Electrically continuous graphene from single crystal copper verified by terahertz conductance spectroscopy and micro four-point probe}.
\newblock \emph{\bibinfo{journal}{Nano Letters}} \textbf{\bibinfo{volume}{14}}, \bibinfo{pages}{6348--6355} (\bibinfo{year}{2014}) .

\bibitem{Cocker_2017}
\bibinfo{author}{Cocker, T.~L.} \emph{et~al.}
\newblock \bibinfo{title}{Microscopic origin of the drude-smith model}.
\newblock \emph{\bibinfo{journal}{Phys. Rev. B}} \textbf{\bibinfo{volume}{96}}, \bibinfo{pages}{205439} (\bibinfo{year}{2017}) .

\bibitem{Schubert_2004}
\bibinfo{title}{Infrared ellipsometry characterization of conducting thin organic films}.
\newblock \emph{\bibinfo{journal}{Thin Solid Films}} \textbf{\bibinfo{volume}{455-456}}, \bibinfo{pages}{295--300} (\bibinfo{year}{2004}).
\newblock \bibinfo{note}{The 3rd International Conference on Spectroscopic Ellipsometry} .

\bibitem{Chen_2019}
\bibinfo{author}{Chen, S.} \emph{et~al.}
\newblock \bibinfo{title}{On the anomalous optical conductivity dispersion of electrically conducting polymers: ultra-wide spectral range ellipsometry combined with a drude–lorentz model}.
\newblock \emph{\bibinfo{journal}{J. Mater. Chem. C}} \textbf{\bibinfo{volume}{7}}, \bibinfo{pages}{4350--4362} (\bibinfo{year}{2019}) .

\bibitem{Lee_1993}
\bibinfo{author}{Lee, K.}, \bibinfo{author}{Heeger, A.~J.} \& \bibinfo{author}{Cao, Y.}
\newblock \bibinfo{title}{Reflectance of polyaniline protonated with camphor sulfonic acid: Disordered metal on the metal-insulator boundary}.
\newblock \emph{\bibinfo{journal}{Phys. Rev. B}} \textbf{\bibinfo{volume}{48}}, \bibinfo{pages}{14884--14891} (\bibinfo{year}{1993}) .

\bibitem{Lee_1995}
\bibinfo{author}{Lee, K.}, \bibinfo{author}{Menon, R.}, \bibinfo{author}{Yoon, C.~O.} \& \bibinfo{author}{Heeger, A.~J.}
\newblock \bibinfo{title}{Reflectance of conducting polypyrrole: Observation of the metal-insulator transition driven by disorder}.
\newblock \emph{\bibinfo{journal}{Phys. Rev. B}} \textbf{\bibinfo{volume}{52}}, \bibinfo{pages}{4779--4787} (\bibinfo{year}{1995}) .

\bibitem{Ando_2006}
\bibinfo{author}{Ando, T.}
\newblock \bibinfo{title}{Screening effect and impurity scattering in monolayer graphene}.
\newblock \emph{\bibinfo{journal}{Journal of the Physical Society of Japan}} \textbf{\bibinfo{volume}{75}}~(7), \bibinfo{pages}{074716} (\bibinfo{year}{2006}) .

\bibitem{Hafez_2020}
\bibinfo{author}{Hafez, H.~A.} \emph{et~al.}
\newblock \bibinfo{title}{Terahertz nonlinear optics of graphene: From saturable absorption to high-harmonics generation}.
\newblock \emph{\bibinfo{journal}{Advanced Optical Materials}} \textbf{\bibinfo{volume}{8}}~(3), \bibinfo{pages}{1900771} .

\bibitem{Gallagher_2019}
\bibinfo{author}{Gallagher, P.} \emph{et~al.}
\newblock \bibinfo{title}{Quantum-critical conductivity of the dirac fluid in graphene}.
\newblock \emph{\bibinfo{journal}{Science}} \textbf{\bibinfo{volume}{364}}~(6436), \bibinfo{pages}{158--162} (\bibinfo{year}{2019}) .

\bibitem{Sensale-Rodríguez_2013}
\bibinfo{author}{Sensale-Rodríguez, B.}, \bibinfo{author}{Yan, R.}, \bibinfo{author}{Liu, L.}, \bibinfo{author}{Jena, D.} \& \bibinfo{author}{Xing, H.~G.}
\newblock \bibinfo{title}{Graphene for reconfigurable terahertz optoelectronics}.
\newblock \emph{\bibinfo{journal}{Proceedings of the IEEE}} \textbf{\bibinfo{volume}{101}}~(7), \bibinfo{pages}{1705--1716} (\bibinfo{year}{2013}) .

\bibitem{Tonouchi_2007}
\bibinfo{author}{Tonouchi, M.}
\newblock \bibinfo{title}{Cutting-edge terahertz technology}.
\newblock \emph{\bibinfo{journal}{Nature Photonics}} \textbf{\bibinfo{volume}{1}}, \bibinfo{pages}{97--105} (\bibinfo{year}{2007}) .

\bibitem{Jepsen_2011}
\bibinfo{author}{Jepsen, P.}, \bibinfo{author}{Cooke, D.} \& \bibinfo{author}{Koch, M.}
\newblock \bibinfo{title}{Terahertz spectroscopy and imaging – modern techniques and applications}.
\newblock \emph{\bibinfo{journal}{Laser \& Photonics Reviews}} \textbf{\bibinfo{volume}{5}}~(1) (\bibinfo{year}{2011}) .

\bibitem{Haug_2004}
\bibinfo{author}{Haug, H.} \& \bibinfo{author}{Koch, S.}
\newblock \emph{\bibinfo{title}{Quantum Theory of the Optical and Electronic Properties of Semiconductors}} Quantum Theory of the Optical and Electronic Properties of Semiconductors.

\bibitem{Kitamura_2015}
\bibinfo{author}{Kitamura, H.}
\newblock \bibinfo{title}{Derivation of the drude conductivity from quantum kinetic equations}.
\newblock \emph{\bibinfo{journal}{European Journal of Physics}} \textbf{\bibinfo{volume}{36}}~(6), \bibinfo{pages}{065010} (\bibinfo{year}{2015}) .

\bibitem{Culcer_2017}
\bibinfo{author}{Culcer, D.}, \bibinfo{author}{Sekine, A.} \& \bibinfo{author}{MacDonald, A.~H.}
\newblock \bibinfo{title}{Interband coherence response to electric fields in crystals: Berry-phase contributions and disorder effects}.
\newblock \emph{\bibinfo{journal}{Phys. Rev. B}} \textbf{\bibinfo{volume}{96}}, \bibinfo{pages}{035106} (\bibinfo{year}{2017}) .

\bibitem{Culcer_2020}
\bibinfo{author}{Culcer, D.}, \bibinfo{author}{Keser, A.~C.}, \bibinfo{author}{Li, Y.} \& \bibinfo{author}{Tkachov, G.}
\newblock \bibinfo{title}{Transport in two-dimensional topological materials: recent developments in experiment and theory}.
\newblock \emph{\bibinfo{journal}{2D Materials}} \textbf{\bibinfo{volume}{7}}~(2), \bibinfo{pages}{022007} (\bibinfo{year}{2020}) .

\bibitem{Hirori_2011}
\bibinfo{author}{Hirori, H.}, \bibinfo{author}{~, A.}, \bibinfo{author}{Blanchard, F.} \& \bibinfo{author}{Tanaka, K.}
\newblock \bibinfo{title}{Single-cycle terahertz pulses with amplitudes exceeding 1 mv/cm generated by optical rectification in linbo3}.
\newblock \emph{\bibinfo{journal}{Applied Physics Letters}} \textbf{\bibinfo{volume}{98}}~(9), \bibinfo{pages}{091106} (\bibinfo{year}{2011}) .

\bibitem{Whelan_2020}
\bibinfo{author}{Whelan, P.~R.} \emph{et~al.}
\newblock \bibinfo{title}{Reference-free thz-tds conductivity analysis of thin conducting films}.
\newblock \emph{\bibinfo{journal}{Opt. Express}} \textbf{\bibinfo{volume}{28}}~(20), \bibinfo{pages}{28819--28830} (\bibinfo{year}{2020}) .

\bibitem{Ivanov_2015}
\bibinfo{author}{Ivanov, I.}, \bibinfo{author}{Bonn, M.}, \bibinfo{author}{Mics, Z.} \& \bibinfo{author}{Turchinovich, D.}
\newblock \bibinfo{title}{Perspective on terahertz spectroscopy of graphene}.
\newblock \emph{\bibinfo{journal}{{EPL} (Europhysics Letters)}} \textbf{\bibinfo{volume}{111}}~(6), \bibinfo{pages}{67001} (\bibinfo{year}{2015}) .

\bibitem{Patterson_2018}
\bibinfo{author}{Patterson, J.~D.} \& \bibinfo{author}{Bailey, B.~C.}
\newblock \emph{\bibinfo{title}{Optical Properties of Solids}}  (\bibinfo{publisher}{Springer International Publishing}, \bibinfo{address}{Cham}, \bibinfo{year}{2018}).

\bibitem{DasSarma}
\bibinfo{author}{Das~Sarma, S.}, \bibinfo{author}{Adam, S.}, \bibinfo{author}{Hwang, E.~H.} \& \bibinfo{author}{Rossi, E.}
\newblock \bibinfo{title}{Electronic transport in two-dimensional graphene}.
\newblock \emph{\bibinfo{journal}{Rev. Mod. Phys.}} \textbf{\bibinfo{volume}{83}}, \bibinfo{pages}{407--470} (\bibinfo{year}{2011}) .

\bibitem{Liu_2018}
\bibinfo{author}{Liu, J.} \& \bibinfo{author}{Lin, I.}
\newblock \emph{\bibinfo{title}{Graphene Photonics}}  (\bibinfo{publisher}{Cambridge University Press}, \bibinfo{year}{2018}).

\bibitem{Das_2011}
\bibinfo{author}{Das~Sarma, S.}, \bibinfo{author}{Adam, S.}, \bibinfo{author}{Hwang, E.~H.} \& \bibinfo{author}{Rossi, E.}
\newblock \bibinfo{title}{Electronic transport in two-dimensional graphene}.
\newblock \emph{\bibinfo{journal}{Rev. Mod. Phys.}} \textbf{\bibinfo{volume}{83}}, \bibinfo{pages}{407--470} (\bibinfo{year}{2011}) .

\bibitem{Levy_2010}
\bibinfo{author}{Levy, N.} \emph{et~al.}
\newblock \bibinfo{title}{Strain-induced pseudo-magnetic fields greater than 300 tesla in graphene nanobubbles}.
\newblock \emph{\bibinfo{journal}{Science}} \textbf{\bibinfo{volume}{329}}~(5991), \bibinfo{pages}{544--547} (\bibinfo{year}{2010}) .

\bibitem{Guinea_2010}
\bibinfo{author}{Guinea, F.}, \bibinfo{author}{Katsnelson, M.~I.} \& \bibinfo{author}{Geim, A.~K.}
\newblock \bibinfo{title}{Energy gaps and a zero-field quantum hall effect in graphene by strain engineering}.
\newblock \emph{\bibinfo{journal}{Nature Physics}} \textbf{\bibinfo{volume}{6}}~(1), \bibinfo{pages}{30--33} (\bibinfo{year}{2010}) .

\bibitem{GuineaF_2010}
\bibinfo{author}{Guinea, F.}, \bibinfo{author}{Geim, A.~K.}, \bibinfo{author}{Katsnelson, M.~I.} \& \bibinfo{author}{Novoselov, K.~S.}
\newblock \bibinfo{title}{Generating quantizing pseudomagnetic fields by bending graphene ribbons}.
\newblock \emph{\bibinfo{journal}{Phys. Rev. B}} \textbf{\bibinfo{volume}{81}}, \bibinfo{pages}{035408} (\bibinfo{year}{2010}) .

\bibitem{Kun_2019}
\bibinfo{author}{Kun, P.} \emph{et~al.}
\newblock \bibinfo{title}{Large intravalley scattering due to pseudo-magnetic fields in crumpled graphene}.
\newblock \emph{\bibinfo{journal}{npj 2D Materials and Applications}} \textbf{\bibinfo{volume}{3}} (\bibinfo{year}{2019}) .

\bibitem{Bhatt_2022}
\bibinfo{author}{Bhatt, M.~D.}, \bibinfo{author}{Kim, H.} \& \bibinfo{author}{Kim, G.}
\newblock \bibinfo{title}{Various defects in graphene: a review}.
\newblock \emph{\bibinfo{journal}{RSC Adv.}} \textbf{\bibinfo{volume}{12}}, \bibinfo{pages}{21520--21547} (\bibinfo{year}{2022}) .

\bibitem{Cullen_2010}
\bibinfo{author}{Cullen, W.~G.} \emph{et~al.}
\newblock \bibinfo{title}{High-fidelity conformation of graphene to ${\mathrm{sio}}_{2}$ topographic features}.
\newblock \emph{\bibinfo{journal}{Phys. Rev. Lett.}} \textbf{\bibinfo{volume}{105}}, \bibinfo{pages}{215504} (\bibinfo{year}{2010}) .

\bibitem{Crossno_2016}
\bibinfo{author}{Crossno, J.} \emph{et~al.}
\newblock \bibinfo{title}{Observation of the dirac fluid and the breakdown of the wiedemann-franz law in graphene}.
\newblock \emph{\bibinfo{journal}{Science}} \textbf{\bibinfo{volume}{351}}~(6277), \bibinfo{pages}{1058--1061} (\bibinfo{year}{2016}) .

\bibitem{Ku_2020}
\bibinfo{author}{Ku, M. J.~H.} \emph{et~al.}
\newblock \bibinfo{title}{Imaging viscous flow of the dirac fluid in graphene}.
\newblock \emph{\bibinfo{journal}{Nature}} \textbf{\bibinfo{volume}{583}}~(7817), \bibinfo{pages}{537--541} (\bibinfo{year}{2020}) .

\bibitem{Sun_2018}
\bibinfo{author}{Sun, Z.}, \bibinfo{author}{Basov, D.~N.} \& \bibinfo{author}{Fogler, M.~M.}
\newblock \bibinfo{title}{Universal linear and nonlinear electrodynamics of a dirac fluid}.
\newblock \emph{\bibinfo{journal}{Proceedings of the National Academy of Sciences}} \textbf{\bibinfo{volume}{115}}~(13), \bibinfo{pages}{3285--3289} (\bibinfo{year}{2018}) .

\bibitem{Meera_2008}
\bibinfo{author}{Parish, M.~M.} \& \bibinfo{author}{Littlewood, P.~B.}
\newblock \bibinfo{title}{Magnetocapacitance in nonmagnetic composite media}.
\newblock \emph{\bibinfo{journal}{Phys. Rev. Lett.}} \textbf{\bibinfo{volume}{101}}, \bibinfo{pages}{166602} (\bibinfo{year}{2008}) .

\bibitem{Gebert_2023}
\bibinfo{author}{Gebert, M.} \emph{et~al.}
\newblock \bibinfo{title}{Passivating graphene and suppressing interfacial phonon scattering with mechanically transferred large-area ga2o3}.
\newblock \emph{\bibinfo{journal}{Nano Letters}} \textbf{\bibinfo{volume}{23}}~(1), \bibinfo{pages}{363--370} (\bibinfo{year}{2023}) .

\bibitem{Park_2008}
\bibinfo{author}{Park, J.} \emph{et~al.}
\newblock \bibinfo{title}{{Bilayer processing for an enhanced organic-electrode contact in ultrathin bottom contact organic transistors}}.
\newblock \emph{\bibinfo{journal}{Applied Physics Letters}} \textbf{\bibinfo{volume}{92}}~(19) (\bibinfo{year}{2008}) .

\bibitem{Saleh_1991}
\emph{\bibinfo{title}{Electro-Optics}}, Ch.~\bibinfo{chapter}{18}, \bibinfo{pages}{696--736} (\bibinfo{publisher}{John Wiley \& Sons, Ltd}).

\bibitem{Whelan_2021}
\bibinfo{author}{Whelan, P.~R.} \emph{et~al.}
\newblock \bibinfo{title}{Case studies of electrical characterisation of graphene by terahertz time-domain spectroscopy}.
\newblock \emph{\bibinfo{journal}{2D Materials}} \textbf{\bibinfo{volume}{8}}~(2), \bibinfo{pages}{022003} (\bibinfo{year}{2021}) .

\bibitem{Whelan_2017}
\bibinfo{author}{Whelan, P.~R.} \emph{et~al.}
\newblock \bibinfo{title}{Robust mapping of electrical properties of graphene from terahertz time-domain spectroscopy with timing jitter correction}.
\newblock \emph{\bibinfo{journal}{Opt. Express}} \textbf{\bibinfo{volume}{25}}~(3), \bibinfo{pages}{2725--2732} (\bibinfo{year}{2017}) .

\bibitem{Binder_2017}
\bibinfo{author}{Binder, R.}
\newblock \emph{\bibinfo{title}{Optical Properties of Graphene}}  (\bibinfo{publisher}{WORLD SCIENTIFIC}, \bibinfo{year}{2017}).

\bibitem{Kim_2008}
\bibinfo{author}{Kim, E.-A.} \& \bibinfo{author}{Neto, A. H.~C.}
\newblock \bibinfo{title}{Graphene as an electronic membrane}.
\newblock \emph{\bibinfo{journal}{Europhysics Letters}} \textbf{\bibinfo{volume}{84}}~(5), \bibinfo{pages}{57007} (\bibinfo{year}{2008}) .

\end{thebibliography}

\clearpage

\includepdf[pages=-]{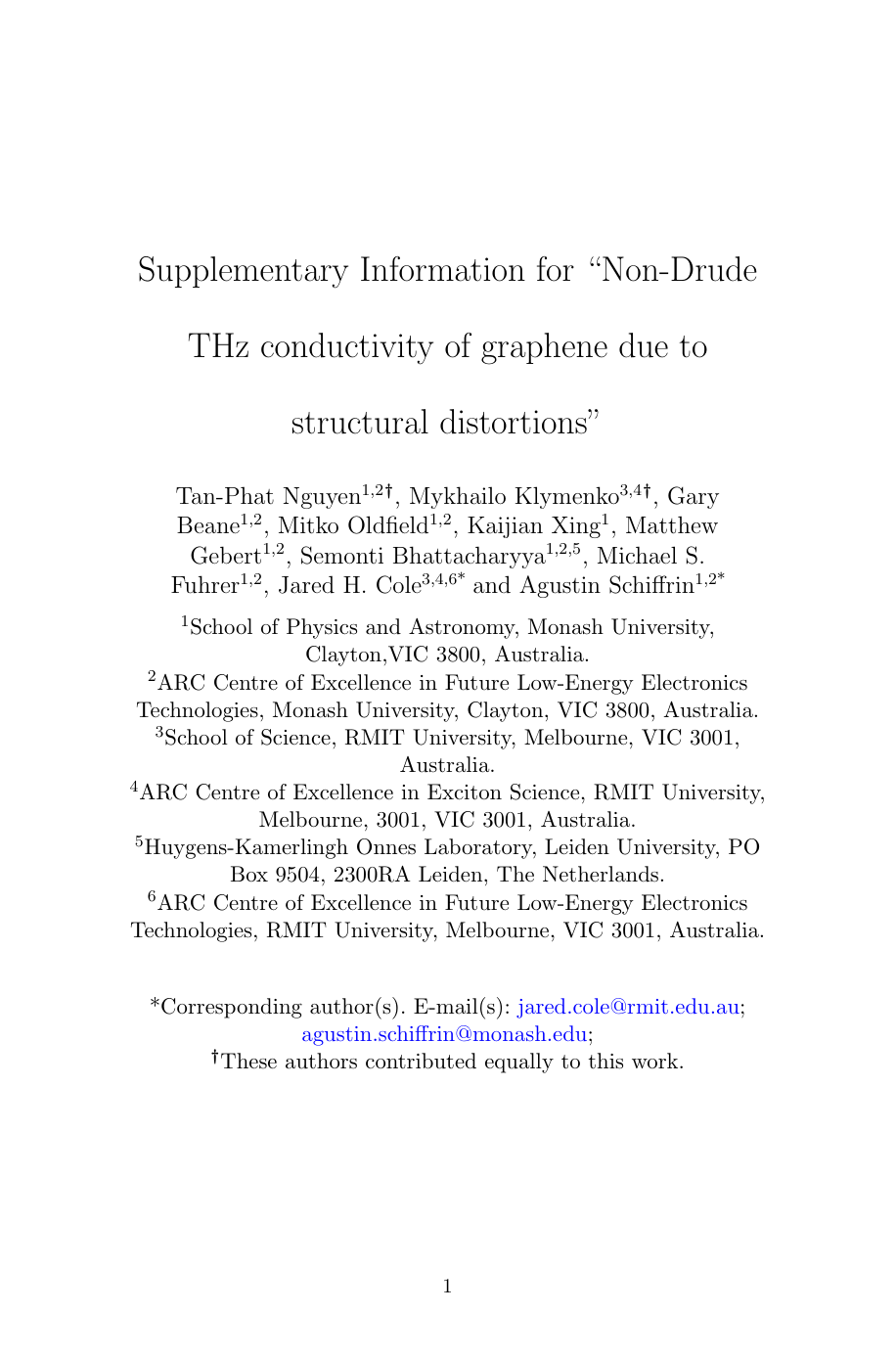}

\end{document}


\title[ ]{Supplementary Information for ``Non-Drude THz conductivity of graphene due to structural distortions''}

\author[1,2]{\fnm{Tan-Phat} \sur{Nguyen}}
\equalcont{These authors contributed equally to this work.}

\author[3,4]{\fnm{Mykhailo} \sur{Klymenko}}
\equalcont{These authors contributed equally to this work.}

\author[1,2]{\fnm{Gary} \sur{Beane}}
\author[1,2]{\fnm{Mitko} \sur{Oldfield}}
\author[1]{\fnm{Kaijian} \sur{Xing}}
\author[1,2]{\fnm{Matthew} \sur{Gebert}}
\author[1,2,5]{\fnm{Semonti} \sur{Bhattacharyya}}
\author[1,2]{\fnm{Michael S.} \sur{Fuhrer}}
\author*[3,4,6]{\fnm{Jared H.} \sur{Cole}}\email{jared.cole@rmit.edu.au}
\author*[1,2]{\fnm{Agustin} \sur{Schiffrin}}\email{agustin.schiffrin@monash.edu}

\affil[1]{\orgdiv{School of Physics and Astronomy}, \orgname{Monash University}, \orgaddress{\city{Clayton},\state{VIC 3800}, \country{Australia}}}

\affil[2]{\orgdiv{ARC Centre of Excellence in Future Low-Energy Electronics Technologies}, \orgname{Monash University}, \orgaddress{\city{Clayton}, \state{VIC 3800}, \country{Australia}}}

\affil[3]{\orgdiv{School of Science}, \orgname{RMIT University}, \orgaddress{\city{Melbourne},  \state{VIC 3001}, \country{Australia}}}

\affil[4]{\orgdiv{ARC Centre of Excellence in Exciton Science}, \orgname{RMIT University}, \orgaddress{\city{Melbourne}, \postcode{3001}, \state{VIC 3001}, \country{Australia}}}

\affil[5]{\orgdiv{Huygens-Kamerlingh Onnes Laboratory}, \orgname{Leiden University}, \orgaddress{\city{PO Box 9504, 2300RA Leiden}, \country{The Netherlands}}}

\affil[6]{\orgdiv{ARC Centre of Excellence in Future Low-Energy Electronics Technologies}, \orgname{RMIT University}, \orgaddress{\city{Melbourne}, \state{VIC 3001}, \country{Australia}}}

\maketitle

\newpage
\section*{Supplementary Note 1. Sample Characterisation}

\begin{figure}
    \centering
    \includegraphics[width=1.0\linewidth]{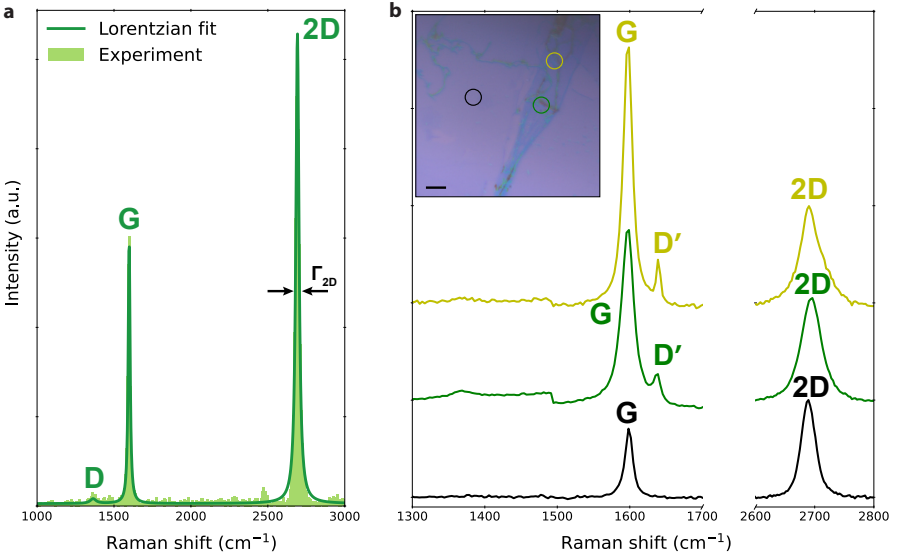}
    \caption{\textbf{Sample characterisation: Raman spectroscopy. a,} Raman spectrum consisting of an average of ${\sim}100$ spectra acquired at different graphene sample locations. \textbf{b,} Raman spectra acquired at different regions of graphene, showing G and 2D peaks, as well as D' peak in the crumpled area (green and yellow). Inset: optical microscopy image of graphene, showing clear clean single-layer (black circle) and darker crumpled bilayer (green and yellow circles) areas; circles indicate locations where Raman spectra were acquired. Scalebar: 10 $\mu$m.}
    \label{fig:Raman}
\end{figure}

Prior to our terahertz (THz) time-domain spectroscopy (THz-TDS) measurements, we characterised our graphene/SiO$_2$/Si samples via Raman spectroscopy and electrical measurements. We performed Raman spectroscopy on graphene at room temperature using a continuous-wave 532-nm-wavelength solid-state laser (with a $100\times$ optical objective, power at sample ${\sim} 0.1$ mW), and a Princeton Instruments Acton Spectra Pro SP-2750 spectrometer. Figure \ref{fig:Raman}a shows a Raman spectrum, corresponding to the average of ${\sim}100$ spectra acquired at different sample locations. This average Raman spectrum exhibits two prominent G (${\sim} 1690$ cm$^{-1}$) and 2D (${\sim} 2700$ cm$^{-1}$) peaks, characteristic of graphene. The full-width at half-maximum (FWHM) of the 2D peak is $\Gamma_\textrm{2D} \approx 29 \pm 2$ cm$^{-1}$, confirming that our sample consists mainly of single-layer graphene \cite{Pisana_2007, Das_2008, Ferrari_2006, Ferrari_2013}. We also observe a small a small D peak (${\sim} 1350$ cm$^{-1}$), indicative of some amount of disorder in the sample \cite{Pisana_2007}. 

Figure \ref{fig:Raman}b shows Raman spectra acquired at different locations of the graphene sample. In particular, spectra acquired at regions that appear darker in optical microscopy image (Fig. \ref{fig:Raman}b inset) exhibit an additional D' peak at ${\sim} 1620$ cm$^{-1}$. Also, the 2D peak for these darker regions is broader than for clearer regions ($\Gamma_\textrm{2D} \approx$ 35 cm$^{-1}$ for the yellow and green spectra compared to $\Gamma_\textrm{2D} \approx$ 25 cm$^{-1}$ for the black spectrum in Fig. \ref{fig:Raman}b). We therefore attribute the clear regions in the optical microscopy image to clean single-layer graphene, and the darker region -- with the additional D' peak and broader 2D peak in the Raman spectra -- to a crumpled strained graphene area (that is, effectively bilayer; see Fig. \ref{fig:foldings}a for schematic of crumpled graphene region) \cite{Couto_2014, Kun_2019, Georgi_2017, LeeSeunghyun_2010}. Such crumpled areas may originate in the graphene growth process and/or during the transfer of graphene from the etching solution onto the substrate \cite{BaiKeKe_2014, JeongSeonYu_2014, ChaeSeungJin_2009, Calado_2012}. Notably, these severely strained, crumpled graphene areas are associated with pseudo-magnetic fields that can be on the order of $\sim$100 T \cite{Couto_2014, Kun_2019, Georgi_2017}.

\begin{figure}
    \centering
    \includegraphics[width=0.5\linewidth]{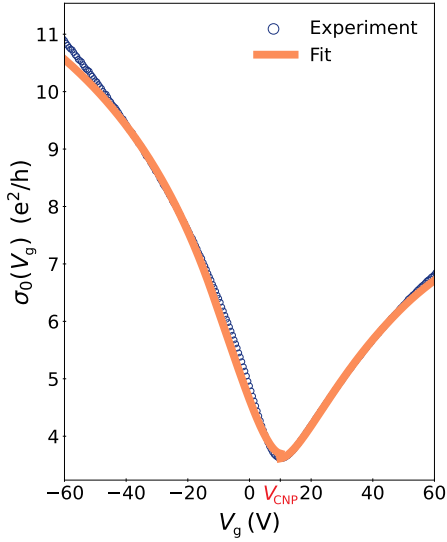}
    \caption{\textbf{Sample characterisation: four-point-probe measurement}. The electrical DC sheet conductivity, $\sigma_0$, as a function of gate voltage $V_{\textrm{g}}$, measured in a four-point-probe configuration. Orange curve: fit curve based on Eq. \ref{SI_eq_dc_sigma}.}
    \label{fig:4probe}
\end{figure}

We measured the electrical DC conductivity of graphene, $\sigma_0$, as a function of gate voltage $V_\textrm{g}$, via a conventional four-point-probe configuration (Fig. \ref{fig:4probe}; see Fig. 1a in the main text for sample schematic). As $V_\textrm{g}$ varies from -60 to 60 V, $\sigma_0$ decreases, reaches a non-zero minimum at $V_\textrm{g} = V_\textrm{CNP} = 10$ V -- that is, the charge neutrality point (CNP) -- and then increases again. $V_\textrm{CNP} = 10$ V indicates graphene is intrinsically $p$-doped.   

The DC conductivity of graphene $\sigma_0\left(V_\textrm{g}\right)$ can be expressed as \cite{Gosling_2021}: 

\begin{equation} \label{SI_eq_dc_sigma}
    \sigma_0 \left(V_\textrm{g}\right) = 
    \begin{cases}
        e \mu_{\textrm{e,h}}\left[ \frac{\delta n}{4} + \frac{n^2\left(V_\textrm{g}\right)}{\delta n} \right],& \text{for } \vert n\left(V_\textrm{g}\right)\vert < \delta n /2\\
        e \mu_{\textrm{e,h}} n\left(V_\textrm{g}\right),              & \text{for } \vert n\left(V_\textrm{g}\right)\vert > \delta n /2
    \end{cases}
\end{equation}
where $\mu_{\textrm{e,h}}$ is the electron/hole mobility, $\delta n$ is a parameter accounting for the non-zero conductivity at the CNP, and $n\left(V_\textrm{g}\right)$ is the effective carrier density given by:

\begin{equation}
    n\left(V_\textrm{g}\right) = \Bigg\vert \left[ \frac{1}{n_c\left(V_\textrm{g}\right)} + \frac{e\mu_{\textrm{e,h}}}{\rho_\textrm{s}} \right]^{-1} + n_0 \Bigg\vert
\end{equation}
Here, $n_0$ is the carrier density of graphene at $V_\textrm{g} = 0$ V (that is, intrinsic doping), $\rho_\textrm{s}$ is a $V_\textrm{g}$-independent resistivity given by short-range scattering in graphene \cite{Morozov_2008}, and $n_c\left(V_\textrm{g}\right) = C V_\textrm{g}/e$ is the geometrical carrier concentration. We estimated $n_c\left(V_\textrm{g}\right)$ using a parallel plate capacitor model, that is, with capacitance $C = \epsilon\epsilon_0A/d$, where $\epsilon = 3.9$ is the dielectric constant of SiO$_2$, and $A = 7 \times 7$ mm$^2$ and $d = 300$ nm are the area and thickness of the SiO$_2$ layer, respectively.

We fit our measured $\sigma_0 \left(V_\textrm{g}\right)$ in Fig. \ref{fig:4probe} to Eq. \eqref{SI_eq_dc_sigma}, separately for $V_\textrm{g} < V_\textrm{CNP} $ (i.e., $p$-doped, with $\mu = \mu_{\textrm{h}}$) and for $V_\textrm{g} > V_\textrm{CNP}$ (i.e., $n$-doped, with $\mu = \mu_{\textrm{e}}$), with $\mu_\textrm{h}$, $\mu_\textrm{e}$, $\delta n$ and $\rho_\textrm{s}$ as fitting parameters. This fitting procedure yielded (see orange curve in Fig. \ref{fig:4probe}) $\mu_\textrm{h} = 2700 \pm 25$ cm$^2$/Vs and $\mu_\textrm{e} = 2200 \pm 15$ cm$^2$/Vs. We can then estimate the Fermi level $E_\textrm{f}$ of graphene at a given $V_\textrm{g}$ as:

\begin{equation} \label{SI_eq_fermi}
     E_\textrm{f}\left(V_\textrm{g}\right) = \hbar v_\textrm{F} \sqrt{\frac{\pi\sigma_0 \left(V_\textrm{g}\right)}{e\mu}}
\end{equation}
where $v_\textrm{F}$ is the graphene Fermi velocity, and with $\mu = \mu_{\textrm{h}}$ for $V_\textrm{g} < V_\textrm{CNP}$ and $\mu = \mu_{\textrm{e}}$ for $V_\textrm{g} > V_\textrm{CNP}$.

\section*{Supplementary Note 2. Terahertz time-domain spectroscopy}

\begin{figure}
    \centering
    \includegraphics[width=1.0\linewidth]{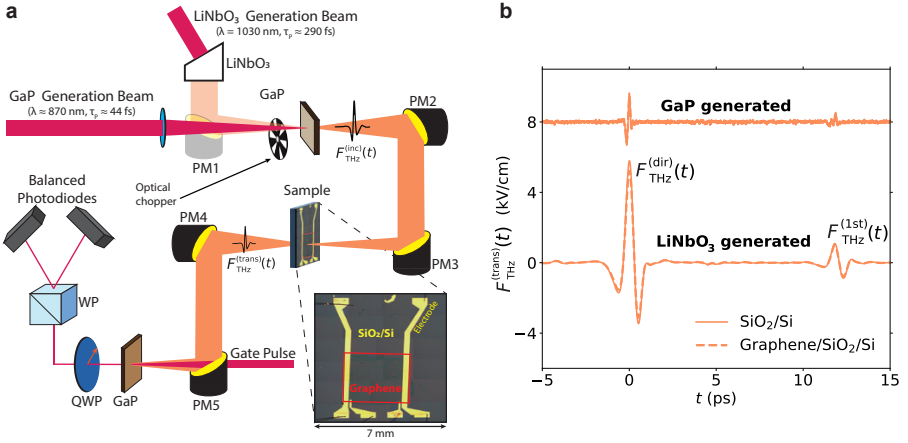}
    \caption{\textbf{THz time-domain spectroscopy (THz-TDS): experimental setup and time-domain THz waveforms. a,} Schematic of experimental setup for generation and detection of few-cycle THz waveforms. PM: off-axis parabolic mirror. QWP: quarter-wave plate. WP: Wollaston prism. \textbf{b,} Instantaneous time-dependent electric field of THz waveforms transmitted through graphene/SiO$_2$/Si (dashed) and reference SiO$_2$/Si sample areas (solid). These THz waveforms were generated via optical rectification of fundamental laser pulses in LiNbO$_3$ and GaP nonlinear crystals, and electro-optically sampled with another GaP crystal. See main text Fig. 1b for spectra of these waveforms.}
    \label{fig:THz-TDS}
\end{figure}

The THz-TDS experimental setup used to determine the complex dynamic sheet conductivity of graphene, $\Tilde{\sigma}(\omega)$,  is illustrated in Fig. \ref{fig:THz-TDS}a. We generated THz waveforms via optical rectification \cite{Hirori_2011} of fundamental laser pulses in both lithium niobate (LiNbO$_3$) and gallium phosphide (GaP) nonlinear crystals. In the LiNbO$_3$ configuration, the fundamental laser pulses were generated by a Yb:KGW laser (Carbide from Light Conversion; laser repetition rate 200 kHz; effective repetition rate 200/3 kHz,i.e., pulse picker 3 was used). In the GaP configuration, the fundamental laser pulses were generated by an optical parametric amplifier (OPA; Orpheus-F from Light Conversion, effective repetition rate 200 kHz,i.e., pulse picker 1 was used) pumped by the Yb:KGW laser.

The generated THz pulses, with instantaneous electric field $F_\textrm{THz}^{\textrm{(inc)}}(t)$,  were collimated and focused onto the sample by two off-axis parabolic mirrors (PM2, 3). The transmitted THz pulses, with instantaneous electric field $F_\textrm{THz}^{\textrm{(trans)}}(t)$, were then focused on another GaP crystal and detected via electro-optic sampling \cite{Saleh_1991}. A linearly polarised gate pulse (from the Yb:KGW laser in the LiNbO$_3$ configuration, or from the OPA in the GaP configuration) propagated through a second GaP crystal, which is birefringent when exposed to the THz electric field. The gate pulse was then transmitted through a quarter-wave plate (QWP), which converted it to circularly or elliptically polarised, depending on the time delay between THz and gate pulses. A Wollaston prism (WP) separated the gate pulse components propagating along the ordinary and extraordinary axes. The difference between the two components, which is proportional to the THz electric field, is detected by two balanced photodiodes and measured using a lock-in amplifier locked at a frequency of 6 kHz of an optical chopper.

Figure \ref{fig:THz-TDS}b illustrates the instantaneous, time-dependent electric field, $F_{\textrm{THz}}^{\textrm{(trans)}}(t)$, of the THz waveforms generated with LiNbO$_3$ and GaP, transmitted through the graphene/SiO$_2$/Si sample, and retrieved via electro-optic sampling. It contains a directly transmitted transient $F_{\textrm{THz}}^{\textrm{(dir)}}(t)$, and a transient $F_{\textrm{THz}}^{\textrm{(1st)}}(t)$ given by the first internal reflection inside the substrate. We derived the complex dynamic conductivity $\Tilde{\sigma}\left( \omega \right)$ of graphene from these transients (see Methods in the main text). All THz-TDS measurements were performed in a nitrogen environment with the sample at room temperature. 

\section*{Supplementary Note 3. Substrate's complex refractive index}

The complex refractive index of the SiO$_2$/Si substrate, $\Tilde{n}_{\textrm{SiO}_2\textrm{/Si}}\left(\omega \right)$, is required to calculate graphene's complex dynamic conductivity $\Tilde{\sigma}\left( \omega \right)$. We calculated $\Tilde{n}_{\textrm{SiO}_2\textrm{/Si}}\left(\omega \right) = \textrm{Re}[\Tilde{n}_{\textrm{SiO}_2\textrm{/Si}}\left(\omega \right)] + i\textrm{Im}[\Tilde{n}_{\textrm{SiO}_2\textrm{/Si}}\left(\omega \right)]$ via:

\begin{equation}
    \textrm{Re}[\Tilde{n}_{\textrm{SiO}_2\textrm{/Si}}\left(\omega \right)] = 1 + \frac{c\Delta\phi(\omega)}{\omega d}
\end{equation}
\begin{equation}
    \textrm{Im}[\Tilde{n}_{\textrm{SiO}_2\textrm{/Si}}\left(\omega \right)] = -\frac{c}{\omega d} \ln \left\{ \frac{\left[\textrm{Re}[\Tilde{n}_{\textrm{SiO}_2\textrm{/Si}}\left(\omega \right)] + 1 \right] ^ 2}{4\textrm{Re}[\Tilde{n}_{\textrm{SiO}_2\textrm{/Si}}\left(\omega \right)]} \Big\vert \Tilde{T}_{\textrm{SiO}_2\textrm{/Si}}^\textrm{(trans)}\left(\omega \right) \Big\vert \right\}
\end{equation}
Here, $\Tilde{T}_{\textrm{SiO}_2/\textrm{Si}}^\textrm{(trans)}\left(\omega \right) = \Tilde{F}_{\textrm{SiO}_2/\textrm{Si}}^\textrm{(trans)}\left(\omega \right) / \Tilde{F}_{\textrm{THz}}^\textrm{(inc)}\left(\omega \right)$ is the SiO$_2$/Si substrate transmission function, where $\Tilde{F}_{\textrm{SiO}_2/\textrm{Si}}^\textrm{(trans)}\left(\omega \right)$ and $\Tilde{F}_{\textrm{THz}}^\textrm{(inc)}\left(\omega \right)$ are the Fourier transforms of the THz waveforms transmitted through and incident onto the SiO$_2$/Si substrate, respectively, measured by THz-TDS; $c$ is the speed of light, $d$ is the SiO$_2$/Si substrate thickness and $\Delta\phi(\omega)=\arg[\Tilde{F}_{\textrm{SiO}_2/\textrm{Si}}^\textrm{(trans)}\left(\omega \right)]-\arg[\Tilde{F}_{\textrm{THz}}^\textrm{(inc)}\left(\omega \right)]$ is the difference between the spectral phases of $\Tilde{F}_{\textrm{SiO}_2/\textrm{Si}}^\textrm{(trans)}\left(\omega \right)$ and $\Tilde{F}_{\textrm{THz}}^\textrm{(inc)}\left(\omega \right)$ (computationally unwrapped by applying the technique reported in Ref. \cite{Jepsen_2019}). 

\begin{figure}
    \centering
    \includegraphics[width=0.65\linewidth]{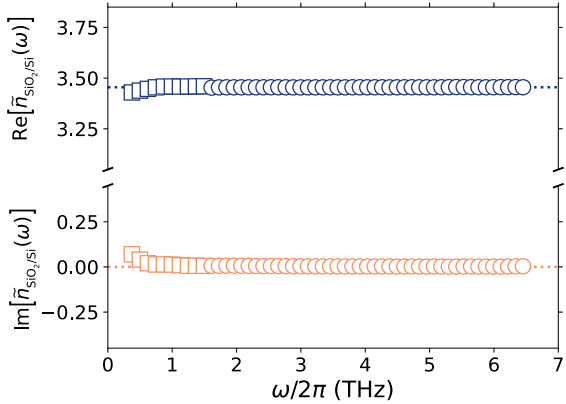}
    \caption{\textbf{Complex refractive index of SiO$_2$/Si substrate}. The Re[$\Tilde{n}_{\textrm{SiO}_2\textrm{/Si}}\left(\omega \right)$] and Im[$\Tilde{n}_{\textrm{SiO}_2\textrm{/Si}}\left(\omega \right)$] components of the substrate refractive index as a function of frequency, obtained from LiNbO$_3$ (square) and GaP (circle markers) THz-generation configurations.}
    \label{fig:substrate}
\end{figure}

Figure \ref{fig:substrate} shows $\textrm{Re}[\Tilde{n}_{\textrm{SiO}_2\textrm{/Si}}\left(\omega \right)]$ and $\textrm{Im}[\Tilde{n}_{\textrm{SiO}_2\textrm{/Si}}\left(\omega \right)]$ obtained using LiNbO$_3$ and GaP THz-generation configurations. The real part of the refractive index is frequency-independent while the imaginary part has infinitesimal values in the entire ${\sim} 0.1 - 7$ THz range. That is, $\Tilde{n}_{\textrm{SiO}_2\textrm{/Si}}\left(\omega \right) \approx n_{\textrm{SiO}_2\textrm{/Si}} \approx 3.45$.

It is worth noting that the remarkable agreement and consistency between the retrieved values of $\Tilde{n}_{\textrm{SiO}_2\textrm{/Si}}\left(\omega \right)$ obtained from the two different LiNbO$_3$ and GaP THz-generation configurations demonstrate the reliability of using these two setups for the accurate retrieval of graphene's THz conductivity $\Tilde{\sigma}(\omega)$.

\section*{Supplementary Note 4. Extended Drude-type models}

In the main text, we show that the conventional Drude model, with  $\Tilde{\sigma}_\textrm{Drude}\left(\omega\right) = \sigma_0 / (1 - i \omega \tau)$, cannot explain our measurements of $\Tilde{\sigma}\left( \omega \right)$ for charge-neutral graphene. In this section, we consider other extensions of the Drude model, i.e., localisation-modified Drude (LMD) model \cite{Lee_1993, Lee_1995}, Drude-Smith (DS) model \cite{Smith_2001, Buron_2014}, Drude-Lorentz (DL) model \cite{Patterson_2018}, and a microscopic graphene-specific model developed by Ando et al. \cite{Ando_2006}. 

The LMD model takes into account charge transport restriction due to disorder-induced weak localisation \cite{Lee_1993, Lee_1995}, with the complex dynamic conductivity given by \cite{Chen_2019}:

\begin{equation}
    \textrm{Re} \left[ \Tilde{\sigma}_\textrm{LMD} \left( \omega \right) \right] = \frac{\sigma_0}{1 + \omega ^2 \tau ^2} \left[ 1 - \frac{C}{(k_\textrm{F} v_\textrm{F})^2 \tau^2} \left(1 - \sqrt{3 \omega \tau} \right) \right]
\end{equation}

\begin{equation}
    \textrm{Im}\left[\Tilde{\sigma}_\textrm{LMD}\left(\omega\right) \right] = \frac{\sigma_0 \omega\tau}{1 + \omega^2 \tau^2} \left[ 1 + \frac{C}{\left( k_\textrm{F}v_\textrm{F}\right)^2 \tau^2} \left(\sqrt{6} - 1 - \sqrt{\frac{3}{\omega \tau}} \right) \right]
\end{equation}
where $C$ is a constant on the order of unity, $k_\textrm{F}$ is the Fermi wave vector and $v_\textrm{F}$ is the Fermi velocity, with $C/\left(k_\textrm{F}v_\textrm{F}\right)^2$ denoting the impact of the charge carrier localisation \cite{Lee_1993}. In this model, $\sigma_0$, $\tau$ and $C$ are fitting parameters.

The DS model assumes that a fraction of the initial carrier velocity -- represented by phenomenological coefficients $C_n$ -- is retained after the $n^\textrm{th}$  elastic scattering event (with $C_n \in [-1; 0]$) \cite{Smith_2001, Buron_2014}. In this DS model, the negative values of $C_n$ can result in the suppression of the low-frequency conductivity. Taking only the first scattering event into account (i.e., $C_n = 0$ for $n >1$), the DS dynamic conductivity is expressed as:

\begin{equation}
    \Tilde{\sigma}_\textrm{DS}\left(\omega\right) =  \frac{\sigma_0}{1 - i \omega \tau} \left[ 1 + \frac{C_1}{1 - i \omega \tau} \right].
\end{equation}

When $C_1 = 0$, the DS model reverts to the original Drude model. Note that the physical origin of the phenomenological parameter $C_1$ can be attributed to, e.g., significant backscattering, and/or diffusive restoring currents; however, its exact physical cause remains unclear \cite{Cocker_2017}. In this DS model, $\sigma_0$, $\tau$ and $C_1$ are treated as fitting parameters. 

Structural vibrations can affect the conductivity of the material, yet they are not taken into consideration by either the LMD or the DS models. The DL model is an extension of the Drude model that takes into account such vibrations phenomenologically, via Lorentz oscillators \cite{Patterson_2018}:

\begin{equation}
    \Tilde{\sigma}_\textrm{DL}\left(\omega\right) = \frac{\sigma_0}{1 - i \omega \tau} + \sum_{j} \frac{i \epsilon_0 \omega A_\textrm{L}^{(j)}}{\omega ^2 - {\omega_\textrm{L}^{(j)}} ^2 + i \omega \gamma_\textrm{L}^{(j)}}
\end{equation}
where $\epsilon_0$ is vacuum permittivity, and $A_\textrm{L}^{(j)}$, $\omega_\textrm{L}^{(j)}$ and $\gamma_\textrm{L}^{(j)}$ are amplitude, resonance frequency and broadening of the Lorentz oscillator $j$ \cite{Schubert_2004}, respectively. In this DL model, $\sigma_0$, $\tau$, $A_\textrm{L}^{(j)}$, $\omega_\textrm{L}^{(j)}$ and $\gamma_\textrm{L}^{(j)}$ serve as fitting parameters. We considered $j=1, 2, 3$, without significant differences of the fit for these different oscillator numbers.

Ando et al. \cite{Ando_2006} further proposed a microscopic model of the graphene dynamic conductivity taking into account carrier scattering by screened charged impurities:

\begin{equation}
    \Tilde{\sigma}_\textrm{Ando}\left(\omega\right) = \frac{e^2v_\textrm{F}^2}{2} \int\limits_{0}^{+\infty} d\epsilon D(E) \frac{\tau(E)}{1 - i\omega \tau(E)} \frac{df(E_\textrm{f}, T_\textrm{el})}{dE}
\end{equation}
where $e$ is the electron charge, $D(E)=2E/\left(\pi \hbar^2 v_\textrm{F}^2 \right)$ is the graphene density of states, $f(E_\textrm{f}, T_\textrm{el})$ is the Fermi-Dirac distribution as a function of Fermi level $E_\textrm{f}$ and electronic temperature $T_\textrm{el}$, and $\tau(E)$ is the energy-dependent carrier relaxation time (governed by scattering due to screened charged impurities, with $\tau \propto E$) \cite{Ando_2006, Das_2011}. In this model, the value of $E_\textrm{f}$ as a function of $V_\textrm{g}$ was determined from the four-point-probe measurements, allowing for a 10\% variation across different $V_\textrm{g}$ values to obtain a best fit (similar to two-component model fit in main text). 




\begin{figure}
    \centering
    \includegraphics[width=1.0\linewidth]{SI_Fig/SI_Fig_Multiple_Fits_Dev49_10.pdf}
    \caption{\textbf{Gate-dependent complex THz conductivity of graphene: comparison between Drude-type fits. a - h,} Real part of $\Tilde{\sigma}\left( \omega \right)$ for different $V_\textrm{g}$ values. \textbf{i - p, } Imaginary part of $\Tilde{\sigma}\left( \omega \right)$. Square and circle markers: experimental data from LiNbO$_3$ and GaP THz-generation configurations, respectively. Shaded areas: $\pm$ standard deviation. Dashed green curves: fit by localisation-modified Drude (LMD) model. Solid magenta curves: fit by Drude-Smith (DS) model.  Dashed blue curves: fit by Drude-Lorentz (DL) model. Solid red curves: fit by Ando model \cite{Ando_2006}. Dashed black curves: fit by Drude model. All Drude-typde models result in very similar fit curves.  
    }
    \label{fig:fittings}
\end{figure}

\begin{figure}
    \centering
    \includegraphics[width=0.8\linewidth]{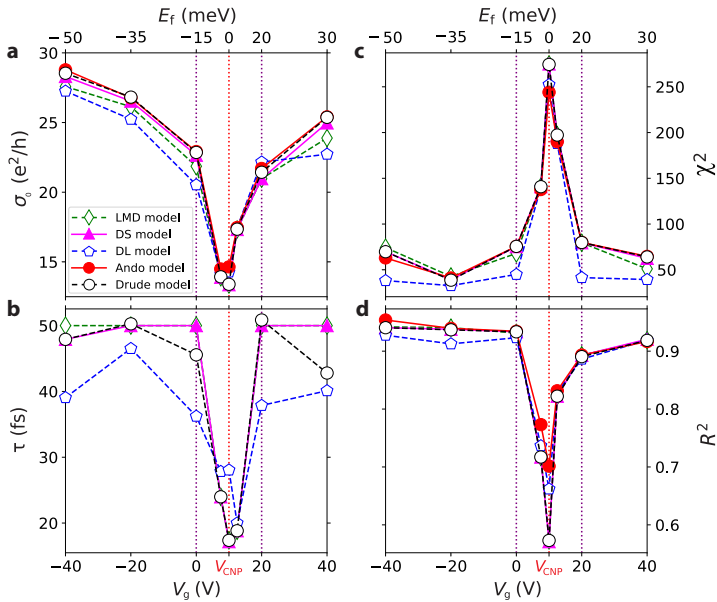}
    \caption{\textbf{Drude-type DC conductivity, carrier relaxation time constant and goodness of fit.} \textbf{a,} DC conductivity $\sigma_0$, \textbf{b,} carrier relaxation time $\tau$, \textbf{c,} Pearson's $\chi^2$ test and \textbf{d,} coefficient of determination $R^2$ as a function of $V_\textrm{g}$ and $E_\textrm{f}$, determined from fitting the measured $\Tilde{\sigma}\left( \omega \right)$ with the conventional Drude (black circles), localisation-modified Drude (LMD, green diamonds), Drude-Smith (DS, magenta triangles), Drude-Lorentz (DL, blue pentagon) and Ando (red circles) models. All Drude-type models yield very similar fit results.
    }
    \label{fig:fitting_results}
\end{figure}

\begin{figure}
    \centering
    \includegraphics[width=0.7\linewidth]{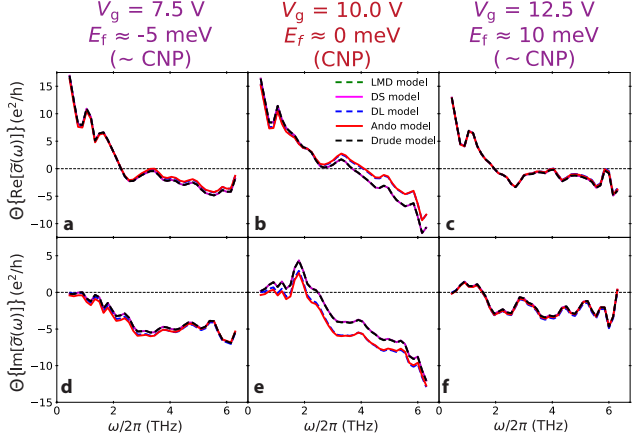}
    \caption{\textbf{Drude-type fit residuals near the charge neutrality point. a-c,} Residuals of $\mathrm{Re}[\Tilde{\sigma}(\omega)]$ fits given by Drude-type models. \textbf{d-f,} Residuals of $\mathrm{Im}[\Tilde{\sigma}(\omega)]$ fits. All Drude-type fits yield similar residuals. 
    }
    \label{fig:residuals}
\end{figure}

For comparison, we fit simultaneously the real and imaginary parts of the THz conductivity of graphene, $\Tilde{\sigma}\left( \omega \right)$, with the Drude, LMD, DS, DL and Ando models using the respective fitting parameters listed above for each model. 

Figure \ref{fig:fittings} shows the experimental $\textrm{Re}\left[\Tilde{\sigma}\left( \omega \right) \right]$ and $\textrm{Im}\left[\Tilde{\sigma}\left( \omega \right) \right]$, as well as the corresponding curves resulting from the Drude-type model fits, for different values of $V_\textrm{g}$ and $E_\textrm{f}$. All Drude-type models are in good agreement with the experimentally measured $\Tilde{\sigma}(\omega)$ for $p$-doped ($E_\textrm{f} \lesssim -15$ meV; see Fig. \ref{fig:fittings}a-c, i-k) and $n$-doped ($E_\textrm{f} \gtrsim 20$ meV; see Fig. \ref{fig:fittings}g-h, o-p) graphene. This agreement is further evidenced by the small Pearson's test values ($\chi^2 < 50$) and high coefficients of determination ($R^2 > 0.9$), as depicted in Fig. \ref{fig:fitting_results}c-d. 

In the case of charge-neutral graphene ($-5 \lesssim E_\textrm{f} \lesssim 10$ meV), none of the considered Drude-type models are able to accurately reproduce both $\textrm{Re}\left[\Tilde{\sigma}\left( \omega \right) \right]$ and $\textrm{Im}\left[\Tilde{\sigma}\left( \omega \right) \right]$ simultaneously (see Fig. \ref{fig:fittings}d-f, l-n). Note that all extended Drude-type models yield fitting curves, as well as values of $\sigma_0$ and $\tau$, that are comparable to the conventional Drude model (Fig. \ref{fig:fittings}a-b).

The inadequacy of the Drude-type models in explaining $\Tilde{\sigma}(\omega)$ for charge-neutral graphene is further evidenced by the substantial increase in $\chi^2$ (with $\chi^2_\textrm{CNP} / \chi^2_\textrm{doped} > 400 \%$), by the decrease of $R^2$ from $>0.9$ to $<0.6$ (Fig. \ref{fig:fitting_results}c-d), and by the large fit residuals (Fig. \ref{fig:residuals}).


\section*{Supplementary Note 5. Derivation of the two-component model}

In this section we present a full derivation of the two-component model for graphene's complex dynamic conductivity. The model is derived based on the time evolution of the density matrix $\rho$ given by the Liouville-von-Neumann equation, $i\hbar\partial_t\rho = [H, \rho]$. Here, a two-band model for a single Dirac cone is considered. Under the irradiation of a THz waveform with instantaneous electric field $F(t)$, the surface current density in the time domain can be expressed by using the density matrix formalism as:

\begin{equation}
    \mathbf{J}(t)=  \frac{\mathbf{e}}{\sqrt{2\pi}} \int d \omega \: \mathrm{e}^{i\omega t} \: \Tilde{F}(\omega) \: \Tilde{\sigma}(\omega) =-(e/L^2)\text{Tr} \left[\rho \: \mathbf{v} \right]
    \label{eq_current_density}
\end{equation}
where $\mathbf{e}$ is the polarisation vector ($\mathbf{e} = e_x\boldsymbol{\kappa}_x + e_y\boldsymbol{\kappa}_y$, with $\{ \boldsymbol{\kappa}_x, \boldsymbol{\kappa}_y \}$ being unit vectors defining a 2D Cartesian coordinate system), $\Tilde{F}(\omega)$ is the Fourier transform of the THz waveform electric field, $\Tilde{\sigma}(\omega)$ is the graphene sheet conductivity (as in the experiment, i.e., in units of conductance $e^2/h$), $e$ is the electron charge, and $L^2$ is the effective area of graphene exposed to the THz waveform. In the vicinity of the Dirac cone and in polar coordinates, the matrix elements of the velocity operator $\mathbf{v}$ read as:

\begin{equation}
    {\mathbf{v}_{\mathbf{k}', \mathbf{k}}^{(m, n)} = v_\textrm{F} \delta \left( \mathbf{k} - \mathbf{k}' \right)
    \begin{cases}
        \lambda(n) \left(\boldsymbol{\kappa}_x \cos \theta  + \boldsymbol{\kappa}_y \sin \theta  \right) , & n = m,\\
        i \lambda(n) \left(\boldsymbol{\kappa}_x \sin \theta  + \boldsymbol{\kappa}_y  \cos \theta \right), & n \neq m,\\
    \end{cases}}
    \label{v_me2}
\end{equation}

Here, $v_\textrm{F}$ is graphene's Fermi velocity, $n, m \in \{c, v\}$ are band indices ($c$: conduction, $v$: valence band), $\lambda(n)=1$ if $n=c$ and $\lambda(n)=-1$ if $n=v$, $\mathbf{k}$ and $\mathbf{k'}$ are electron wavevectors, and $\theta$ represent polar angles. 

We consider graphene's electronic density matrix $\rho$, whose (unitless) elements are associated with the occupation probability and polarisation of graphene's electronic states $\vert \mathbf{k}, n \rangle$. By decomposing $\rho$ into diagonal  ($\Bar{f}$), non-diagonal intraband ($f$)  and non-diagonal interband ($\pi$) components, $\rho = \Bar{f} + f + \pi$, and considering the system's Hamiltonian (see Methods in the main text), the quantum kinetic equations can be written as:

\begin{subequations}
\begin{align}
    \partial_t f_{n, \mathbf{k}', \mathbf{k}} &= i \Delta \omega_{\mathbf{k}, \mathbf{k}'}^{(n,n)} f_{n, \mathbf{k}', \mathbf{k}} +i \left(\Bar{f}_{n,\mathbf{k}'} - \Bar{f}_{n,\mathbf{k}} \right)  \Omega_{\mathbf{k}', \mathbf{k}}^{(n,n)}  + \Pi_{\mathbf{k}',\mathbf{k}}^{(n,n)},\\
    \partial_t \pi_{\mathbf{k}',\mathbf{k}} &= i \Delta \omega_{\mathbf{k}, \mathbf{k}'}^{(c,v)} \pi_{\mathbf{k}', \mathbf{k}} +i \left(\Bar{f}_{c,\mathbf{k}'} - \Bar{f}_{v,\mathbf{k}} \right)  \Omega_{\mathbf{k}',\mathbf{k}}^{(c,v)}  + \Pi_{\mathbf{k}', \mathbf{k}}^{(c,v)},\\
    \partial_t \Bar{f}_{n,\mathbf{k}} &= \lambda(n)\sum_{\mathbf{k}'} \text{Im}\left[ \Omega_{\mathbf{k}}^{(c,v)}  \pi_{\mathbf{k}}^*\right] + \lambda(n)\sum_{\mathbf{k}'} \text{Im}\left[ \Omega_{\mathbf{k}, \mathbf{k}'}^{(n,n)}  f_{n, \mathbf{k}', \mathbf{k}}^*\right] + \Pi_{\mathbf{k},\mathbf{k}}^{(n,n)},
\end{align}
\label{sbe}
\end{subequations}

Here, $\Omega_{\mathbf{k}', \mathbf{k}}^{(n,m)} = eF(t)\left[\mathbf{e} \cdot \mathbf{d}_{\mathbf{k}', \mathbf{k}}^{(n,m)} \right] / \hbar$ is the Rabi frequency, $\mathbf{d}_{\mathbf{k}', \mathbf{k}}^{(n,m)}$ is the dipole moment matrix element, $\Delta \omega_{\mathbf{k}, \mathbf{k}'} = \left( E_{c, \mathbf{k}} - E_{v,\mathbf{k}'} \right) / \hbar$ is the transition angular frequency, and $\Pi_{\mathbf{k}',\mathbf{k}}^{(n,m)}$ is the scattering term associated with a scattering potential $V$:

\begin{equation}
\Pi_{\mathbf{k}',\mathbf{k}}^{(n,m)}= -\frac{i}{\hbar} \left[V, \rho \right] =-\frac{i}{\hbar}\sum_{j,\mathbf{k}''} \left[ V_{\mathbf{k}',\mathbf{k}''}^{(n,j)}\rho_{\mathbf{k}'',\mathbf{k}}^{(j,m)} - V_{\mathbf{k}'',\mathbf{k}}^{(j,m)}\rho_{\mathbf{k}',\mathbf{k}''}^{(n,j)}\right],
\label{Cap_pi}
\end{equation}

Assuming that the perturbation from the incident THz electric field is sufficiently weak to maintain the carrier distribution in the Dirac cone near its thermal equilibrium value, the density matrix $\rho(t)$ can be expressed as:

\begin{equation}
    \rho(t) = \rho^{(0)} + \rho^{(1)}(t)
\end{equation}
where $\rho^{(0)}=\Bar{f}^{(0)}$ represents the thermal equilibrium Fermi-Dirac distribution, while $\rho^{(1)}(t)=\Bar{f}^{(1)}(t)+f^{(1)}(t)+\pi^{(1)}(t)$ corresponds to the first-order time-dependent density matrix. We neglect higher-order perturbations; that is, we assume we are in the linear regime. The time evolution of $\rho(t)$ becomes \cite{Liu_2018}:

\begin{equation}
    i\hbar \partial_t \rho^{(1)}(t) = \left[H^{(0)}, \rho^{(1)}(t) \right] + \left[H^{(1)}, \rho^{(0)} \right].
\end{equation}
where $H^{(0)}$ is zeroth-order Hamiltonian (independent of incident field) and $H^{(1)} = H_\textrm{I}$ is the first-order Hamiltonian describing the interaction between graphene and incident THz electric field (Hamiltonian is defined in the Methods of the main text). Note that $H^{(0)} \neq H_0$, and that $H^{(0)}$ contains $H_{\mathrm{scatt}}$ accounting for scattering of electrons by the scattering potential $V$.  

In Eq. \eqref{sbe}c, the first two terms are of second-order with respect to the incident field, thus they do not contribute to the linear response of the system and can be neglected \cite{Lindberg}. The system of Eqs. \eqref{sbe}a-c can be rewritten as:

\begin{subequations}
\begin{align}
    \partial_t f_{n,\mathbf{k}', \mathbf{k}}^{(1)} &= i \Delta \omega_{\mathbf{k}, \mathbf{k}'}^{(n,n)} f_{n,\mathbf{k}', \mathbf{k}}^{(1)} +i \left(\Bar{f}_{n,\mathbf{k}'}^{(0)} - \Bar{f}_{n,\mathbf{k}}^{(0)} \right)  \Omega_{\mathbf{k}', \mathbf{k}}^{(n,n)}  + \Pi_{\mathbf{k}',\mathbf{k}}^{(n,n)(1)},\\
    \partial_t \pi_{\mathbf{k}', \mathbf{k}}^{(1)} &= i \Delta \omega_{\mathbf{k}, \mathbf{k}'}^{(c,v)} \pi_{\mathbf{k}', \mathbf{k}}^{(1)} +i \left(\Bar{f}_{c,\mathbf{k}'}^{(0)} - \Bar{f}_{v,\mathbf{k}}^{(0)} \right) \Omega_{\mathbf{k}, \mathbf{k}}^{(c,v)}  + \Pi_{\mathbf{k}',\mathbf{k}}^{(c,v)(1)},\\
    \partial_t \Bar{f}_{n,\mathbf{k}}^{(1)} &= \Pi_{\mathbf{k},\mathbf{k}}^{(n,n)(1)}.
\end{align}
\label{sbe1}
\end{subequations}
where $\Pi_{\mathbf{k'},\mathbf{k}}^{(n,m)(1)}$ is defined by Eq. \eqref{Cap_pi} where $\rho$ is replaced by $\rho^{(1)}$.

Note that the applied field does not induce changes in the density matrix diagonal elements in Eq. \eqref{sbe1}c in the first order, however, it can contribute here indirectly via the scattering terms. 

The second term in Eq. \eqref{sbe1}a reads:

\begin{equation}
    i \left(\Bar{f}_{n,\mathbf{k}'}^{(0)} - \Bar{f}_{n,\mathbf{k}}^{(0)} \right)  \Omega_{\mathbf{k}', \mathbf{k}}^{(n,n)} =  \frac{ieF(t)}{\hbar} \frac{v_\textrm{F}\left(\Bar{f}_{n,\mathbf{k}'}^{(0)} - \Bar{f}_{n,\mathbf{k}}^{(0)} \right)}{\Delta \omega_{\mathbf{k}, \mathbf{k}'}^{(n,n)}} \frac{\mathbf{e} \cdot \mathbf{k}}{\lvert \mathbf{k} \rvert}
\end{equation}

In the linear regime, and without considering scattering, the incident THz electric field itself is not capable of providing significant momentum transfer. Therefore, the first-order microscopic intraband polarisation $f_{n,\mathbf{k}', \mathbf{k}}^{(1)}$ is non-zero only when $\mathbf{k}' \rightarrow \mathbf{k}$. Taking this limit,  we obtain:

\begin{equation}
    \lim_{\mathbf{k}' \rightarrow \mathbf{k}} \frac{v_\textrm{F}\left(\Bar{f}_{n,\mathbf{k}'}^{(0)} - \Bar{f}_{n,\mathbf{k} }^{(0)} \right) }{\Delta \omega_{\mathbf{k}, \mathbf{k}'}^{(n,n)}} = \lim_{\mathbf{k}' \rightarrow \mathbf{k}}  \frac{v_\textrm{F}\left(\Bar{f}_{n,\mathbf{k'}}^{(0)} - \Bar{f}_{n,\mathbf{k}}^{(0)} \right) }{v_\textrm{F} \left(\lvert \mathbf{k} \rvert - \lvert \mathbf{k}' \rvert \right)}  = \nabla_{\mathbf{k}} \Bar{f}_{n,\mathbf{k}}^{(0)}.
\end{equation}


Therefore, 

\begin{equation}
    i \left(\Bar{f}_{n,\mathbf{k}'}^{(0)} - \Bar{f}_{n,\mathbf{k}}^{(0)} \right)  \Omega_{\mathbf{k}', \mathbf{k}}^{(n,n)} \approx eF(t)\left( \mathbf{e} \cdot \nabla_{\mathbf{k}} \Bar{f}_{n,\mathbf{k}}^{(0)} \right)
\end{equation}

The third term in Eq. \eqref{sbe1}a can be approximated as \cite{Kitamura_2015}:

\begin{equation}
     \lim_{\mathbf{k}' \rightarrow \mathbf{k}}  \Pi_{\mathbf{k}',\mathbf{k}}^{(n,n)(1)} = -\frac{f_{n,\mathbf{k}}^{(1)}}{\tau}
\end{equation}
where constant $\tau$ is the intraband carrier momentum relaxation time constant. 

The scattering term in Eq. \eqref{sbe1}c reads:

\begin{equation}
     \begin{split}
        \Pi_{\mathbf{k},\mathbf{k}}^{(n,n)(1)} &=-\frac{i}{\hbar}\sum_{j,\mathbf{k}'} \left[ V_{\mathbf{k},\mathbf{k}'}^{(n,j)}\rho_{\mathbf{k}',\mathbf{k}}^{(j,n)(1)} - V_{\mathbf{k}',\mathbf{k}}^{(j,n)}\rho_{\mathbf{k},\mathbf{k}'}^{(n,j)(1)} \right] =\\
        &-\frac{1}{\hbar} \text{Im} \left[ V_{\mathbf{k},\mathbf{k}}^{(n,n)}\Bar{f}_{n, \mathbf{k}}^{(1)} \right]-\frac{1}{\hbar} \text{Im} \left[ V_{\mathbf{k},\mathbf{k}}^{(n,n)}f_{n, \mathbf{k}}^{(1)} \right]-\frac{1}{\hbar} \text{Im} \left[ V_{\mathbf{k},\mathbf{k}}^{(c,v)}\pi_{\mathbf{k}, \mathbf{k}}^{(1)} \right]\\
        &-\frac{i}{\hbar}\sum_{j,\mathbf{k}' \neq \mathbf{k}} \left[ V_{\mathbf{k},\mathbf{k}'}^{(n,j)}\rho_{\mathbf{k}',\mathbf{k}}^{(j,n)(1)} - V_{\mathbf{k}',\mathbf{k}}^{(j,n)}\rho_{\mathbf{k},\mathbf{k}'}^{(n,j)(1)} \right] 
     \end{split}
\end{equation}

Note that the product $V_{\mathbf{k},\mathbf{k}}^{(n,n)}\Bar{f}_{n, \mathbf{k}}^{(1)}$ is real, i.e. $\text{Im} \left[ V_{\mathbf{k},\mathbf{k}}^{(n,n)}\Bar{f}_{n, \mathbf{k}}^{(1)} \right]=0$. Also, we assume that: (i) scattering involving electronic transitions within a same band is significantly more probable than scattering involving interband transitions, i.e., $V_{\mathbf{k},\mathbf{k}}^{(n,n)} \gg V_{\mathbf{k},\mathbf{k}}^{(c,v)}$; (ii) in the considered few-THz frequency range, the intraband response remains larger than the interband response, i.e., $f_{n, \mathbf{k}}^{(1)} > \pi_{\mathbf{k}, \mathbf{k}}^{(1)}$ \cite{Emani_2015}. This leads to:

\begin{equation}
     \begin{split}
        \Pi_{\mathbf{k},\mathbf{k}}^{(n,n)(1)} = -\frac{1}{\hbar} \text{Im} \left[ V_{\mathbf{k},\mathbf{k}}^{(n,n)}f_{n, \mathbf{k}}^{(1)} \right]-\frac{i}{\hbar}\sum_{j,\mathbf{k}' \neq \mathbf{k}} \left[ V_{\mathbf{k},\mathbf{k}'}^{(n,j)}\rho_{\mathbf{k}',\mathbf{k}}^{(j,n)} - V_{\mathbf{k}',\mathbf{k}}^{(j,n)}\rho_{\mathbf{k},\mathbf{k}'}^{(n,j)} \right] 
     \end{split}
     \label{scatterings_for_diag}
\end{equation}

It has been previously shown in  Ref. \cite{Kitamura_2015} that the sum in the last term in the right-hand side of Eq. \eqref{scatterings_for_diag} can be neglected when scattering remains moderate. With these approximations, the kinetic equations Eqs. \eqref{sbe1}a-c become:

\begin{subequations}
\begin{align}
    \partial_t f_{n,\mathbf{k}}^{(1)} &=  \frac{ e F(t)}{\hbar}\left(\mathbf{e} \cdot \nabla_{\mathbf{k}} \Bar{f}_{n,\mathbf{k}}^{(0)} \right)   -\frac{f_{n,\mathbf{k}}^{(1)}}{\tau},\\
    \partial_t \pi_{\mathbf{k}', \mathbf{k}}^{(1)} &= i \Delta \omega_{\mathbf{k}, \mathbf{k}'}^{(c,v)} \pi_{\mathbf{k}', \mathbf{k}}^{(1)} +i \left(\Bar{f}_{c,\mathbf{k}'}^{(0)} - \Bar{f}_{v,\mathbf{k}}^{(0)} \right) \Omega_{\mathbf{k}, \mathbf{k}}^{(c,v)}  +  \Pi_{\mathbf{k}', \mathbf{k}}^{(c,v)(1)},\\
    \partial_t \Bar{f}_{n,\mathbf{k}}^{(1)} & =  -\frac{1}{\hbar} \text{Im} \left[ V_{\mathbf{k},\mathbf{k}}^{(n,n)}f_{n, \mathbf{k}}^{(1)} \right].
    \end{align}
\label{sbe2}
\end{subequations}


The scattering potential $V$ causes the relaxation of the non-equilibrium distribution function. As an ansatz, we therefore propose a solution for $\Bar{f}_{n,\mathbf{k}}^{(1)}(t)$ in Eq. \eqref{sbe2}c in the form of $\Bar{f}_{n,\mathbf{k}}^{(1)} (t) = a(t)e^{-\alpha t}$:

\begin{equation}
    -\alpha \, a(t)e^{-\alpha t} + \left[ \partial_t a(t) \right]e^{-\alpha t} =  -\frac{1}{\hbar} \text{Im} \left[ V_{\mathbf{k},\mathbf{k}}^{(n,n)}f_{n, \mathbf{k}}^{(1)} \right].
    \label{eq_osc}
\end{equation}

Assuming that $a(t)$ varies slowly in comparison to the fast decay of the first-order distribution function $\Bar{f}_{n,\mathbf{k}}^{(1)}$ (where an out-of-equilibrium first-order distribution  $\Bar{f}_{n,\mathbf{k}}^{(1)}$ decays to equilibrium typically on timescales of 10-100 fs \cite{Brida_2013, Johannsen_2013, Winzer_2010}, i.e., typical characteristic frequencies of 10-100 THz) -- that is, with $\alpha$ larger than $\sim$7 THz, the frequency of the fastest oscillating Fourier component of the incident THz waveform -- the second term in the left-hand side of Eq. \eqref{eq_osc} can be neglected. As a result, $\alpha=V_{\mathbf{k},\mathbf{k}}^{(n,n)}/\hbar$, and we obtain:



\begin{subequations}
\begin{align}
    \partial_t f_{n,\mathbf{k}}^{(1)} &=  \frac{ e F(t)}{\hbar}\left(\mathbf{e} \cdot \nabla_{\mathbf{k}} \Bar{f}_{n,\mathbf{k}}^{(0)} \right)   -\frac{f_{n,\mathbf{k}}^{(1)}}{\tau},\\
    \partial_t \pi_{\mathbf{k}', \mathbf{k}}^{(1)} &= i \Delta \omega_{\mathbf{k}, \mathbf{k}'}^{(c,v)} \pi_{\mathbf{k}', \mathbf{k}}^{(1)} +i \left(\Bar{f}_{c,\mathbf{k}'}^{(0)} - \Bar{f}_{v,\mathbf{k}}^{(0)} \right) \Omega_{\mathbf{k}, \mathbf{k}}^{(c,v)}  +  \Pi_{\mathbf{k}', \mathbf{k}}^{(c,v)(1)},\\
    \Bar{f}_{n,\mathbf{k}}^{(1)} & =  \text{Im} \left[ f_{n, \mathbf{k}}^{(1)} \right].
    \end{align}
\label{sbe3}
\end{subequations}

Note that Eq. \eqref{sbe3}c is reminiscent of the relationship between the retarded Green's function and the lesser Green's function (electron correlation function) according to Lehmann's representation \cite{Bruus_2004}. 

Equation \eqref{sbe3}a can be solved analytically:

\begin{equation}
     f_{n,\mathbf{k}}^{(1)}(t) = \frac{e}{\hbar} \left( \mathbf{e} \cdot \nabla_{\mathbf{k}}\Bar{f}_{n,\mathbf{k}}^{(0)} \right) \int\limits_{-\infty}^{t} dt'  F(t') e^{ \frac{t' - t}{\tau} } 
     \label{distr_func1}
\end{equation}

In the frequency domain, Eq. \eqref{distr_func1} reads:

\begin{equation}
    f_{n,\mathbf{k}}^{(1)}(\omega) = \frac{\tau e \Tilde{F}(\omega) \left( \mathbf{e} \cdot \nabla_{\mathbf{k}}\Bar{f}_{n,\mathbf{k}}^{(0)} \right)}{\hbar \left(1 - i\omega \tau \right)} 
    \label{distr_func2}
\end{equation}


Assuming that the electromagnetic field is linearly polarised along the $x$-axis, this expression can be rewritten in the polar coordinates as follows:

\begin{equation}
    f_{n,\mathbf{k}}^{(1)}(\omega) = \frac{\tau e \Tilde{F}(\omega) \partial_k \Bar{f}_{c,k}^{(0)} \cos \theta}{\hbar \left(1 - i\omega \tau \right)}.
    \label{distr_func3}
\end{equation}

Let us now consider Eq. \eqref{sbe3}b when $\mathbf{k}' \rightarrow \mathbf{k}$. The scattering term can be decomposed as:

\begin{subequations}
\begin{align}
   \Pi_{\mathbf{k},\mathbf{k}}^{(c,v)(1)} &= -\frac{i}{\hbar}\left[ V_{\mathbf{k},\mathbf{k}}^{(c,v)}\Bar{f}_{v,\mathbf{k}}^{(1)} - \Bar{f}_{c,\mathbf{k}}^{(1)} V_{\mathbf{k},\mathbf{k}}^{(c,v)} \right] \\  
   &-\frac{i}{\hbar}\sum_{\mathbf{k}'} \left[ V_{\mathbf{k},\mathbf{k}'}^{(c,c)}\pi^{(1)}_{\mathbf{k}',\mathbf{k}} - \pi^{(1)}_{\mathbf{k},\mathbf{k}'} V_{\mathbf{k}',\mathbf{k}}^{(v,v)} \right] \\ 
    &-\frac{i}{\hbar}\sum_{\mathbf{k}'} \left[ V_{\mathbf{k},\mathbf{k}'}^{(c,v)}\rho_{\mathbf{k}',\mathbf{k}}^{(v,v)(1)} - \rho_{\mathbf{k},\mathbf{k}'}^{(c,c)(1)} V_{\mathbf{k}',\mathbf{k}}^{(c,v)} \right] 
\end{align}
\label{dephase}
\end{subequations}

The first term, \eqref{dephase}a, depends linearly on the scattering potential $V$. For disordered graphene, the spatial variation of this scattering potential is random, and can take both positive and positive values, i.e., its real-space integral is close to zero. The second term, \eqref{dephase}b, is of second-order with respect to the scattering potential and can result in constructive or destructive interference with coherent interband processes. The term \eqref{dephase}c contributes to dephasing and can be accounted for via the dephasing time approximation:

\begin{equation}
    \Pi_{\mathbf{k},\mathbf{k}}^{(c,v)(1)} \approx - \frac{i}{\hbar}\sum_{\mathbf{k}'} \left[ V_{\mathbf{k},\mathbf{k}'}^{(c,c)}\pi^{(1)}_{\mathbf{k}',\mathbf{k}} - \pi^{(1)}_{\mathbf{k},\mathbf{k}'}V_{\mathbf{k}',\mathbf{k}}^{(v,v)}  \right] -  \gamma \pi^{(1)}_{\mathbf{k}, \mathbf{k}},
    \label{eq_scat_nondiagonal}
\end{equation}
where $\gamma$ is a  phenomenological quantity describing the dephasing rate of the interband polarisation \cite{Haug_2004}. With this approximation, the solution of Eq. \eqref{sbe3}b in the frequency domain for $\mathbf{k}=\mathbf{k}'$ reads:

\begin{equation}
    \begin{split}
        \pi_{\mathbf{k}}^{(1)}(\omega) =& \frac{ie \Tilde{F}(\omega)\left(\mathbf{e} \cdot \mathbf{d}_{\mathbf{k}}\right) \left(\Bar{f}_{c,\mathbf{k}}^{(0)} - \Bar{f}_{v,\mathbf{k}}^{(0)} \right)}{\hbar \left(\omega - \Delta \omega_{\mathbf{k}, \mathbf{k}} + i\gamma  \right)} -\\
        &\frac{i\sum_{\mathbf{k}'} \left[ V_{\mathbf{k},\mathbf{k}'}^{(c,c)} \pi_{\mathbf{k}', \mathbf{k}}^{(1)}(\omega) - \pi_{\mathbf{k}, \mathbf{k}'}^{(1)}(\omega)V_{\mathbf{k}',\mathbf{k}}^{(v,v)}  \right] 
        }{\hbar \left(\omega - \Delta \omega_{\mathbf{k}, \mathbf{k}} + i\gamma  \right)}
    \end{split}
    \label{polarization}
\end{equation}
where $\mathbf{e} \cdot \mathbf{d}_{\mathbf{k}}=(e/k)\left(e_x \sin \theta + e_y \cos \theta \right)$.

The second term on the right-hand side of Eq. \eqref{polarization} is dependent on $\pi_{\mathbf{k}', \mathbf{k}}^{(1)}(\omega)$, which is yet to be determined. For this calculation of $\pi_{\mathbf{k}', \mathbf{k}}^{(1)}(\omega)$ [given by Eq. \eqref{sbe3}b], we simplify the scattering term $\Pi_{\mathbf{k}',\mathbf{k}}^{(c,v)(1)}$ by just keeping the term Eq. \eqref{dephase}a, which depends linearly on the scattering potential $V$ \cite{Kitamura_2015}. From this approximation and from Eq. \eqref{sbe3}b, we obtain in the frequency domain for $\textbf{k} \neq \textbf{k}'$: 

\begin{equation}
        \pi_{\mathbf{k}', \mathbf{k}}^{(1)}(\omega) = - \frac{iV_{\mathbf{k}', \mathbf{k}}^{(c,v)} \left[\Bar{f}_{c,\mathbf{k}'}^{(1)}(\omega) - \Bar{f}_{v,\mathbf{k}}^{(1)}(\omega) \right]}{\hbar \left(\omega - \Delta \omega_{\mathbf{k}', \mathbf{k}} + i\gamma'  \right)}
    \label{polarization_nondiag}
\end{equation}
where $\gamma'$ is the dephasing rate for non-momentum-conserving interband coherences. Note that $\gamma \neq \gamma'$.   

Substituting Eq. \eqref{distr_func2} into Eq. \eqref{sbe3}c and then into Eq. \eqref{polarization_nondiag}, we obtain:

\begin{equation}
    \pi_{\mathbf{k}', \mathbf{k}}^{(1)}(\omega) = \frac{e\Tilde{F}(\omega)V_{\mathbf{k}', \mathbf{k}}^{(c,v)} \left( \partial_k \Bar{f}_{c,k'}^{(0)} \cos \theta - \partial_k \Bar{f}_{v,k}^{(0)} \cos \theta \right) }{\hbar^2 \left(\omega - \Delta \omega_{\mathbf{k}', \mathbf{k}} + i\gamma'  \right)}\text{Im} \left[ \frac{\tau}{1 - i\omega \tau } \right]
    \label{nondiag_pi}
\end{equation}

We simplify Eq. \eqref{polarization} by defining the factor $$\Theta=\sum_{\mathbf{k}'} \left[ V_{\mathbf{k},\mathbf{k}'}^{(c,c)}\pi_{\mathbf{k}',\mathbf{k}}^{(1)} - \pi_{\mathbf{k},\mathbf{k}'}^{(1)} V_{\mathbf{k}',\mathbf{k}}^{(v,v)} \right]$$

The sum over $\mathbf{k}$ can be transformed into an integral via:

\begin{equation}
    \sum_{\mathbf{k}} \rightarrow \frac{L^2}{(2 \pi)^2} \int\limits_0^{2 \pi} d \theta \int\limits_0^{\infty} dk \, k
    \label{k_sum}
\end{equation}

This leads to the following expression for $\Theta$:

\begin{equation}
\begin{split}
    \Theta &= \Tilde{F}(\omega)  \text{Im} \left[ \frac{\tau}{1-i \omega \tau } \right] \frac{L^2}{(2 \pi)^2} \int\limits_0^{2\pi} d \theta'  \int\limits_0^{\infty} dk' k' \times \\  
    &\left\{   \frac{eV_{\mathbf{k},\mathbf{k}'}^{(c,c)}V_{\mathbf{k}',\mathbf{k}}^{(c,v)}}{\hbar^2}   \frac{\left[ \partial_{k'} \Bar{f}_{c,k'}^{(0)} \cos \theta' - \partial_{k} \Bar{f}_{v,k}^{(0)} \cos \theta \right]}{\omega - \Delta \omega_{k, k'} + i\gamma'}- \right.\\
    & \left.\frac{eV_{\mathbf{k},\mathbf{k}'}^{(c,v)} V_{\mathbf{k}',\mathbf{k}}^{(v,v)} }{\hbar^2}   \frac{\left[\partial_{k} \Bar{f}_{c,k}^{(0)} \cos \theta - \partial_{k'} \Bar{f}_{v,k'}^{(0)} \cos \theta' \right]}{\omega - \Delta \omega_{k', k} + i\gamma'} \right\}
\end{split}
\end{equation}

We assume that, in real space, the scattering potential $V$ varies very abruptly as a function of position (i.e., within $\sim$1 nm, i.e., several times the lattice constant of graphene), that is, in  reciprocal space, it varies smoothly as a function of $k$ in the vicinity of the Dirac cone (maintaining, however, the angular dependence given by the Bloch functions in graphene):

\begin{equation}
\begin{split}
    \Theta &= \Tilde{F}(\omega) \text{Im} \left[ \frac{\tau}{1-i \omega \tau } \right] \frac{L^2}{(2 \pi)^2} \frac{e}{\hbar^2}  \times \\  
    &\left\{ \int\limits_0^{2\pi} d \theta'  V_{\theta,\theta'}^{(c,c)} V_{\theta',\theta'}^{(c,v)}   \int\limits_0^{\infty} dk' k'    \frac{\left[ \partial_{k'} \Bar{f}_{c,k'}^{(0)} \cos \theta' - \partial_{k} \Bar{f}_{v,k}^{(0)} \cos \theta \right]}{\omega - \Delta \omega_{k, k'} + i\gamma'} \right.\\
    & \left. - \int\limits_0^{2\pi} d \theta' V_{\theta,\theta'}^{(c,v)} V_{\theta',\theta}^{(v,v)}   \int\limits_0^{\infty} dk' k'   \frac{\left[\partial_{k} \Bar{f}_{c,k}^{(0)} \cos \theta - \partial_{k'} \Bar{f}_{v,k'}^{(0)} \cos \theta' \right]}{\omega - \Delta \omega_{k', k} + i\gamma'} \right\}
\end{split}
\label{Sigma11}
\end{equation}



The scattering potential matrix elements in Eq. \eqref{Sigma11} can be associated with either a scalar scattering potential or vector scattering potential. For any scalar potential $V(\textbf{r})$, e.g., given by charged impurities \cite{Ando_2006}, these scattering potential matrix elements are:

\begin{equation}
    V_{\mathbf{k}, \mathbf{k}'}^{(n, m)} = \langle {\lambda', \mathbf{k}'} \vert {V(\mathbf{r}) \mathbf{I}} \vert {\lambda, \mathbf{k}} \rangle =\left[ e^{i \left(\theta' - \theta \right)} + \lambda(n) \lambda(m) \right] V_{\mathbf{k}, \mathbf{k}'}
    \label{eq_scalar_pot}
\end{equation}
where $\mathbf{I}$ is the 2 $\times$ 2 identity matrix and $V_{\mathbf{k}, \mathbf{k}'} = (1/L^2) \int d \mathbf{r} V(\mathbf{r}) e^{i \left(\mathbf{k} - \mathbf{k}' \right)\mathbf{r}}$.

Note that the expression
\begin{equation}
    V_{\mathbf{k}, \mathbf{k}'}^{(n,m)} V_{\mathbf{k}', \mathbf{k}}^{(m, m)} = i  V_{\mathbf{k}, \mathbf{k}'}^2 \left( \sin \theta' \cos \theta - \cos \theta' \sin\theta \right)
    \label{scalar_product1}
\end{equation}
is symmetric with respect to the exchange of band indices.

Conversely, for interactions of electrons with a vector scattering potential $\textbf{A}(\mathbf{r}) = A^{(x)} (\mathbf{r}) \: \boldsymbol{\kappa}_x + A^{(y)} (\mathbf{r}) \: \boldsymbol{\kappa}_y$ (for example, associated with out-of-plane pseudo-magnetic fields in graphene caused by significant strain or structural deformations such as folds, bumps, crumpled areas \cite{Kun_2019}), the scattering potential matrix elements are given by:
\begin{equation}
\begin{split}
    V_{\mathbf{k}, \mathbf{k}'}^{(n, m)} &=  e v_\textrm{F}\langle n, \mathbf{k}' \vert \boldsymbol{\sigma} \cdot \mathbf{A}  \vert m, \mathbf{k} \rangle \\ 
    &= \frac{e v_\textrm{F}}{2} \left[\lambda(m) e^{i \theta} \left(A_{\mathbf{k}, \mathbf{k}'}^{(x)} -iA_{\mathbf{k}, \mathbf{k}'}^{(y)} \right)  + \lambda(n) e^{-i \theta'} \left(A_{\mathbf{k}, \mathbf{k}'}^{(x)} + iA_{\mathbf{k}, \mathbf{k}'}^{(y)} \right) \right] 
\end{split}
        \label{mat_el_vec_pot}
\end{equation}
where $\boldsymbol{\sigma}$ is the vector of Pauli matrices, and

\begin{equation}
    A_{\mathbf{k}, \mathbf{k}'}^{(x,y)}=\frac{1}{L^2}\int d\mathbf{r} A^{(x,y)}(\mathbf{r}) e^{i(\mathbf{k}- \mathbf{k}')\mathbf{r}}.
\end{equation}

In this case, the products $V_{\mathbf{k}, \mathbf{k}'}^{(n,m)} V_{\mathbf{k}', \mathbf{k}}^{(m,m)}$ become:

\begin{equation}
    \begin{split}
    V_{\mathbf{k}, \mathbf{k}'}^{(n, m)} V_{\mathbf{k}', \mathbf{k}}^{(m, m)} &= \lambda(n) \frac{i e ^2v_\textrm{F}^2}{2} \times \\
    &\left[\left( \vert A_{\mathbf{k}, \mathbf{k}'}^{(x)} \vert ^2  -  \vert A_{\mathbf{k}, \mathbf{k}'}^{(y)} \vert ^2  \right) \left( \sin \theta' \cos\theta + \cos\theta' \sin\theta  \right) -  \right. \\
    &\left. 2A_{\mathbf{k}, \mathbf{k}'}^{(x)} A_{\mathbf{k}, \mathbf{k}'}^{(y)} \left( \cos \theta' \cos\theta  - \sin \theta' \sin \theta \right) \right].
    \end{split}
    \label{product1}
\end{equation}

This product  $V_{\mathbf{k}, \mathbf{k}'}^{(n, m)} V_{\mathbf{k}', \mathbf{k}}^{(m, m)}$  is antisymmetrical with respect to the interchange of the band indices. 

In the case of scalar scattering potentials, Eq. \eqref{scalar_product1}, where $V_{\mathbf{k}, \mathbf{k}'}^{(c,c)} V_{\mathbf{k}', \mathbf{k}}^{(c,v)} =V_{\mathbf{k}', \mathbf{k}}^{(v,v)} V_{\mathbf{k}, \mathbf{k}'}^{(v,c)}$, the subtraction in the curly brackets in Eq. \eqref{Sigma11} is close to zero at the CNP due to symmetry between electrons and holes. Conversely, for vector scattering potentials, $V_{\mathbf{k}, \mathbf{k}'}^{(c,c)} V_{\mathbf{k}', \mathbf{k}}^{(c,v)} = -V_{\mathbf{k}', \mathbf{k}}^{(v,v)} V_{\mathbf{k}, \mathbf{k}'}^{(v,c)}$, which leads to the flip of the sign of one of the terms in the curly brackets in Eq. \eqref{Sigma11}. Therefore, only scattering given by a vector scattering potential can significantly affect the interband transition probability, and hence the dynamic conductivity associated with such transitions, in the vicinity of the CNP. 

For a vector scattering potential, the product $ V_{\mathbf{k}, \mathbf{k}'}^{(n, m)} V_{\mathbf{k}', \mathbf{k}}^{(m, m)}$depends on a factor of either $\cos \theta'$ or $\sin \theta'$, see Eq. \eqref{product1}. The integrant in Eq. \eqref{Sigma11} contains terms either with $\cos \theta'$ or $\cos \theta$ . Consequently, terms in the curly brackets of Eq. \eqref{Sigma11} can be either proportional to $\cos\theta'$ or to $\cos^2\theta'$. When integrating over $\theta'$, the former case results in zero. Hence, Eq. \eqref{Sigma11} can be rewritten as:  

\begin{equation}
\begin{split}
    \Theta &= \Tilde{F}(\omega) \text{Im} \left[ \frac{\tau}{1-i \omega \tau } \right] \frac{L^2}{(2 \pi)^2} \frac{e}{\hbar^2} \times \\  
    &\left\{ \int\limits_0^{2\pi} d \theta'  V_{\theta,\theta'}^{(c,c)} V_{\theta',\theta}^{(c,v)}   \int\limits_0^{\infty} dk' k'    \frac{\partial_{k'} \Bar{f}_{c,k'}^{(0)} \cos \theta'}{\omega - \Delta \omega_{k, k'} + i\gamma'} \right.\\
    & \left. + \int\limits_0^{2\pi} d \theta' V_{\theta,\theta'}^{(c,v)} V_{\theta',\theta}^{(v,v)}   \int\limits_0^{\infty} dk' k'   \frac{\partial_{k'} \Bar{f}_{v,k'}^{(0)} \cos \theta'}{\omega - \Delta \omega_{k', k} + i\gamma'} \right\}
\end{split}
\label{Sigma1}
\end{equation}

Taking the symmetry of the scattering vector potential, Eq. \eqref{product1}, into account, Eq. \eqref{Sigma1} becomes:

\begin{equation}
\begin{split}
    \Theta &=  \frac{\Tilde{F}(\omega) L^2}{(2 \pi)^2} \frac{e}{\hbar^2} \text{Im}  \left[ \frac{\tau}{1-i \omega \tau } \right]  
    \int\limits_0^{2\pi} d \theta'  V_{\theta,\theta'}^{(c,c)} V_{\theta',\theta}^{(c,v)} \cos \theta'   \int\limits_0^{\infty} dk' k'    \frac{\partial_{k'} \Bar{f}_{c,k'}^{(0)} -\partial_{k'} \Bar{f}_{v,k'}^{(0)} }{\omega - \Delta \omega_{k, k'} + i\gamma'}
\end{split}
\label{Sigma2}
\end{equation}

Assuming $\gamma' \rightarrow 0$ \cite{Kitamura_2015, Haug_2004}, we obtain 

\begin{equation}
\lim_{\gamma\to0} \frac{i}{\left(\omega - \Delta \omega_{k,k'} + i\gamma  \right)} = iP\frac{1}{\omega - \Delta \omega_{k,k'}} + \delta \left( \omega - \Delta \omega_{k,k'} \right)
    \label{dirac}
\end{equation}
where $P$ denotes the principal value of an integral under which this relation is used. Only considering the real part \cite{Kitamura_2015, Haug_2004} Eq. \eqref{Sigma2} becomes:  

\begin{equation}
        \Theta =  \frac{e \pi k \Tilde{F}(\omega)}{v_F \hbar^2}  \text{Im} \left[ \frac{\tau}{1-i \omega \tau } \right] \left[ \partial_{k} \Bar{f}_{c,k}^{(0)} - \partial_{k} \Bar{f}_{v,k}^{(0)} \right]  \int\limits_0^{2\pi} d \theta'  V_{\theta,\theta'}^{(c,c)} V_{\theta',\theta}^{(c,v)} \cos \theta'      
    \label{Sigma}
\end{equation}

Substituting Eq. \eqref{Sigma} into Eq. \eqref{polarization}, and then Eq. \eqref{distr_func2} and Eq. \eqref{polarization} into Eq. \eqref{eq_current_density}, we obtain the following conductivity: 

\begin{equation}
    \Tilde{\sigma}(\omega)= \Tilde{\sigma}_\textrm{intra}\left( \omega \right)  + \Tilde\sigma_\textrm{inter}^{(\textrm{o})} \left( \omega \right) + \Tilde{\sigma}_\textrm{inter}^{(\textrm{s})}\left( \omega \right)
    \label{eq_sigma}
\end{equation}
where $\Tilde{\sigma}_\textrm{intra}\left( \omega \right)$ is the intraband dynamic conductivity, $\Tilde\sigma_\textrm{inter}^{(\textrm{o})} \left( \omega \right)$ is the direct (i.e., momentum-conserving) interband dynamic conductivity, and $\Tilde{\sigma}_\textrm{inter}^{(\textrm{s})}\left( \omega \right)$ is the indirect (i.e., non-momentum-conserving) scattering-assisted interband dynamic conductivity. The first two can be obtained by assuming that the incident THz waveform electric field is linearly polarised along the $x$-axis; using Eqs. \eqref{eq_current_density}-\eqref{v_me2} we get:

\begin{equation}
    \Tilde{\sigma}_\textrm{intra}(\omega) = 
    \frac{e^2 v_\textrm{F} }{2 \hbar L^2} \sum_{n,\mathbf{k}} \frac{\tau  \partial_{k}\Bar{f}_{n,k}^{(0)}}{1 - i\omega \tau}  \cos^2 \theta
\label{sbe1_sol2}
\end{equation}

\begin{equation}
    \Tilde{\sigma}_\textrm{inter}^{(\textrm{o})}(\omega)  = 
    \frac{ie^2 v_\textrm{F} }{2 \hbar  L^2} \sum_{\mathbf{k}, \eta=\pm1}  \frac{\eta\left(\Bar{f}_{c,k}^{(0)} - \Bar{f}_{v,k}^{(0)} \right)}{k \left(\omega - \eta \Delta \omega_{k} + i\gamma  \right)} \sin^2 \theta
    \label{curr_ter_o}
\end{equation}

\begin{equation}
\begin{split}
\Tilde{\sigma}_\textrm{inter}^{(\textrm{s})}(\omega) &= \frac{ie^2}{\hbar^2} \text{Im} \left[ \frac{\tau}{1 - i\omega \tau} \right] \\
&\times \sum_{\mathbf{k}, \eta=\pm1} \left[ \frac{\eta k\left( \partial_{k} \Bar{f}_{c,k}^{(0)} - \partial_{k} \Bar{f}_{v,k}^{(0)} \right)   }{\hbar\left(\omega - \eta \Delta \omega_k + i\gamma  \right)} \right] \int\limits_0^{2\pi} d \theta'  V_{\theta,\theta'}^{(c,c)} V_{\theta',\theta}^{(c,v)} \cos \theta'  \sin \theta 
\end{split}
\label{curr_ter_s}
\end{equation}


The integral over $\theta'$ in \eqref{curr_ter_s}  results in:

\begin{equation}
\Tilde{\sigma}_\textrm{inter}^{(\textrm{s})}(\omega) = \frac{ie^2v_\textrm{F}}{\hbar L^2} \frac{\Bar{v}^2 L^2}{\hbar v_\textrm{F}} \text{Im} \left[ \frac{\tau}{1 - i\omega \tau } \right] \sum_{\mathbf{k}, \eta=\pm1} \left[ \frac{\eta k\left( \partial_{k} \Bar{f}_{c,k}^{(0)} - \partial_{k} \Bar{f}_{v,k}^{(0)} \right)   }{\hbar\left(\omega - \eta \Delta \omega_k + i\gamma  \right)} \right]  \sin^2 \theta 
\label{curr_ter_s3}
\end{equation}

where
\begin{equation}
    \Bar{v}^2 = \frac{e^2 v_\textrm{F}^2}{2} \left[ \vert A_{\mathbf{k}=\mathbf{k}'=0}^{(x)}\vert ^2  -  \vert A_{\mathbf{k}=\mathbf{k}'=0}^{(y)} \vert ^2  \right].
    \label{v-bar}
\end{equation}

The quantity $\Bar{v}$ has units of energy and is associated with the spatially averaged square of the vector scattering potential of disordered graphene. This interpretation provides a connection with the autocorrelation function of the vector potential via the Wiener–Khinchin theorem \cite{hristopulos2020random}.

By using polar coordinates \cite{Haug_2004} and again transforming discrete sums over $\mathbf{k}$ into integrals:

\begin{subequations}
\begin{align}
    \Tilde{\sigma}_\textrm{intra}\left( \omega \right) &= \frac{e^2v_{\textrm{F}}}{4\pi\hbar} \int\limits_0^{\infty} d k  \frac{k \tau}{1 - i\omega \tau } \left( \partial_{k} \Bar{f}_{c,k}^{(0)} - \partial_{k} \Bar{f}_{v,k}^{(0)} \right) \\
    \Tilde\sigma_\textrm{inter}^{(\textrm{o})}(\omega) &= \frac{ie^2 v_\textrm{F}^2}{4 \pi \hbar} 
    \int\limits_0^{\infty} dk\, k \,  \frac{\left(\Bar{f}_{c,k}^{(0)} - \Bar{f}_{v,k}^{(0)} \right)}{ \Delta \omega_k^2 - \left(\omega  + i\gamma  \right)^2} \\
    \Tilde\sigma_\textrm{inter}^{(\textrm{s})}(\omega) &= \frac{ie^2 v_\textrm{F}^2}{4 \pi \hbar} 
    \int\limits_0^{\infty} dk\, k^2 \,  \frac{\Bar{v}^2 \cdot \Gamma(\omega, k, \tau) \cdot \left( \partial_{k} \Bar{f}_{c,k}^{(0)} - \partial_{k} \Bar{f}_{v,k}^{(0)} \right)}{ \Delta \omega_k^2 - \left(\omega  + i\gamma  \right)^2}
\end{align}
\label{eq_conductivities}
\end{subequations}

Here, 
\begin{equation}
\Gamma(\omega, k, \tau) = \frac{k L^2}{\hbar^2 v_\textrm{F}}\text{Im} \left[\frac{\tau}{1-i \omega\tau} \right] 
\end{equation}
is the self-energy associated with scattering-assisted interband transitions.

We used Eqs. \eqref{eq_conductivities}a-c to fit our experimental measurements of $\Tilde{\sigma}(\omega)$ (see Figs. 2, 3 of main text), with the Fermi-Dirac distribution $\Bar{f}_{n,k}^{(0)}$ at room temperature determined using the Fermi level $E_\textrm{f}$ obtained from four-point-probe measurements for different gate voltages $V_\textrm{g}$ (see SI Supplementary Note 1; we allowed for a 10\% variation of $E_\textrm{f}$ across different values of $V_\textrm{g}$ to obtain a best fit). We assumed $\tau(V_\textrm{g})=\alpha \sqrt{n(V_\textrm{g})}$ \cite{Das_2011}, where $n(V_\textrm{g})$ is the $V_\textrm{g}$-dependent carrier concentration determined via four-point-probe measurements. We used $\Bar{v}$ and $\alpha$ as global fit parameters (i.e., same for all $V_\textrm{g}$, allowing for a 10\% variation of $\alpha$ across different values of $V_\textrm{g}$), and $\gamma$ as a local fit parameter (i.e., varying as a function of $V_\textrm{g}$). Note that $\Tilde{\sigma}_\textrm{intra}\left( \omega \right)$ follows a Drude-like trend (see Fig. 2 of main text). 

Equation \eqref{eq_conductivities}a represents Drude-type intraband conductivity. The position of poles in Eq. \eqref{eq_conductivities}b coincides with the positions of poles for  Eq. \eqref{eq_conductivities}c. For both equations, they are located in the lower complex half-plane. Therefore, the expression of $\Tilde{\sigma}(\omega)$ given by Eq. \eqref{eq_conductivities}a-c satisfies the Kramers-Kronig relations. 

The scattering-assisted interband transitions in graphene are attributed to the presence of a vector scattering potential $\mathbf{A}$. The $\Gamma  \left(\omega, k, \tau\right)$ function describes the interplay between intra- and interband dynamics that result from the presence of such a vector scattering potential, and which cannot be associated with a scalar potential (see above). Such a coherent coupling between intra- and interband transitions is enabled in the linear regime by the vector scattering potential which hybridises states that possess different pseudo-spins. Note that this process is characteristic of semimetals and narrow-gap semiconductors, where interband transition energies are on the order of $h/\tau$. It has been established \cite{Kun_2019} that a vector scattering potential enables intravalley backscattering involving flipping of the electron pseudo-spin; such pseudo-spin-flipping intravalley backscattering is forbidden in graphene in which vector potentials are absent. Additionally, we have established that such processes involving a vector potential driving intravalley backscattering affect and hence can enhance effectively interband transitions. Note that the quantity $\Bar{v}^2 \Gamma$ is unitless, and can be considered as a vertex correction \cite{Bruus_2004, Rammer_2018} to the electron-photon coupling given by the scattering of electrons by the vector potential.

Now, where can such a vector scattering potential arise from in the specific case of graphene, in particular when no external magnetic field is applied? In the case of graphene, significant strain and structural deformations (e.g., folding, crumples, wrinkles, ripples, point defects), which can occur during growth or transfer \cite{BaiKeKe_2014, JeongSeonYu_2014, ChaeSeungJin_2009, Calado_2012, ChenJiaHao_2009}, can give rise to significant gauge fields $\mathbf{A}$, associated with pseudo-magnetic fields $\mathbf{B}=\nabla \times \mathbf{A}$ \cite{Rainis, Kun_2019}. That is, we hypothesise that such pseudo-magnetic fields in graphene, resulting from significant structural deformations, can affect $\Tilde{\sigma}(\omega)$ via scattering driven by $\mathbf{A}$.

\begin{figure}
    \centering
    \includegraphics[width=0.9\linewidth]{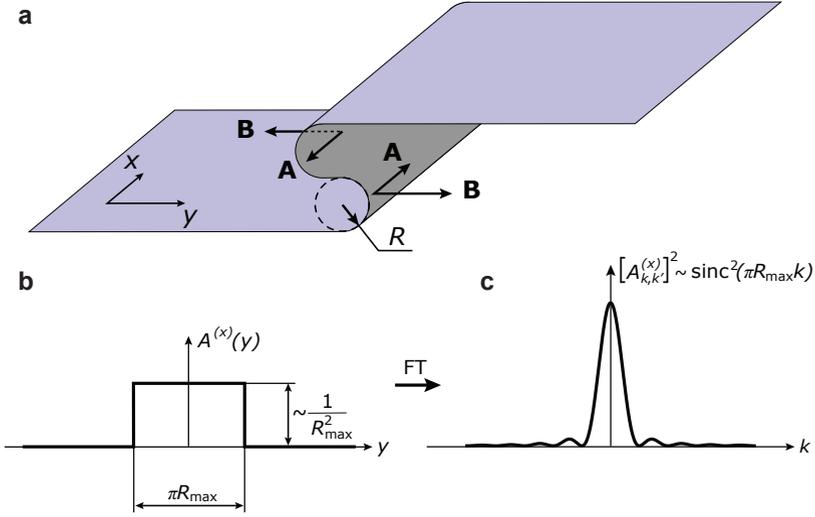}
    \caption{\textbf{Example of crumpled graphene area. a,} Schematic of crumpled area of graphene, with two folds with a radius of curvature ${\sim}R_{\textrm{max}}$. \textbf{b,} Magnitude of gauge field $\mathbf{A}(\mathbf{r})$ as a function of coordinate $y$: the gauge field is non-null and constant within a $y$-span of $\pi R_{\textrm{max}}$. \textbf{c,} Squared modulus of the Fourier decomposition of $A^{(x)} \left( y \right)$ as a function of wave vector $k$, proportional to $\mathrm{sinc}^{2}(\pi R_{\mathrm{max}} k)$.}
    \label{fig:foldings}
\end{figure}

As an example of such a structural deformation, here we consider the specific case of the effect of a crumpled area on graphene's $\Tilde{\sigma}(\omega)$. We model such a crumpled area as a pair of half-circular folds (see Fig. \ref{fig:foldings}a). Such folding can give rise to a gauge field $\mathbf{A}(\mathbf{r})=A^{(x)}(y) \: \mathbf{e}_x$, with \cite{Rainis, Kun_2019}:
 \begin{equation}
     A^{(x)} \left( y \right) = \frac{3 \epsilon_{\pi \pi} }{8ev_F} \frac{a^2}{R^2\left( y \right)},
 \end{equation}
where, for the sake of this example, we use parameters $\epsilon_{\pi \pi}=3$ eV, C-C bond length $a=1.42$ \AA, and a maximum fold radius of curvature $R(y) \approx R_{\textrm{max}} \approx$ 4 \AA\ \cite{Kim_2008, Annett2016, Rainis, Kun_2019}. In this case, $A^{(x)} \left( y \right) = 0$ except for $y$ coordinates associated with the location of the folds; at these locations, we assume that $A^{(x)} \left( y \right) \propto {1}/{R_{\textrm{max}}^2}$ is constant, for $y$ within a real-space span of $\pi R_{\textrm{max}}$ (see Fig. \ref{fig:foldings}b).

It is important to note that in real systems, $R(y)$ varies smoothly as a function of $y$; otherwise, $\mathbf{B}=\nabla \times \mathbf{A}$ is ill-defined. The assumption of a constant $R(y)$ at the fold is an approximation that simplifies computations of the scattering potential. We can estimate the maximum magnitude of $\mathbf{B}$, $B^{(z)}_{\textrm{max}}$ (along direction $z$ orthogonal to the graphene sheet), by assuming that $R(y)$ goes from $R_{\textrm{max}}$ to zero at the fold edge within one lattice constant $\sqrt{3}a$. By applying a finite difference approximation for $\nabla \times \mathbf{A}$, we obtain:


 \begin{equation}
     B^{(z)}_{\textrm{max}} = \frac{\sqrt{3} \epsilon_{\pi \pi} }{8 ev_F} \frac{a}{R_{\textrm{max}}^2} = 192 \text{ T}.
 \end{equation}

This estimated value of $B^{(z)}_{\textrm{max}}$ is consistent with previous studies on graphene \cite{Kun_2019}.

Assuming a random distribution of folds in our graphene sample, with folds aligned with equal probability along any axis, the vector potential Fourier components can be averaged over angles $\theta$ and  $\theta'$, and expressed (i.e., Taylor-expanded) around $\mathbf{k},\: \mathbf{k}'=0$ as follows:
\begin{equation}
    \begin{split}
            A_{\mathbf{k}, \mathbf{k}'}^{(x)}=&\frac{1}{L^2} \int d\mathbf{r} A^{(x)}(\mathbf{r}) e^{i(\mathbf{k}- \mathbf{k}')\mathbf{r}}=\\
            &\frac{3\pi^2}{4L^2} \frac{\epsilon_{\pi \pi} a^2 l}{R_{\textrm{max}}} \text{sinc}(\pi R_{\textrm{max}} \vert \mathbf{k} - \mathbf{k}' \vert) \approx \frac{3\pi^2}{4L^2} \frac{\epsilon_{\pi \pi} a^2 l}{R_{\textrm{max}}},
    \end{split}
\end{equation}
where $l$ is the overall effective fold length throughout the considered graphene area.

By substituting this expression in Eq. \eqref{v-bar}, we can establish a relationship between the effective length of folds in our graphene sample, $l$, and parameter $\Bar{v}$. From the fitting of our experimental $\Tilde{\sigma}(\omega)$ for charge-neutral graphene with the expression of $\Tilde{\sigma}(\omega)$ given by our two-component model [Eqs. \eqref{eq_conductivities}a-c], we estimated $\Bar{v} \approx 4.63 \cdot 10^{-6}$ eV, which gives us an estimate of $l \approx 8.28$ mm. We can also estimate $l$ experimentally via optical microscopy and Raman spectroscopy of our graphene sample (Fig. \ref{fig:Raman}), for an area of graphene of ${\sim}1$ mm$^2$  irradiated by the THz waveforms in our THz-TDS experiments; we obtain an experimental estimate of $l \approx 2.7$ mm. We assert that the qualitative agreement between these estimates of $l$ validates our two-component model, and that the latter provides a plausible physical explanation of our experiments. Importantly, note that this considered case of a graphene fold is an arbitrary example of a plausible cause for the vector scattering potential $\mathbf{A}$; as mentioned above, this vector scattering potential $\mathbf{A}$ could be the result of many other types of deformations or defects.

\begin{figure}[hbt!]
    \centering
    \includegraphics[width=0.6\linewidth]{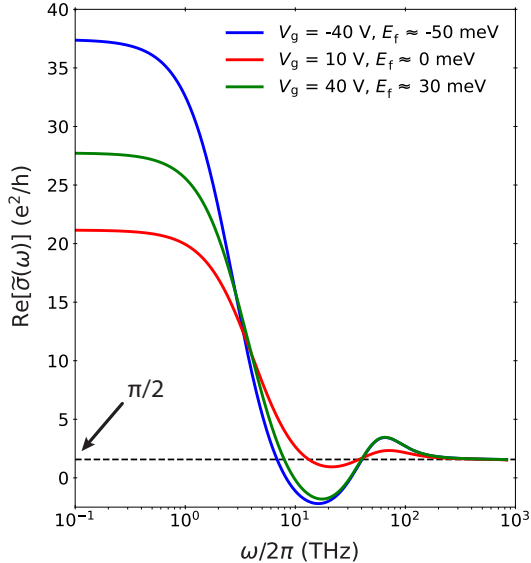}
    \caption{\textbf{Real part of two-component model dynamic conductivity: from sub-THz to optical frequencies.}  for different gate voltages ($V_\textrm{g}$). Curves calculated by using the two-component model Eqs. \ref{eq_conductivities}a-c, for different gate voltages $V_\textrm{g}$, for parameters $\tau$, $\Bar{v}$ and $\gamma$ obtained from fitting of experimental data. Regardless of $V_\textrm{g}$ values, Re$\left[ \Tilde{\sigma}(\omega) \right]$converges to constant optical value of $\pi e^2/2h$.}
    \label{fig:cond_optical}
\end{figure}

To provide further validation for the two-component model, we calculated  $\Tilde{\sigma}(\omega)$ using Eqs. \eqref{eq_conductivities}a-c with values of $\tau$, $\Bar{v}$ and $\gamma$ obtained by fitting our experimental data (see above), for $V_\mathrm{g} =$ -40, 10 and 40V, for sub-THz to optical (${\sim} 10^3$ THz) frequencies. Figure \ref{fig:cond_optical} shows the calculated Re$[\Tilde{\sigma}(\omega)]$. Independently of the Fermi level $E_{\mathrm{f}}$, Re$[\tilde{\sigma}(\omega)]$ converges to a constant value of $\pi e^2/2h$ for $\omega/2\pi \gtrsim 100$ THz, that is, in the optical region, where direct interband transitions dominate, consistent with previous studies \cite{Das_2011, Dawlaty_2008, Hwang_2007, MakKinFai_2008}. Note that this calculated Re$[\Tilde{\sigma}(\omega)]$ is also consistent with measurements in the mid- and far-IR \cite{Ren_2012, Horng_2011}. This further corroborates the validity and reliability of the two-component model in describing the complex dynamic conductivity of graphene for a broad range of doping levels and frequencies.

\section*{Supplementary Note 6. Contribution of scattering-assisted interband transitions to complex dynamic conductivity}

In the main text and Supplementary Note 4, we concluded that Drude-type models cannot fully describe the measured $\Tilde{\sigma}(\omega)$ in the $\sim$0.1 - 7 THz range for charge-neutral graphene. As discussed in the main text and above, the complex dynamic conductivity of graphene is given by contributions of intraband, $\Tilde{\sigma}_\textrm{intra}(\omega)$, and interband, $\Tilde{\sigma}_\textrm{inter}(\omega)$, electronic transitions. As outlined in the two-component model section above, the interband transitions include direct (momentum-conserving) transitions [with contribution $\Tilde{\sigma}_\textrm{inter}^{(\textrm{o})}(\omega)$] and scattering-assisted indirect transitions [involving momentum transfer, with contribution $\Tilde{\sigma}_\textrm{inter}^{(\textrm{s})}(\omega)$]; see Eqs. \eqref{eq_conductivities}a-c.

\begin{figure}[hbt!]
    \centering
    \includegraphics[width=1.0\linewidth]{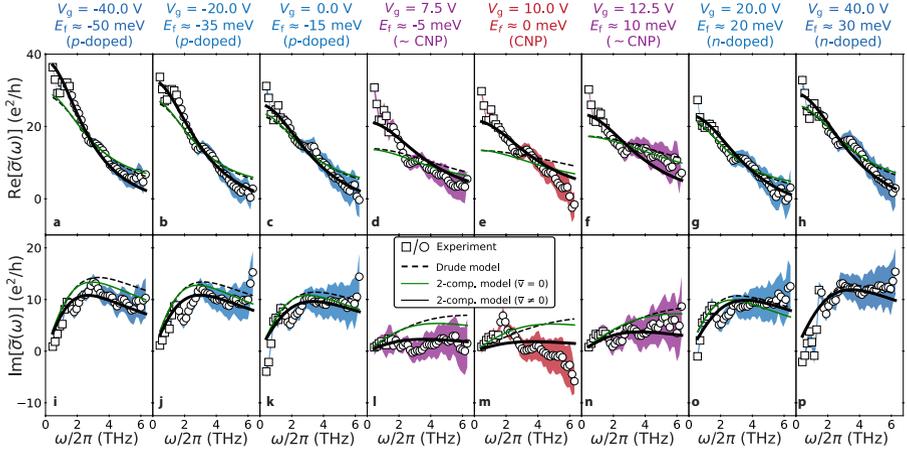}
    \caption{\textbf{Gate-dependent complex THz conductivity of graphene: comparison between Drude model and two-component model. a - h,} Real part of $\Tilde{\sigma}\left( \omega \right)$ for different $V_\textrm{g}$ values. \textbf{i - p, } Imaginary part of $\Tilde{\sigma}\left( \omega \right)$. Square and circle markers: experimental data from LiNbO$_3$ and GaP THz-generation configurations, respectively. Shaded areas: $\pm$ standard deviation. Dashed black curves: fit by conventional Drude model. Solid green curves: fit by two-component model without contribution of indirect scattering-assisted interband transitions ($\Bar{v} = 0$). Black curves: fit by complete two-component model with contribution of indirect scattering-assisted interband transitions ($\Bar{v} \neq 0$). Complete two-component model provides a better fit.
    }
    \label{fig:fittings_intra_inter}
\end{figure}

\begin{figure}[hbt!]
    \centering
    \includegraphics[width=0.75\linewidth]{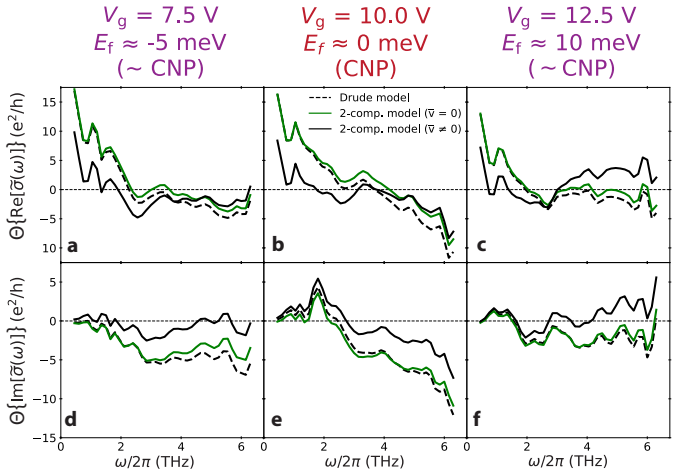}
    \caption{\textbf{Residuals of fits near the charge neutrality point. a-c,} Residuals of $\mathrm{Re}[\Tilde{\sigma}(\omega)]$ fits. \textbf{d-f,} Residuals of $\mathrm{Im}[\Tilde{\sigma}(\omega)]$ fits. Dashed black curves: fit by conventional Drude model. Solid green curves: fit by two-component model without contribution of indirect scattering-assisted interband transitions ($\Bar{v} = 0$). Black curves: fit by complete two-component model with contribution of indirect scattering-assisted interband transitions ($\Bar{v} \neq 0$).
    }
    \label{fig:residuals_intra_inter}
\end{figure}

\begin{figure}[hbt!]
    \centering
    \includegraphics[width=0.8\linewidth]{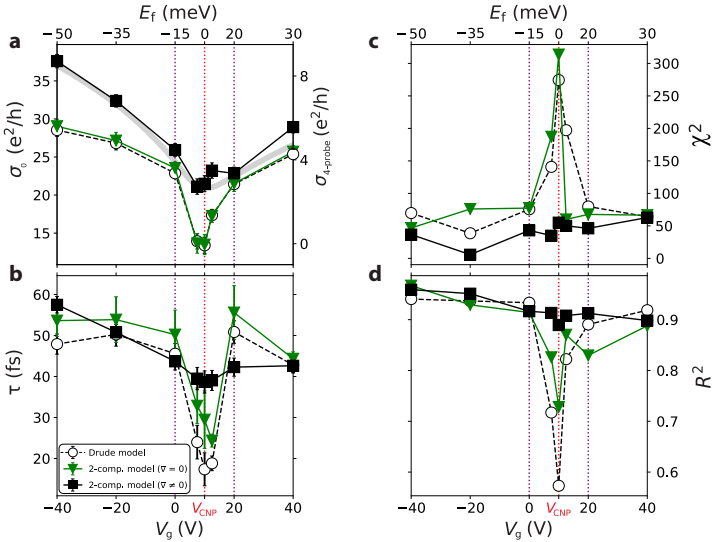}
    \caption{\textbf{DC conductivity, carrier relaxation time constant and goodness of fit.} \textbf{a,} DC conductivity $\sigma_0$, \textbf{b,} carrier relaxation time constant $\tau$, \textbf{c,} Pearson's $\chi^2$ test and \textbf{d,} coefficient of determination $R^2$ as a function of $V_\textrm{g}$ and $E_\textrm{f}$, determined from fitting the measured $\Tilde{\sigma}\left( \omega \right)$ with the conventional Drude (black circles) and two-component models (without and with contribution of the scattering-assisted interband transitions;  green triangles andblack squares, respectively).
    }
    \label{fig:fitting_results_intra_inter}
\end{figure}

Here we show that only accounting for direct interband transitions, i.e., $\Tilde\sigma_\textrm{inter}^{(\textrm{s})}(\omega)=0$ with $\Bar{v} = 0$, is not able to reproduce our experimental data. Figure \ref{fig:fittings_intra_inter} shows the experimental $\textrm{Re}\left[\Tilde{\sigma}\left( \omega \right) \right]$ and $\textrm{Im}\left[\Tilde{\sigma}\left( \omega \right) \right]$ for different gate voltages $V_\mathrm{g}$, as well as fit curves given by the conventional Drude model and the two-component model (with both $\Bar{v} = 0$ and $\Bar{v} \neq 0$, i.e., without and with contributions from scattering-assisted interband transitions, respectively). See above for details on fitting procedure. Residuals, DC conductivities $\sigma_0$, intraband relaxation time constants $\tau$, Pearson's $\chi^2$ test and coefficient of determination $R^2$ resulting from these fits are shown in Figs. \ref{fig:residuals_intra_inter}, \ref{fig:fitting_results_intra_inter}. 

For doped graphene, all considered models provide similar fits (see Fig. \ref{fig:fittings_intra_inter}a-c, g-h, i-k and o-p). For charge-neutral graphene (Fig. \ref{fig:fittings_intra_inter}d-f, l-n), the Drude model and the incomplete two-component model with $\Tilde{\sigma}_\textrm{inter}^{(\textrm{s})}(\omega) =0$ (i.e., $\Bar{v}=0$) cannot precisely capture both $\textrm{Re}\left[\Tilde{\sigma}\left( \omega \right) \right]$ and $\textrm{Im}\left[\Tilde{\sigma}\left( \omega \right) \right]$ simultaneously; only the complete two-component including indirect scattering-assisted interband transitions (i.e., $\Bar{v} \neq 0$ and $\Tilde{\sigma}_\textrm{inter}^{(\textrm{s})}(\omega) \neq 0$). This is evident from the larger fit residuals for the Drude and incomplete two-component models in comparison to the full two-component model (Fig. \ref{fig:residuals_intra_inter}), as well as from the values of Pearson's $\chi^2$ test and coefficient of determination $R^2$ near the CNP (Fig. \ref{fig:fitting_results_intra_inter}c-d).


\section*{Supplementary Note 7. Consistency of complex dynamic conductivity measurements}

\begin{figure}[hbt!]
    \centering
    \includegraphics[width=1.0\linewidth]{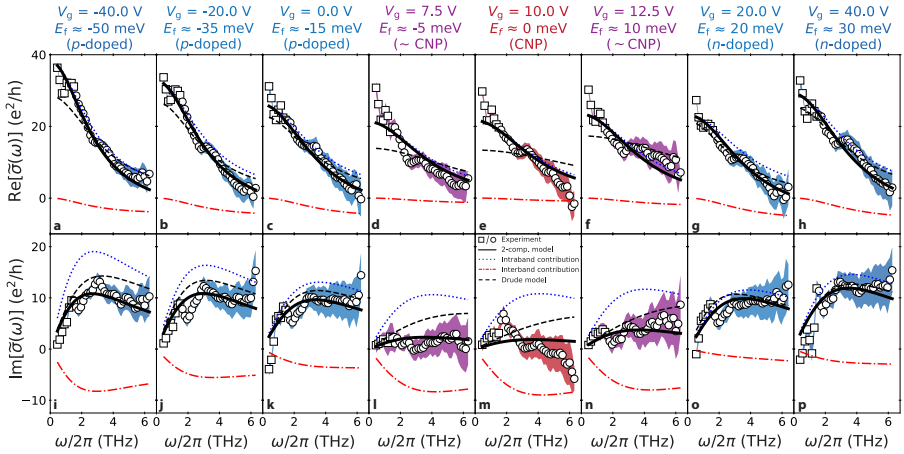}
    \caption{\textbf{Gate-dependent complex THz conductivity of graphene. a - h,} Real part of $\Tilde{\sigma}\left( \omega \right)$ for full $V_\textrm{g}$ range considered. \textbf{i - p}, Imaginary part of $\Tilde{\sigma}\left( \omega \right)$. Square and circle markers: experimental data from LiNbO$_3$ and GaP THz-generation configurations, respectively. Solid curve: fit by two-component model. Dotted (dashed-dotted) curve: intraband (interband) contribution. Dashed curve: fit by Drude model. Shaded areas: $\pm$ standard deviation. Panels \textbf{a, d-f, h-i, l-n, p} are shown in Fig. 2 of the main text. 
    }
    \label{fig:SI_sigma_fitting}
\end{figure}

In Fig. 2 of the main text, we show the complex dynamic conductivity of graphene $\Tilde{\sigma}\left( \omega \right)$, for five representative gate voltages $V_\textrm{g}$ and corresponding Fermi levels $E_\textrm{f}$. Figure \ref{fig:SI_sigma_fitting} shows $\Tilde{\sigma}\left( \omega \right)$ for the full $V_\textrm{g}$ range considered. The DC conductivity $\sigma_0$, electron intraband relaxation time constant $\tau$, Pearson's $\chi^2$ test and coefficient of determination $R^2$ resulting from the two-component model fit of these $\Tilde{\sigma}\left( \omega \right)$ data are shown in Figs. 3a-c of the main text. We then calculated parameters $\beta_{\textrm{intra}}$ and $\beta_{\textrm{inter}}$ (main text Fig. 3d) for gate voltages $V_\textrm{g}$ at which $\Tilde{\sigma}\left( \omega \right)$ in Fig. \ref{fig:SI_sigma_fitting} were acquired.

\begin{figure}
    \centering
    \includegraphics[width=1.0\linewidth]{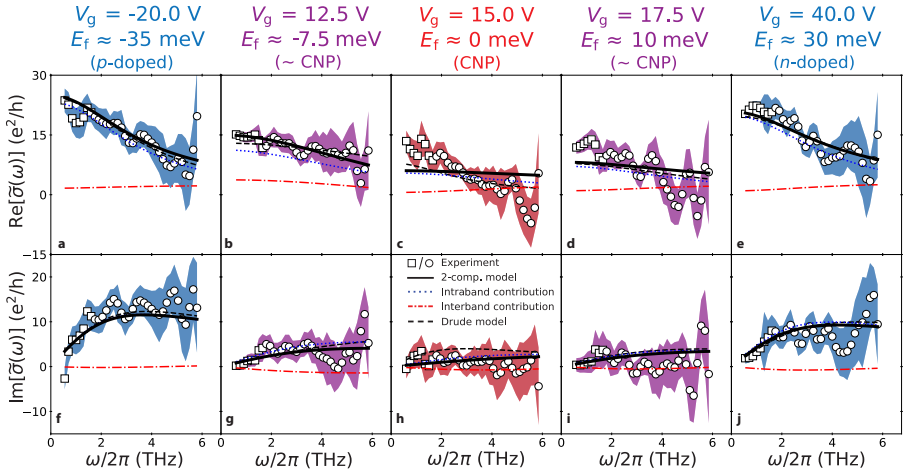}
    \caption{Same as Fig. \ref{fig:SI_sigma_fitting}, for additional device Dev. 2.
    }
    \label{fig:SI_sigma_Dev45}
\end{figure}

For consistency, we performed THz time-domain spectroscopy (THz-TDS) and retrieved $\Tilde{\sigma}\left( \omega \right)$ for other graphene/SiO$_2$/Si devices, in addition to the device for which data are reported in the main text. These additional devices 2 (Dev. 2) and 3 (Dev. 3) were fabricated by the same method, and showed similar carrier mobilities (${\sim} 2000$ cm$^2$/Vs), as device 1 (Dev. 1) of the main text (see Methods). As for Dev. 1 in Fig. 2 of the main text, we observed for these Devs. 2 and 3 the suppression of $\mathrm{Im}[\Tilde{\sigma}\left( \omega \right)]$ when $E_\textrm{f} \approx 0$ (Figs. \ref{fig:SI_sigma_Dev45}, \ref{fig:SI_sigma_Dev29}), with the two-component model providing a better fit of the experimental data for charge-neutral graphene (significantly smaller $\chi^2$ and larger $R^2$ in Fig. \ref{fig:SI_Fit_Results}) than Drude-type models . Therefore, we conclude that our experimental observations and their explanation by the two-component model shed light on general physical properties of gate-controlled graphene, that are not specific to a particular device.

\begin{figure}[hbt!]
    \centering
    \includegraphics[width=1.0\linewidth]{SI_Fig/SI_Fig_Conductivity_Dev29_6.pdf}
    \caption{Same as Fig. \ref{fig:SI_sigma_fitting}, for Dev.3.
    }
    \label{fig:SI_sigma_Dev29}
\end{figure}

\begin{figure}[hbt!]
    \centering
    \includegraphics[width=0.68\linewidth]{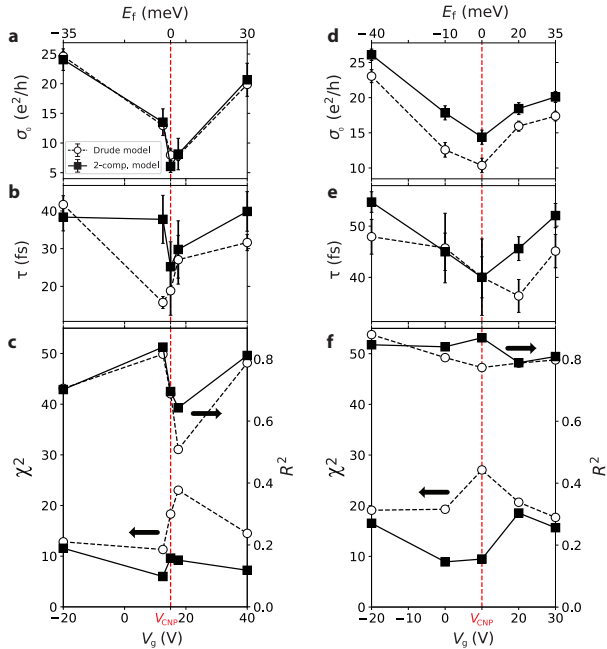}
    \caption{\textbf{Two-component fit vs. Drude fit for devices Dev. 2 and 3. a, b,} DC conductivity $\sigma_0$ and intraband relaxation time constant $\tau$ as a function of $V_\textrm{g}$ for Dev. 2, determined from fitting experimental $\Tilde{\sigma}\left( \omega \right)$ (Fig. \ref{fig:SI_sigma_Dev45}) with $\Tilde{\sigma}_\textrm{Drude}\left( \omega \right)$ (circles) and $\Tilde{\sigma}_\textrm{2-comp.}\left( \omega \right)$ (squares). Error bars: $\pm$ standard deviation of fits. \textbf{c, } Pearson's $\chi^2$ test (left-axis) and coefficient of determination $R^2$ (right-axis) as a function of $V_\textrm{g}$, for Drude (circles) and two-component (squares) model fits. \textbf{d - f,} Same as (a - c), for Dev. 3 (Fig. \ref{fig:SI_sigma_Dev29}). Near the charge neutrality point (i.e., $V_\textrm{g} \approx V_\textrm{CNP})$, the two-component model yields better fits, with smaller $\chi^2$ and larger $R^2$.
    }
    \label{fig:SI_Fit_Results}
\end{figure}

\section*{Supplementary Note 8. Complex dynamic conductivity retrieved with LiNbO$_3$ THz-generation configuration: direct vs. 1st-reflection retrieval}

In the ${\sim}0.1 - 1.5$ THz range, with THz waveforms generated with the LiNbO$_3$ configuration, the graphene complex dynamic conductivity $\Tilde{\sigma}(\omega)$ can be retrieved by measuring either the directly transmitted or the first-reflected THz transient, as discussed in the Methods of the main text. This section compares these two retrieval methods.

\begin{figure}
    \centering
    \includegraphics[width=0.7\linewidth]{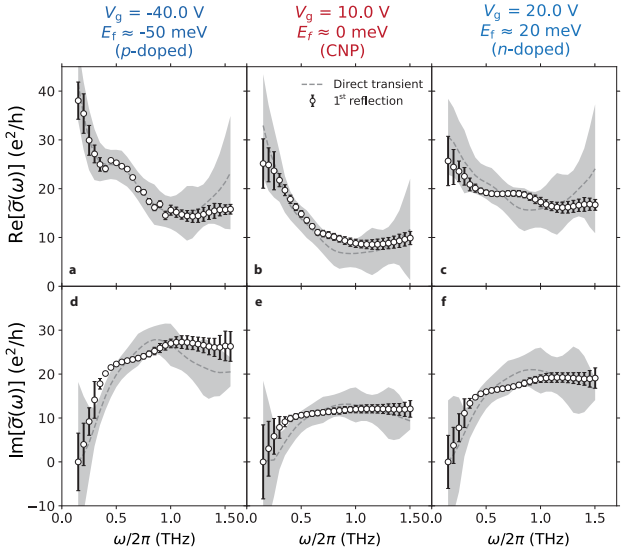}
    \caption{\textbf{Complex dynamic conductivity in the $\sim$0.1 - 1.5 THz range: direct vs. 1st-reflection retrieval.} \textbf{a-c,} Re$\left[\Tilde{\sigma}(\omega)\right]$ and \textbf{d-f,} Im$\left[\Tilde{\sigma}(\omega)\right]$ of graphene for different gate voltages $V_\textrm{g}$ and corresponding Fermi levels $E_\textrm{f}$, retrieved via LiNbO$_3$ THz-generation configuration measuring either the directly transmitted (grey dashed curve) or first-reflected (white circles) THz transient. Grey shaded areas and error bars: $\pm$ standard deviation. The two retrieval techniques result in comparable values of $\Tilde{\sigma}(\omega)$, with 1st-reflection technique showing significantly smaller standard deviation.} 
    \label{fig_retrieval}
\end{figure}

Figure \ref{fig_retrieval} shows the real and imaginary parts of $\Tilde{\sigma}(\omega)$ in the frequency range of ${\sim} 0.1 - 1.5$ THz for different gate voltages $V_\textrm{g}$ and Fermi levels $E_\textrm{f}$ ($p$-doped to charge-neutral to $n$-doped graphene), retrieved via the LiNbO$_3$ THz-generation configuration (see SI Supplementary Note 2).

The values of Re$\left[\Tilde{\sigma}(\omega)\right]$ and Im$\left[\Tilde{\sigma}(\omega)\right]$ retrieved from both methods are very comparable, with values obtained from the directly transmitted THz transient falling within the standard deviation of $\Tilde{\sigma}(\omega)$ retrieved via the first-reflection method. Notably, the standard deviation of $\Tilde{\sigma}(\omega)$ retrieved via the first-reflection technique is significantly smaller than that obtained from measuring the directly transmitted THz waveform, within the full ${\sim}0.1 - 1.5$ THz range and for all considered gate voltages. Consequently, the retrieval based on the first-reflection THz transient was chosen for determining $\Tilde{\sigma}(\omega)$ within the ${\sim} 0.1 - 1.5$ THz range, as discussed in the Methods section.

\newpage
\bibliography{bib}